\documentclass[twocolumn,english,aps,amsmath,amssymb,reprint,nofootinbib,floatfix,showkeys]{revtex4-2}

\usepackage[table]{xcolor}
\usepackage{graphicx}
\usepackage{dcolumn}
\usepackage{bm}
\usepackage[colorlinks = true,citecolor=blue,linkcolor = blue,urlcolor=blue]{hyperref}

\usepackage[capitalise]{cleveref}
\usepackage{graphicx}	
\usepackage{amsmath}	
\usepackage{amssymb}	
\usepackage{wasysym}    
\usepackage{mathtools}  
\usepackage{aas_macros}
\usepackage{array,multirow,graphicx}  
\usepackage{multirow}

\newcommand{\Hunit}{km\,s$^{-1}$\,Mpc$^{-1}$}

\renewcommand{\Cref}[1]{\cref{#1}}

\begin{document}

\title{Measuring the baryon fraction using galaxy clustering}





\author{Alex Krolewski}
 \email{alex.krolewski@uwaterloo.ca}
 \author{Will J. Percival}%
  \affiliation{Waterloo Centre for Astrophysics, University of Waterloo, Waterloo, ON N2L 3G1, Canada}
  \affiliation{Department of Physics and Astronomy, University of Waterloo, Waterloo, ON N2L 3G1, Canada}
  \affiliation{Perimeter Institute for Theoretical Physics, 31 Caroline St. North, Waterloo, ON NL2 2Y5, Canada}

\date{\today}

\begin{abstract}
The amplitude of the baryon signature in galaxy clustering depends on the cosmological baryon fraction. 
We consider two ways to isolate this signal in galaxy redshift surveys. First, we extend standard template-based Baryon Acoustic Oscillation (BAO) models to include the amplitude of the baryonic signature by splitting the transfer function into baryon and cold dark matter components with freely varying proportions.
Second, we include the amplitude of the split as an extra parameter in Effective Field Theory of Large Scale Structure (EFT) models of the full galaxy clustering signal. We find similar results from both approaches. For the Baryon Oscillation Spectroscopic Survey (BOSS) data we find $f_b\equiv\Omega_b/(\Omega_b + \Omega_c)=0.173\pm0.027$ for template fits post-reconstruction, $f_b=0.153\pm0.029$ for template fits pre-reconstruction, and $f_b=0.154\pm0.022$ for EFT fits, with an estimated systematic error of 0.013 for all three methods. Using reconstruction only produces a marginal improvement for these measurements. Although significantly weaker than constraints on $f_b$ from the Cosmic Microwave Background, these measurements rely on very simple physics and, in particular, are independent of the sound horizon. In a companion paper we show how they can be used, together with Big Bang Nucleosynthesis measurements of the physical baryon density and geometrical measurements of the matter density from the Alcock-Paczynski effect, to constrain the Hubble parameter. While the constraints on $H_0$ based on density measurements from BOSS are relatively weak, measurements from DESI and Euclid will lead to errors on $H_0$ that can differentiate the cosmological and distance ladder $H_0$ measurements.
\end{abstract}

\maketitle


\section{Introduction \label{sec:intro}}

The flat $\Lambda$CDM model of cosmology is now constrained to such high precision \cite{Planck:2020} that it is not the values of the parameters that are of primary interest, but instead the consistency and details of the physical processes that lead to the observations. Thus, the most interesting questions in cosmology today are resolving tensions such as the Hubble tension and understanding the physics of dark matter and dark energy. The Hubble tension refers to the difference between recent measurements of the Hubble constant \cite{Riess-Hubble} using a local distant ladder formed from Cepheids and supernovae, which gives the measurement $H_{0}=73.04\pm1.04$\,\Hunit. This is in strong tension with the recent Planck measurement $H_{0}=67.37\pm0.54$\,\Hunit\ from the Cosmic Microwave Background \cite{Planck:2020}. Likewise, the Baryon Acoustic Oscillations (BAO) standard ruler measures a low value of $H_{0}=67.35\pm0.97$\,\Hunit\ when calibrating the sound horizon used as a standard ruler with Big Bang Nucleosynthesis (BBN) observations \citep{Cooke2016}.
This difference is not currently understood: it either points to new physics, such as Early Dark Energy \cite{Hill20}, or to an unknown systematic problem with one or more sets of data. New methods to measure the Hubble constant are important to help to distinguish between these options.

In a companion paper \cite{krolewski}, we introduce a new method to measure the Hubble constant without including either CMB power spectra or calibrating the BAO standard ruler with a known value of the sound horizon. 
This allows us to test modifications to the early expansion history such as Early Dark Energy, which change the sound horizon.
The method is based on comparing the physical matter density $\Omega_mh^2$ obtained from measurements of cosmological abundances against geometrical measurements, which depend on the ratio of the matter density to the critical value $\Omega_m$. Although the CMB provides a strong constraint on the physical density, the desire to solve the Hubble tension makes the importance of independent measurements paramount. We therefore consider measuring $\Omega_mh^2$ by combining BBN constraints on $\Omega_bh^2$ with a measurement of $\Omega_b/\Omega_m$ obtained from the amplitude of baryonic features in the galaxy power spectrum. Note that we make a distinction between the BAO and the signature of baryons, which includes both the BAO and the change in shape of the power spectrum caused by the presence of baryons \cite{Eisenstein-Hu}.  We use the full baryon signature to extract the baryon fraction from a galaxy redshift survey. The measurements are used in our companion paper to set constraints on the Hubble parameter. 

The signature of baryons was first seen in the 2-degree Field Galaxy Redshift Survey (2dFGRS), using early data \cite{Percival01}. The analysis relied on using both the change in shape of the power due to baryons and the BAO, which both provide information about the baryon fraction. The BAO became clearer as more galaxy redshifts were obtained by the 2dFGRS \cite{cole05} and the Sloan Digital Sky Survey (SDSS) \cite{Eisenstein05}. The detection of BAO from galaxy redshift surveys passed the 5$\sigma$ threshold with the Baryon Oscillation Spectroscopic Survey (BOSS) \cite{Anderson12,Alam17}. For galaxy redshift surveys, the baryon signature is one of the cleanest physical signals. Traditionally (e.g. \cite{Alam-eBOSS:2021}), BAO observations have focused on measuring the projected BAO positions along and across the line of sight, which depend on $D_H/r_d$ and $D_M/r_d$ respectively, where $D_H(z) \equiv c/H(z)$; $D_M(z) \equiv (1+z) D_A(z)$ is the comoving angular diameter distance $D_\mathrm{M}(z)$; and $r_d$ is the comoving sound horizon at the baryon drag epoch. The sound horizon $r_d$ has a complicated dependence on the baryon density, matter density and Hubble parameter, but it is extremely well-predicted given our knowledge of cosmological parameters and assuming the standard cosmological model. 
When comparing BAO at different redshifts (or to the CMB) the relative positions probe ratios of $D_A(z)$ and $H(z)$ removing the sound horizon dependence. BAO are either measured using a template fit or as a component of full model fits to the 2-point clustering signal. We now consider the extension of both methods to include measuring the baryon fraction.


The clustering of baryons and dark matter are traditionally described using transfer functions. These give the distribution of perturbations as a function of size and time, for any energy-density component in the universe. Fluctuations in the matter field can be described by the combined transfer function $T(k)$, combining baryon $T_b(k)$ and dark matter $T_c(k)$ transfer functions,
\begin{equation}  \label{eq:transfer}
    T(k)=\frac{\Omega_b}{\Omega_{bc}}T_b(k)+\frac{\Omega_c}{\Omega_{bc}}T_c(k)\,,
\end{equation}
weighted by their normalised average energy densities (where $\Omega_{bc}$ refers to the matter density excluding massive neutrinos). The dependence on the baryon fraction in Eq.~\ref{eq:transfer} has a very simple physical meaning, and offers the chance to measure this quantity without relying on complicated physical models. 
Performing a full fit with galaxy clustering data and extracting the constraint on $\Omega_b/\Omega_m$ from all physical processes that contribute to the clustering signal, would potentially bring in information from the location of the BAO feature, complicating the interpretation of the measurement. We instead parameterise the balance between transfer functions, isolating the split between energy-density components in this parameter, and marginalize over all other parameters required to model the power spectrum or correlation function.

\begin{figure*}
    \centering
    \includegraphics[scale=0.53]{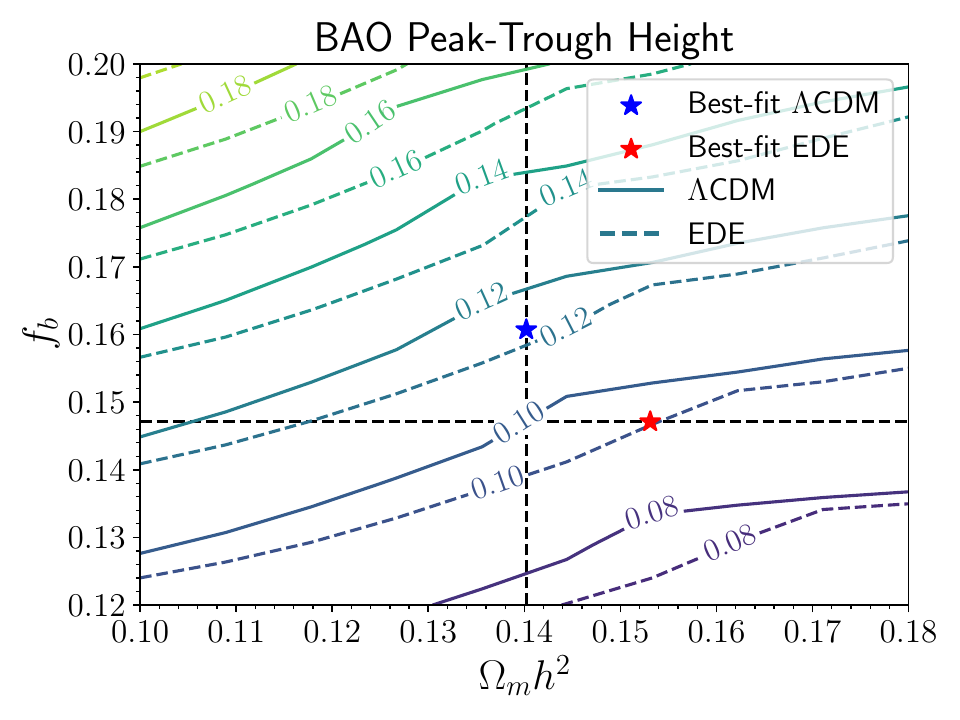}
    \includegraphics[scale=0.53]{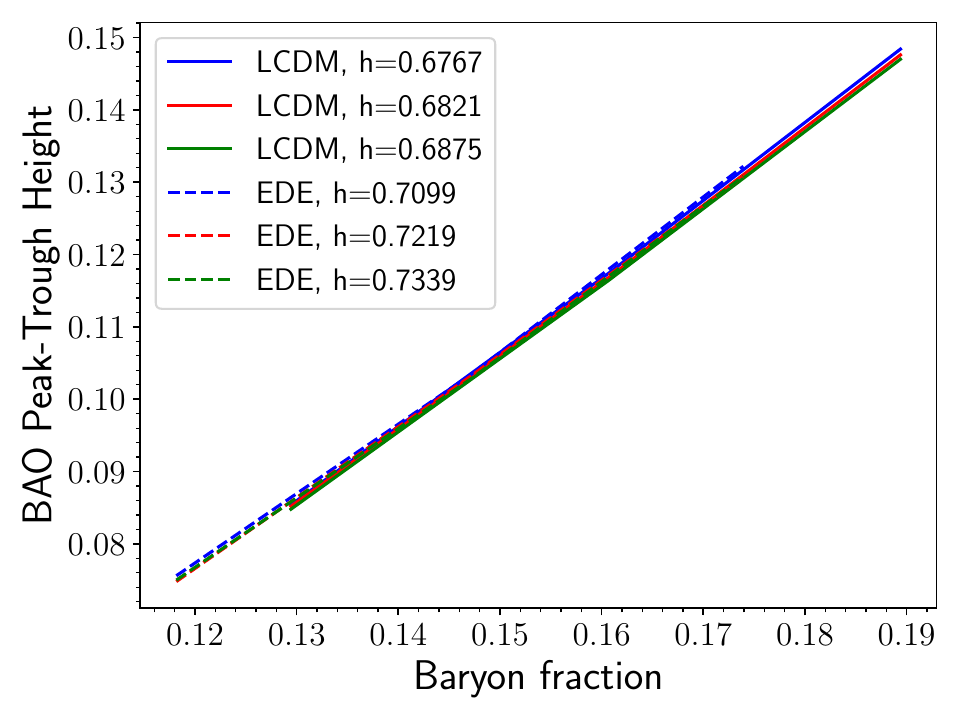}
    \caption{The BAO amplitude can measure the baryon fraction independent of whether the true underlying cosmology is $\Lambda$CDM or Early Dark Energy. \textit{Left:} Contours of the fractional first peak-to-trough height in the linear power spectrum as a function of the baryon fraction $f_b$ and the matter density $\Omega_m h^2$ (following Fig.~5 in \cite{Eisenstein-Hu}), in both a $\Lambda$CDM (solid) and Early Dark Energy (dashed) cosmology. 
    If the universe truly follows an EDE cosmology (red star, best-fit from \cite{Hill20}), we should expect a peak-to-trough height of 0.10. In our method, we fit the data with a $\Lambda$CDM model with $\Lambda$CDM parameters (best fit from \cite{Hill20}) shown with the blue star. 
    The value of $f_b$ that we would infer from our method is the intersection of the best-fit $\Lambda$CDM $\Omega_m h^2$ (vertical line) and the $\Lambda$CDM (solid) 0.10 contour. This lies exactly at the correct $f_b$ of the EDE model (horizontal dashed line).
    \textit{Right:} A further demonstration of this point, plotting the baryon fraction vs.\ the peak-trough height for $\Lambda$CDM and EDE cosmologies with a variety of different values of $H_0$ around the best-fit. Again the relationship is nearly identical, showing that the baryon fraction can be measured independently of whether $\Lambda$CDM or EDE is the correct cosmological model at early times.
    \label{fig:lcdm_vs_ede_schematic}}
\end{figure*}

The transfer functions are initially very different for baryons and dark matter, a consequence of the different behaviour before photon decoupling. The post-decoupling baryon transfer function can be modelled
\begin{equation}
    T_b(k)=\alpha_b\frac{\sin(kr_d)}{kr_d}\cal{D}(k)\,,
\end{equation}
in the limit $kr_d>>1$ \cite{Eisenstein-Hu}.
$\cal{D}(k)$ represents the effects of Silk damping, and $\alpha_b$ is a suppression factor arising from the adiabatic decrease in the sound speed, and the growth suppression between matter-radiation equality and the drag epoch before velocities decay due to cosmological expansion (see equations 14 \& 15 in \cite{Eisenstein-Hu}). 
For cosmological models close to the Planck best-fit, $\alpha_b$ 
is primarily determined by the baryon fraction, though with a small secondary dependence on $\Omega_mh^2$. We discuss this further in Section~\ref{sec:cosmo_degeneracies}.

Crucially, in this paper we find that the amplitude of the baryon signature can be recovered independently of the presence of Early Dark Energy. Fig.~\ref{fig:lcdm_vs_ede_schematic} shows this to be the case using phenomenological measurements from the linear power spectrum. Following Fig.~5 in \cite{Eisenstein-Hu}, we measure the fractional change in $P(k)/P_{\textrm{sm}}(k)$ between the first BAO peak and first trough as a function of the baryon fraction and $\Omega_m h^2$. If Early Dark Energy were the correct model of the universe, the contours of peak-trough height would be shifted (solid to dashed contours in Fig.~\ref{fig:lcdm_vs_ede_schematic}). However, if we analyze the data with a $\Lambda$CDM model (which has a different best-fit $\Omega_m h^2$ from EDE) the observed peak-trough height maps to the correct baryon fraction for the EDE model. This is clearly shown in the right panel, where we vary $\Omega_b h^2$ at fixed $\Omega_{cdm} h^2$ and $h$, and show that the relationship between the baryon fraction and BAO peak-trough height is the same in either $\Lambda$CDM or EDE.

As the Universe evolves, the baryon and dark matter fields both grow through gravity driven by the combined density of both components. While the transfer functions thus become more similar, the overall growth is scale-independent in the matter dominated regime without baryon velocity effects, so only the amplitude of the summed transfer function $T(k)$ changes, not the shape. The baryon fraction is expected to be constant in redshift in standard models. Thus, when we fit $T(k)$ to data to measure the relative amplitude of baryonic and dark matter components, there is a choice in the $T_b(k)$ \&~$T_c(k)$ we use in the fit: we can select these at any redshift, adjusting the amplitude, and still use Eq.~\ref{eq:transfer}. We therefore have an opportunity to optimise their difference and increase the fidelity with which we can measure $\Omega_b/\Omega_m$. Most of the signal constraining $\Omega_b/\Omega_m$ will come from the amplitude of the BAO in the combined transfer function, which enter through $T_b(k)$, compared to the smooth broad-band signal that predominantly comes from the CDM transfer function $T_c(k)$. However, there is also some information from the different broad-band shapes of $T_b(k)$ \&~$T_c(k)$. This is discussed further in Section~\ref{sec:gamma-B}. In general, we measure $f_b$ with larger statistical and systematic errors than the BAO position. The systematic errors arise because of degeneracies with other features in the transfer functions controlled by nuisance parameters with poorly-constrained posteriors. They will all be reduced by the inclusion of better data.

To compare with the measurement of the physical matter density, we also need a geometrical measurement of $\Omega_m$. This can be obtained from the Alcock-Paczynski effect (AP; \cite{Alcock:1979}). Consider a spherical object (or scale) that is sufficiently large that it expands with the expansion of the Universe. Along the line of sight to the object, we will measure a change in redshift $\Delta z$ from front to back, and across the line-of-sight we will measure an angular extent $\Delta\theta$. If we know that the object is spherical (or isotropic), we can translate these measurements into a cosmological constraint on $D_H/D_M$. Such a ratio of distances does not depend on the Hubble parameter, and for a flat $\Lambda$CDM model only depends on $\Omega_m$. We can use the BAO feature to provide the isotropic "object", or we can also use voids \cite{Lavaux:2012}. In general, these measurement will be correlated with the constraints on $f_b$, and so we will have to measure and fit these simultaneously.

After introducing the method, we fit to data from the Sloan Digital Sky Survey (SDSS; \cite{York00}) Baryon Oscillation Spectroscopic Survey (BOSS; \cite{Dawson13}). We consider fits to the data both with and without reconstruction \cite{eisenstein07}, a technique designed to sharpen the BAO features by moving overdensity in such a way as to remove non-linear motions. Using the correlation functions presented in \cite{Ross2017}, we are able to extract the baryon fraction with a systematic error that is larger than the BAO position, but still well within 1$\sigma$. 

We discuss extracting the amplitude of the baryonic signature in Section~\ref{sec:BAO-amp}, introducing new methods for full fits and template fits in Sections~\ref{sec:full-fits} \&~\ref{sec:template-fits} respectively. These are tested against mock catalogues in Section~\ref{sec:mock-tests} with the results and a comparison of methods presented in Section~\ref{sec:mock-results}. In Section~\ref{sec:baofit_data}, we describe the galaxy redshift surveys we use and present the results of the fits. Finally we present our conclusions in Section~\ref{sec:conclusions}.

\section{Measuring the baryon fraction}
\label{sec:BAO-amp}

\subsection{A parameterized power spectrum}
\label{sec:gamma-B}

In order to measure the baryon fraction we need to make some changes to the standard model for the power spectrum. We isolate the relative contributions of the component transfer functions by including an extra parameter $\gamma_\text{B}$. $\gamma_\text{B}$
is an estimator for the baryon fraction, $f_b$, as it is
designed to mimic the effect of changing the baryon fraction when combining baryon and dark matter transfer functions. 
Note that, because galaxy clustering responds to the power spectrum of baryons and CDM only \cite{Villaescusa-Navarro14}, $\gamma_B$ is an estimator of the baryon to baryon-plus-CDM density, excluding neutrinos, although the difference is negligible at the current and future precision expected on $\gamma_{\text{B}}$.

$\gamma_\text{B}$ is thus substituted into the linear transfer function (Eq.~\ref{eq:transfer}) and enters into the linear power spectrum
\begin{multline}  \label{eq:gamma_bao}
    P_{\textrm{lin},\gamma}(k) = A_s\left(\frac{kh}{0.05}\right)^{n_s-1}  h^4 \frac{k}{2\pi^2} 
    \big[\gamma_\text{B}T_b'(k) \\
        +(1-\gamma_\text{B})T_c'(k)\big]^2 D(z)^2\,.
\end{multline}
where both $k$ and the linear power spectrum are in units of $h$\,Mpc$^{-1}$ and $h^{-3}$\,Mpc$^{3}$, respectively.

We need to determine the functions $T_b'$ and $T_c'$ to use in this expression with care. 
For example, let's suppose that we model $T_b'$ and $T_c'$ as the low redshift baryon and CDM transfer functions that one would obtain from a Boltzmann code such as CLASS \citep{CLASS} or CAMB \citep{CAMB,Howlett12}. 
The transfer functions include the growth of perturbations, which means that the BAO oscillations in $T_b(z=0)$ are suppressed. They are also introduced into the CDM transfer function due to the growth of structure, as the growth depends on both components. Thus, the low redshift baryon and CDM transfer functions are nearly identical, and will not work to split the baryonic and dark matter features.
We instead need to use functions that maximise the differential features between the two transfer functions that can be used to measure the relative components, such as only having the oscillations in one of the terms.

We cannot simply use the transfer functions at very high redshift, for example at the drag epoch $z_d \sim 1060$, because the acoustic oscillations do not immediately stop travelling at $z_d$, but rather keep propagating outwards until $z\sim600$, though at much reduced speed.\footnote{As seen in the animations at \url{https://scholar.harvard.edu/files/deisenstein/files/acoustic_anim.gif}, \url{https://lweb.cfa.harvard.edu/~deisenst/acousticpeak/acoustic_physics.html}} Consequently, at this epoch, the BAO have a different shape and position than at the lower redshifts that we want to model. An alternative approach would be to use the transfer functions at $z\sim 600$ when the baryon pulse finally stops travelling. However, by this epoch, the dark matter has already begun to clump around the baryon perturbations, polluting the CDM transfer function with a bit of the BAO signal, again diluting the oscillatory signal---the fit would not be optimal in terms of the information differentiating $T_b'$ and $T_c'$.

Rather than using transfer functions directly from CLASS or CAMB at a particular epoch, we use the fitting function from Eisenstein and Hu \citep{Eisenstein-Hu} (section 3.1) for $T_c'$. We choose to use their CDM transfer function (their equation~17) rather than the zero-baryon transfer function (their equation~29). We define the baryon transfer function as the (appropriately scaled) difference between the density weighted sum of the baryon and CDM $z=0$ transfer functions from CLASS and the Eisenstein and Hu transfer function $T_c'$
\begin{equation}
    T_b^{'}(k) = \left(T_{cb}^{\textrm{CLASS}}(k)
     - \frac{\Omega_c}{\Omega_c + \Omega_b} T_{c}^{\rm EH}(k)\right) \frac{\Omega_c + \Omega_b}{\Omega_b}\,.
\end{equation}
While this choice is clearly better than using $z = 0$ or high redshift transfer functions, it may not be the optimal choice, as it is limited by numerical inaccuracies in the Eisenstein and Hu fitting formulas. Improving these fitting formulas could improve the recovery of $\gamma_B$ and is left for future work.

For our template fits, we use the transfer functions and densities for the fiducial cosmology at which the template is created. 
For our full fits we apply this equation to form the linear model used as the starting point for the perturbative calculations for each cosmology tested (hence $T_b'$ and $T_c'$ become cosmology dependent). The low-redshift baryonic signature is retained in $T_b'$, but with a split in signal between $T_b'$ and $T_c'$ as if we had taken high-redshift transfer functions where the two transfer functions are maximally different. This relies on the shape of the summed transfer function remaining invariant with time, as expected in the linear regime. We expect any deviations from this to be small. 

As shown in Fig.~\ref{fig:transfers}, this choice means that $T_c'$ and $T_b'$ differ in both their overall shape and the presence of oscillations in $T_b'$. Thus, their separation captures two physical effects: the oscillations from the baryon-photon fluid, and the suppression of power on scales smaller than the sound horizon. This is because dark matter fluctuations can begin to grow at equality, whereas baryon fluctuations continue to oscillate until the pressure drops to zero at the drag epoch. In contrast, if we had set $T_c'$ to the zero-baryon transfer function $T_0$ instead of $T_c$ in \cite{Eisenstein-Hu}, the $T_c'$-$T_b'$ split would only capture the oscillations, not the effect of baryons on the shape. This distinction doesn't matter much for the template-based fits for the BAO position, but we achieve tighter constraints (and less degeneracy with the $\Lambda$CDM cosmological parameters) in the full-shape fits including the baryon amplitude with this approach (see Section~\ref{sec:full-fits}).

Our analysis assumes that the bias of the galaxies that we observe is the same when comparing the galaxy density field to either the baryon or dark matter fields, such that we only have a single set of bias parameters relating galaxy and matter fields. This is common to all full-shape fits to date. In practice, the differential velocity and (more importantly) density of the baryon field compared to the dark matter field at early times could lead to a difference in the bias with these fields $b_{\delta_{bc}}$, $b_{v_{bc}}$ \cite{Barkana11,Schmidt16}, dependent on the physics of galaxy formation. We should expect such a bias to be scale-dependent and it could be removed in an analysis such as ours by including the appropriate template \cite{ChenHowlett24}. However, as there is currently no evidence that such a bias has an appreciable effect on the galaxies analysed and no preferred model for such a bias, we do not include it in our current analysis.

\begin{figure*}
    \centering

    \includegraphics[scale=0.6]{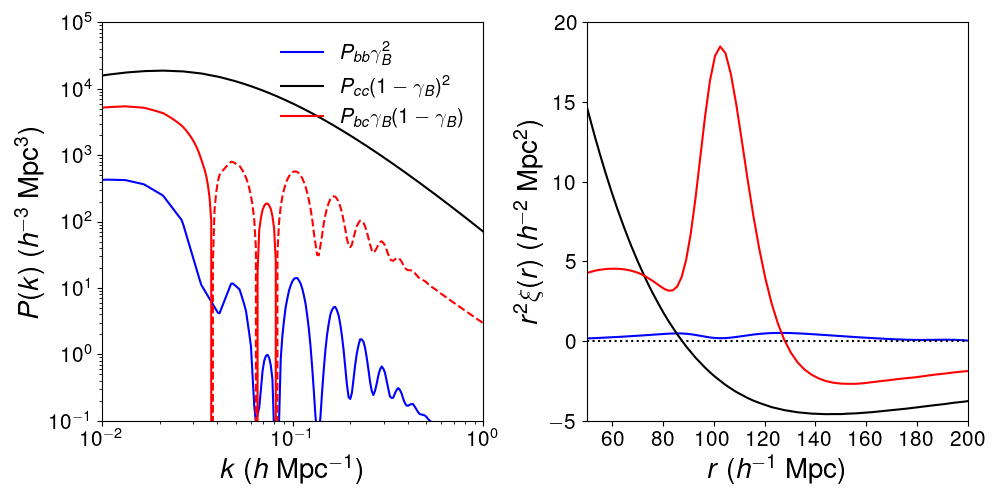}
    \caption{Power spectra and correlation functions of the templates. Dotted lines indicate a negative power spectrum. The dominant oscillatory component arises from the baryon-cold dark matter cross power spectrum.}
    \label{fig:my_label}
\end{figure*}

\begin{figure}
\includegraphics[scale=0.5]{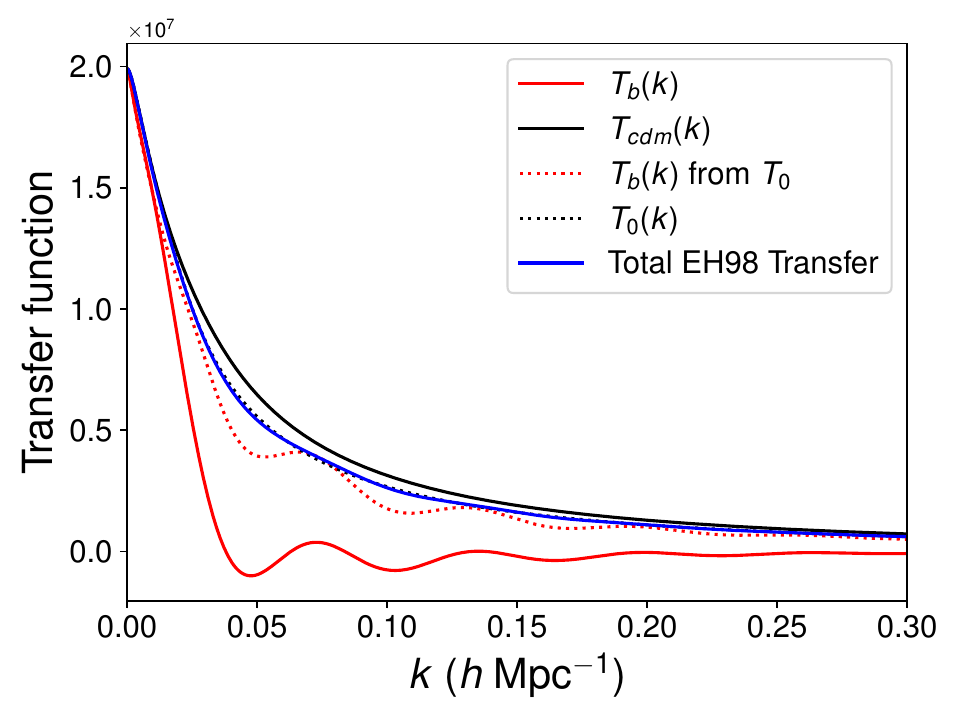}
\caption{Transfer functions from \cite{Eisenstein-Hu}. We show our fiducial choice (smooth transfer function from \cite{Eisenstein-Hu} $T_c'$, their Eq.~17, and $T_b'$ subsequently generated to match the overall power spectrum) in solid black and red lines. In this choice, $T_b'$ contains both the oscillatory component and the overall shape suppression from baryons suppressing the early growth of structure on small-scale modes that are within the horizon before recombination. An alternative transfer function would be $T_0$ (\cite{Eisenstein-Hu} Eq.~29; black dotted line) which matches the shape of the total \cite{Eisenstein-Hu} transfer function (solid blue line). However this choice leads to weaker constraints on the baryon fraction that are more degenerate with the other cosmological parameters.
\label{fig:transfers}}
\end{figure}

\subsection{Full-shape fits} \label{sec:full-fits}

We first consider extracting the information on $\gamma_\text{B}$ as part of a fit of a full model designed to model the shape of the power spectrum as well as the BAO component. Using Eq.~\ref{eq:gamma_bao} to model the linear power spectrum, we include $\gamma_\text{B}$ as an extra parameter within full-shape $\Lambda$CDM EFTofLSS \cite{Baumann12,Carrasco12} fits, extending the \textsc{CLASS-PT} code \cite{Philcox_Ivanov,Chudaykin20,Ivanov19,Philcox20,Philcox21,Ivanov21}.\footnote{We modify the publicly available likelihood, \url{https://github.com/oliverphilcox/full_shape_likelihoods}, which calls the CLASS-PT code itself \url{https://github.com/Michalychforever/CLASS-PT}.}
The power spectrum is modelled using one-loop perturbation theory supplemented by ultraviolet counterterms to account for short-scale non-local physics, a symmetry-based bias expansion, and infrared resummation to correctly capture long-wavelength modes and the shape of the BAO. \textsc{CLASS-PT} is similar to the methods of \cite{Chen21,DAmico20} and was shown to give consistent results in a blinded mock challenge \cite{Nishimichi20}. Treating the transfer function split and the non-linear correction separately is reasonable as we expect the transition from linear to non-linear to depend only on the matter (joint CDM+baryon) distribution.

Schematically, the power spectrum multipoles are the sums of several terms
\begin{multline}
P_\ell(k) = P_\ell^{\textrm{tree}}(k) + P_\ell^{\textrm{1-loop}}(k) + P_\ell^{\textrm{counterterm}}(k) \\ + P_\ell^{\textrm{stochastic}}(k)
\end{multline}
where the tree-level power spectrum is the linear power spectrum with IR resummation for the BAO, the 1-loop power spectrum is an integral over products of linear power spectra, the counterterms scale as $k^2 P_{\textrm{lin}}(k)$ and the stochastic terms account for shot noise and scale as constants plus $k^2$ and $k^2 \mu^2$ corrections.

We modify \textsc{CLASS-PT} to insert $\gamma_\text{B}$ into the linear power spectrum multipoles using Eq.~\ref{eq:gamma_bao} before the 1-loop integrals. \textsc{CLASS-PT} performs its own wiggle-no-wiggle decomposition of the power spectrum, by sine transforming the power spectrum and removing coefficients with frequencies within the BAO range \citep{Chudaykin20}. This is necessary for IR resummation (i.e.\ exponentially damping the wiggle portion of the power spectrum), which is required to accurately model the BAO \citep{Blas16}. However, this split is done at the power spectrum level, whereas we prefer to split the baryon signal at the transfer function level; hence we do not use the built-in wiggle-no-wiggle split to model the BAO amplitude.

The model contains nuisance parameters accounting for galaxy bias ($b_1$, linear; $b_2$, quadratic; $b_{\mathcal{G}_2}$, tidal; $b_{\Gamma_3}$, third-order), counterterms ($c_0$, monopole; $c_2$; quadrupole; $c_4$, hexadecapole; $\tilde{c}$, Fingers-of-God), and stochastic terms ($P_{\textrm{shot}}$, $a_0$, $a_2$) \cite{Ivanov22}. The stochastic terms are equal to
\begin{multline}
    P_{\textrm{stochastic}}(k) =\\ \frac{1}{\bar{n}} \left[1 + P_{\textrm{shot}} + a_0 \left(\frac{k}{k_{\textrm{NL}}}\right)^2 + a_2 \mu^2 \left(\frac{k}{k_{\textrm{NL}}}\right)^2 \right]
\end{multline}
We adopt the following priors:
\begin{equation}
b_1 \in \textrm{flat}[0,4]
\end{equation}
\begin{equation}
b_2 \sim \mathcal{N}(0, 1^2)
\end{equation}
\begin{equation}
b_{\mathcal{G}_2} \sim \mathcal{N}(0, 1^2)
\end{equation}
\begin{equation}
b_{\Gamma_3} \sim \mathcal{N}\left(\frac{23}{42}(b_1-1), 1^2\right)
\end{equation}
\begin{equation}
\frac{c_0}{(\textrm{Mpc/$h$})^2} \sim \mathcal{N}(0, 15^2)
\end{equation}
\begin{equation}
\frac{c_2}{(\textrm{Mpc/$h$})^2} \sim \mathcal{N}(30, 15^2)
\end{equation}
\begin{equation}
\frac{c_4}{(\textrm{Mpc/$h$})^2} \sim \mathcal{N}(0, 15^2)
\end{equation}
\begin{equation}
\frac{\tilde{c}}{(\textrm{Mpc/$h$})^4} \sim \mathcal{N}(500, 250^2)
\end{equation}
\begin{equation}
P_{\textrm{shot}} \sim \mathcal{N}(0, 0.5^2)
\end{equation}
\begin{equation}
a_0 \sim \mathcal{N}(0, 0.5^2)
\end{equation}
\begin{equation}
a_2 \sim \mathcal{N}(0, 0.5^2)
\end{equation}
These priors are inspired by those used in \cite{Ivanov22} and \cite{Philcox_Ivanov}, but not identical; in particular, the prior widths on the counterterms and stochastic parameters have been halved
compared to the priors in \cite{Ivanov22} and \cite{Philcox_Ivanov}.\footnote{Note that the stochastic priors in the publicly available likelihood \url{https://github.com/oliverphilcox/full_shape_likelihoods} have half the width of the priors reported in Eq.~12 of \cite{Philcox_Ivanov}.}
We modify the priors to reduce the dependence on the prior width and improve parameter recovery for the tests described in Section~\ref{sec:nbody_tests_full_shape}. The resulting narrower priors are more similar to the more restrictive ``West-Coast'' (WC) priors of \cite{DAmico20b} (see \cite{Simon23} for a comprehensive discussion of the impact of nuisance parameter priors on $\Lambda$CDM parameter constraints), especially for $c_0$ and $a_2$. Translating the WC priors into the \textsc{CLASS-PT} parameter set \cite{DAmico21,Simon23,Ivanov22}, we find they correspond to priors of $\mathcal{N}(0,12.97)$ on $c_0$, $\mathcal{N}(0, 46.2)$ on $c_2$, $\mathcal{N}(0, 56.0)$ on $c_4$, fixing $\tilde{c}$ to zero, $\mathcal{N}(-1,2^2)$ on $P_{\textrm{shot}}$, fixing $a_0$ to zero, and $\mathcal{N}(0, 0.64^2)$ on $a_2$.

We assume a Gaussian likelihood, with the covariance computed
from the power spectrum multipoles and post-reconstruction Alcock-Paczynski parameters measured on 999 Patchy mocks, using the publicly released covariances of \citep{Philcox20}. 
Parameters entering the model linearly ($b_{\Gamma_3}$, $c_0$, $c_2$, $c_4$, $\tilde{c}$, $P_{\textrm{shot}}$, $a_0$, and $a_2$) are analytically marginalized \cite{Philcox21b}. The model is stretched by the distance ratio (Hubble parameter or angular diameter distance) between distances in the sample cosmology and the BOSS fiducial cosmology.

We vary and marginalize over the $\Lambda$CDM parameters controlling the shape of the power spectrum [$h$, $\ln(10^{10}A_s)$, $\Omega_m$, $n_s$]. We impose flat and uninformative priors on all of these parameters, and fix $\omega_b = 0.02237$, the mean value from Planck alone from \cite{Planck:2020}. We consider all of these parameters to be nuisance parameters, as we do not want any information about the full shape of the galaxy power spectrum. We then add three additional parameters $\omega_b^{\textrm{dens}}$, $\Omega_m^{\textrm{dens}}$ and $h^{\textrm{dens}}$, using the superscript ``dens'' to indicate that the information comes from our density-based cosmological split. These parameters control the AP ratio $\alpha_{\perp}/\alpha_{\parallel}$ ($\Omega_m^{\textrm{dens}}$), the amplitude of the baryonic signature ($\gamma_\text{B} = \omega_b^{\textrm{dens}}/(\Omega_m^{\textrm{dens}} (h^{\textrm{dens}})^2)$), and a simple Gaussian BBN prior centered at the true value of $\omega_b = 0.023$ for the Nseries mocks (see Sec.~\ref{sec:mock-tests} for a full description of these mocks) with width 0.00037. 
When there is no ambiguity, the quoted value of $h$ refers $h^{\textrm{dens}}$ unless stated differently.
The neutrino mass is fixed, either to zero (for Nseries) or to the minimal value, $\Sigma_{m_\nu} = 0.06$ eV, on data.

\subsection{Template fits} \label{sec:template-fits}

Standard BAO extraction methods using the correlation function or power spectrum multipole moments fit a model with terms that are designed to allow marginalization over broad-band shape and amplitude information. The idea is that we wish to isolate the BAO signal without having to model the broadband shape (allowing for a less model-dependent fit than the full-shape method). 
To add the baryon signature parameter, we follow 
a similar path to that adopted for full fits as discussed in the previous section, and simply replace the linear power spectrum in existing templates with Eq.~\ref{eq:gamma_bao}. For example, consider the model used by \cite{Ross2017} to isolate the BAO signal:
\begin{equation}
    \xi_{0, {\rm mod}}(s) = B_0\xi_{0}(s,\alpha_\perp,\alpha_\parallel) + A_{0}(s) 
\end{equation}
\begin{multline}
    \xi_{2, {\rm mod}}(s) = \frac{5}{2}\left(B_2\xi_{\mu2}(s,\alpha_{\perp},\alpha_{||}) - B_0\xi_0(s,\alpha_{\perp},\alpha_{||})\right) \\ + A_{2}(s)\,.
\end{multline}
The parameters $B_i$ sets the overall amplitude of the monopole and quadrupole (i.e.\ the combination of galaxy bias and power spectrum amplitude), while 
\begin{equation}  \label{eq:poly-terms}
    A_i(s) = \frac{a_{i,1}}{s^2}+\frac{a_{i,2}}{s}+a_{i,3}\,,
\end{equation}
is designed to isolate the BAO signal from broadband shape changes by fitting and removing any changes in the shape of the template. 
The bias parameters $B_i$ and $a_{i,j}$, commonly called the polynomial parameters, are treated as nuisance parameters in the fit, allowing the model to:
\begin{enumerate}
  \item fit smooth differences between template and data allowing for non-linear, bias and redshift space distortion (RSD) effects,
  \item fit smooth changes in clustering caused by observational systematics, such as the effects of stellar density,
  \item fit any effects from reconstruction when the model is used to fit post-reconstruction data,
  \item for different $\alpha$, ensure that the smooth template (with potentially the wrong fiducial cosmology) is still able to fit the dilated clustering statistics, ensuring that the $\alpha$ are only constrained by the BAO.
\end{enumerate} 
We expect that these parameters will remove some of the smooth signal differentiating between baryon and dark matter clustering components.

In the standard BAO model \citep[e.g.][referred to as SBRS after its authors]{SBRS}, $\xi_0$ and $\xi_2$ are linear models calculated by transforming the anisotropic power spectrum
\begin{equation}
    P(k, \mu) = C^2(k, \mu) \left[\Delta P(k) D^2(k, \mu) + P_{\textrm{smooth}}(k)\right]
    \label{eqn:standard_model_pkmu}
\end{equation}
where $C(k, \mu)$ includes the Kaiser + Finger of God RSD term \cite{Kaiser}, $D(k, \mu)$ is the anisotropic damping, $P_{\textrm{smooth}}(k)$ is the smooth power spectrum model (in our case, calculated using the fitting formulae of \citep{Eisenstein-Hu}), and $\Delta P(k)$ is the difference between the linear power spectrum calculated using CAMB \cite{CAMB,Howlett12} and the smooth power spectrum. Eq.~\ref{eqn:standard_model_pkmu} is then integrated over $\mu$ (weighted by Legendre polynomials) to calculate the power spectrum multipoles, Hankel transformed to real space, shifted by $\alpha_{\parallel}$ and $\alpha_{\perp}$, and fitted to data.

\begin{figure*}
    \includegraphics[scale=0.6]{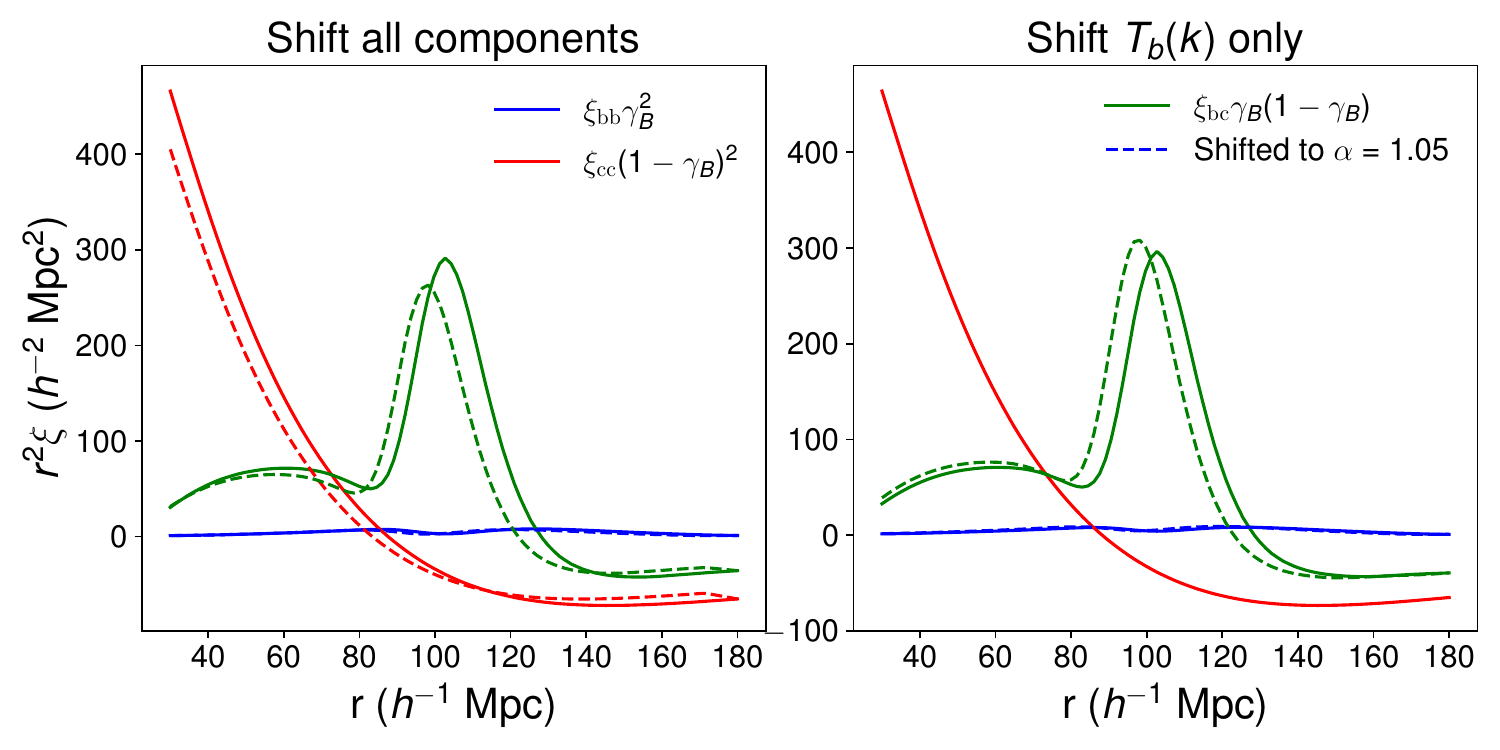}
\caption{\textit{Left}: all components of the model are shifted by $\alpha = 1.05$ (except the polynomial terms, which are not shown). This is the default model used in previous work. \textit{Right}: only $T_b(k)$ is shifted
by $\alpha$. Note that shifting only the transfer function $T_b(k)$ is not the same as shifting the wiggly component of the power spectrum, because
$P_w(k)$ comes from the product of $T_b(k)$ (which is shifted) and $T_c(k)$ (which is not).
\label{fig:shifting}}
\end{figure*}

In order to incorporate the new parameter $\gamma_\text{B}$ introduced in Section~\ref{sec:gamma-B} in this model, we make the following changes. 

\begin{enumerate}

\item At the heart of the new model is using Eq.~\ref{eq:gamma_bao} when we calculate $P(k)$, giving a template that depends on $\gamma_\text{B}$. As in the standard template-based method, the terms do not need to be calculated for every model to be tested, but we need to store them in component form - three components corresponding to $T_b'T_b'$, $T_c'T_c'$ and $T_b'T_c'$, to allow for variation of $\gamma_\text{B}$. 

\item We make use of the split of the linear power spectrum into transfer functions to isolate the BAO scaling, applying the scaling parameters $\alpha_{||}$ and $\alpha_\perp$ to $T_b'$ only. These therefore scale both the BAO and baryon shape change in $T_b'$, and consequently the physical interpretation is complicated, although we expect the nuisance broadband and polynomial terms to reduce the theory dependency such that it is close to $r_d$ as expected for the BAO. In our companion paper \cite{krolewski}, we do not use the derived $\alpha$'s, instead using the results from the literature.

\item  We add two new broadband scaling parameters, $q_{||,BB}$ and $q_{\perp,BB}$, which are distinct from $\alpha_{||}$ and $\alpha_\perp$. As Fig.~\ref{fig:shifting} shows,
$\alpha_{||}$ and $\alpha_\perp$ only shift $T_b'$, and all other functions of $k$
are shifted by $q_{||,BB}$ and $q_{\perp,BB}$.
This shifting must necessarily be performed in Fourier space, since the two transfer functions in the cross-term $T_b^{'}(k) T_c^{'}(k)$ are shifted by different parameters. These broadband scaling parameters are only sensitive to the distance scale. 

\item $\gamma_\text{B}$ is correlated with the broad-band polynomial terms, and we have to change these terms to accommodate this. We find that, in particular, the $1/s^2$ term is quite degenerate with the BAO amplitude. For the pre-reconstruction fits, we require a constant term for both the monopole and quadrupole, whereas for the post-reconstruction fits, we require a constant for only the monopole. These constant terms are motivated by the shape of the difference between the weighted and unweighted correlation functions (Fig.\ 6 in \cite{Ross2017}), and by changes in the goodness of fit when adding polynomials in fits to the data.
Future work will benefit from the new broadband marginalization introduced in \cite{ChenHowlett24}.

\item We use the beyond-linear model of \cite{Ding18} for pre-reconstruction and post-reconstruction templates, with a more sophisticated treatment of the post-reconstruction damping and $k^2$-dependent derivative bias (essentially, an EFT counterterm) to account for nonlinearities. Including the derivative bias allows us to marginalize over degeneracies between the BAO amplitude and nonlinear or higher-bias terms (which also scale as $k^2 P_{\textrm{lin}}$), although in an approximate way compared to the full EFT fits.

\item 
We vary the damping parameters:
$\Sigma_{fog}$ for the Finger of God term; and $\Sigma_{xy}$ ($\Sigma_{sm}$) for the nonlinear BAO damping pre-reconstruction (post-reconstruction).
We show how these parameters affect $P(k)$ below, in Eqs.~\ref{eqn:csq_prerecon} and~\ref{eqn:cnw_postrecon} for pre- and post-reconstruction.
The post-reconstruction nonlinear damping is complex, and we vary a single parameter $\Sigma_{sm}$, the effective smoothing scale from reconstruction, to control the damping
(\Cref{eqn:sigma_sm,eqn:sigma_dd,eqn:sigma_ds,eqn:sigma_ss}). 
The pre-reconstruction nonlinear damping is simpler, following a Gaussian form controlled by a single parameter $\Sigma_{xy}$ (Eq.~\ref{eqn:prerecon_damping}); the parameters of the model used by the BOSS team \cite{SBRS} are $\Sigma_\perp = \Sigma_{xy}$, $\Sigma_{||} = (1 + f) \Sigma_{xy}$ where $f$ is the growth rate.



\end{enumerate}

We then proceed in the usual fashion: define the multipole moments $P_\ell(k)$; Hankel transform to $\xi_\ell(s)$, and fit $\xi_0$ and $\xi_2$ (noting the rather convoluted definition of $B_2$). We elaborate on the reasons for these changes in the sections below. We also carefully show how changing each assumption changes the results on the BOSS data away from the results found in \cite{Ross2017}. We use the Nseries mocks to test the effect of the changes and whether they bias results.






\subsubsection{Pre-reconstruction template fits}

The pre-reconstruction beyond-linear ``EFT1'' model of \cite{Ding18} is very similar to the standard model of SBRS \cite{SBRS}, except for the extra derivative bias $b_\partial$.
We use Eq.~\ref{eqn:standard_model_pkmu} with
\begin{multline}
    C^2(k, \mu) = \left( (b_1 + f \mu^2)^2 + b_\partial \frac{k^2}{k_L^2} (b_1 + f \mu^2)\right) \\ \times \frac{1}{(1 + k^2 \mu^2 \Sigma_{\textrm{fog}}^2 / 2)^2}
\label{eqn:csq_prerecon}
\end{multline}
and
\begin{multline}
    D^2(k, \mu) = \exp\big[-k^2(1-\mu^2)\Sigma_{xy}^2/2 \\ - k^2 \mu^2(1+f)^2\Sigma_{xy}^2/2\big]
    \label{eqn:prerecon_damping}
\end{multline}

We fix $k_L$ to 1\,$h$\,Mpc$^{-1}$. The damping term is very similar to the SBRS damping, except with $\Sigma_{||}$ and $\Sigma_{\perp}$ explicitly tied together through the factor of $1+f$. We allow $b_\partial$ to vary within a uniform prior between -250 and 250. We find that a wider prior on $b_\partial$ leads to poor recovery of the BAO amplitude and AP parameters when changing the input true cosmology (but fixing the fiducial template cosmology), but a narrower prior on $b_\partial$ biases fits on the $N$-body mocks.

\subsubsection{Post-reconstruction template fits}

As for our pre-reconstruction fits, we use the ``EFT1'' model of \cite{Ding18} as our fiducial model.
The post-reconstruction EFT1 model of \cite{Ding18} contains two key differences from the model of \cite{SBRS}. First, it models the power spectrum as the sum of contributions from the displaced galaxies, shifted randoms, and shifted randoms cross-correlated with displaced galaxies. Rather than imposing a single damping on $\Delta P_{L}$, the individual terms are separately damped. This effectively leads to two separate sets of damping parameters, transitioning from $\Sigma_{ds}$ on large scales where the $ds$ term dominates to $\Sigma_{dd}$ on small scales where $dd$ dominates. Second, the EFT1 model includes a derivative bias, in addition to the standard linear bias, to model nonlinearities and higher-order bias. That is, $b_1 \rightarrow b_1 + k^2 b_{\partial}$. This yields a term $\propto k^2 P_{L}(k)$, that is essentially identical to the counterterms in the EFT of Large-Scale structure (exactly equal for SPT \citep[e.~g.\ Eqn.\ 2.3 in][]{Chudaykin20}, and nearly equal for LPT, where the counterterm $\propto k^2 P_{\textrm{zel}}(k)$ \cite{Chen21}). This model is not truly perturbative in that it drops the higher-order bias and 1-loop terms that contribute at equivalent order, but the amplitude of the derivative bias can be tuned to approximately match these contributions.

As in the linear model, we start by splitting the power spectrum into wiggle and no-wiggle components. The wiggle component is multiplied by the standard Kaiser RSD term (reduced by the ``Rec-Iso'' reconstruction scheme, i.e.\ reconstruction nearly eliminates the quadrupole) times a Finger-of-God term. The new element here is the additional derivative bias to encode beyond-linear physics. We fix $k_L = 1$ $h$\,Mpc$^{-1}$. To conform to the notation of the standard method, we explicitly divide out all factors of the linear bias, $b_1$, and will later scale the monopole and quadrupole separately by $B_0$ and $B_2$. We therefore replace the divided $b_1$ by $B_0^{0.5}$. The Finger-of-God damping is only applied to the 
no-wiggle term, not to the oscillatory term, which is damped by $D(k,\mu)$.

\begin{multline}
    P(k, \mu) = \left[C_{\textrm{nw}}^2 + \frac{b_{\partial}}{B_0^{0.5}} \left(\frac{k}{k_L}\right)^2 \left(1 + \frac{f\mu^2}{B_0^{0.5}} (1 - S) \right)\right. \\ \left.\left(\frac{1}{1 + k^2 \mu^2 \Sigma_{\textrm{fog}}^2/2}\right)^2\right] P_{\textrm{nw}}(k) + C^2 \Delta P_{\textrm{lin}}(k) 
\end{multline}

\begin{equation}
    S = \exp{\left(-0.5 k^2 \Sigma_{\textrm{sm}}^2\right)}
    \label{eqn:sigma_sm}
\end{equation}

\begin{multline}
    C_{\textrm{nw}} = C_{\textrm{Kaiser}} C_{\textrm{fog}} \\= \left( 1 + \frac{f \mu^2}{B_0^{0.5}} (1 - S) \right) \frac{1}{1 + k^2 \mu^2 \Sigma_{\textrm{fog}}^2/2}
    \label{eqn:cnw_postrecon}
\end{multline}

The BAO term then consists of three pieces, all scaling the wiggle power spectrum, each with their own damping:

\begin{equation}
    C^2(k, \mu) = \delta P_{dd}(k,\mu) - 2 \delta P_{ds}(k,\mu) + \delta P_{ss}(k,\mu)
\end{equation}

\begin{multline}
    \delta P_{dd}(k,\mu) = \exp\bigg(-k^2 (1 + (2 + f) f \mu^2) \Sigma_{dd}^2\bigg) \\ \left(1 - \frac{S}{B_0^{0.5}} + \frac{f \mu^2}{B_0^{0.5}}(1-S)\right)^2 + \\
   \frac{b_{\partial}}{B_0^{0.5}} \left(1 - \frac{S}{B_0^{0.5}} + \left(\frac{k}{k_L}\right)^2\frac{f\mu^2}{B_0^{0.5}}(1-S)\right)
\end{multline}

\begin{multline}
    \delta P_{ds}(k,\mu) = - \exp\bigg(-k^2 (1 + f\mu^2) \Sigma_{sd}^2\bigg) \\ \left(\frac{1}{B_0^{0.5}} - \frac{S}{B_0} + \frac{f\mu^2}{B_0}(1-S) + \frac{1}{2} \frac{b_{\partial}}{B_0} \left(\frac{k}{k_L}\right)^2 \right) S
\end{multline}

\begin{equation}
    \delta P_{ss}(k,\mu) = \exp\bigg(-k^2 \Sigma_{ss}^2\bigg) \frac{S^2}{B_0}
\end{equation}

The damping parameters are given by:

\begin{equation}
    \Sigma^2_{dd}(q,z) = \frac{1}{3} \int \frac{dp}{2\pi^2} (1 - j_0(qp) ) (1 - S(p))^2 P_L(p,z)
    \label{eqn:sigma_dd}
\end{equation}

\begin{multline}
    \Sigma^2_{ds}(q,z) = \frac{1}{3} \int \frac{dp}{2 \pi^2} \left( \frac{1}{2} \left(S(p)^2 + (1 - S(p))^2\right)\right. \\  - j_0(qp)(1-S(p)) S(p)\bigg) P_L(p, z)
    \label{eqn:sigma_ds}
\end{multline}

\begin{equation}
    \Sigma^2_{ss}(q,z) = \frac{1}{3} \int \frac{dp}{2\pi^2} (1 - j_0(qp)) S(p)^2 P_L(p,z)
    \label{eqn:sigma_ss}
\end{equation}

They are calculated from a fixed template linear matter power spectrum,
and evaluated with $q = 110$\,$h^{-1}$\,Mpc.

$\Delta P_{\textrm{lin}}$ is defined in a similar way
as for the linear model. We explicitly write it in terms of the primordial
power spectrum and the growth factor:

\begin{multline}
    \Delta P_{\textrm{lin}}(k,z) = \left[\left(\gamma_\text{B} T_b(k) + (1-\gamma_\text{B})T_c^{EH}(k)\right)^2 -\right. \\ \left.(1-\gamma_\text{B})^2 (T_c^{EH}(k))^2 \right] 2 \pi^2 h^4 \left(\frac{k h}{k_{\textrm{piv}}}\right)^{n_s } A_s D(z)^2
\end{multline}
allowing us to cancel the third term arising from $P_{\textrm{smooth}}$:
\begin{multline}
    \Delta P_{\textrm{lin}}(k,z) = \left[\gamma_\text{B}^2 T_b^2(k) + 2 \gamma_\text{B} (1-\gamma_\text{B})T_c^{EH}(k) \right] 2 \pi^2 h^4 \\ \left(\frac{k h}{k_{\textrm{piv}}}\right)^{n_s } A_s D(z)^2
\end{multline}

Finally, we scale $k$ and $\mu$ following \cite{Ballinger96,Beutler16BAO}
\begin{equation}
    k' = \frac{k}{\alpha_\perp} \left[1 + \mu^2 \left(\frac{1}{F^2}-1\right)\right]^{1/2}
\end{equation}
\begin{equation}
    \mu' = \frac{\mu}{F}\left[1 + \mu^2 \left(\frac{1}{F^2}-1\right)\right]^{1/2}
\end{equation}
and allow for overall scalings of the monopole and quadrupole
with free parameters $B_0$ and $B_2$. We place a prior on these parameters
following the same approach as \cite{Ross2017}: first fitting $B_0$ from 50 to 80\,$h^{-1}$\,Mpc (fixing $B_2$ equal to $B_0$ in this fit),
and then putting a prior on $\log{B_0}$ and $\log{B_2}$ equal to the fitted value and with width 0.4.

\section{Mock validation tests}  \label{sec:mock-tests}

We conduct two sets of mock validation tests. For each test, we perform a full Bayesian analysis, deriving credible intervals for the cosmological parameters of interest after marginalising over nuisance parameters. First, we consider the results when fitting to realistic N-body mocks matching the parameters of the BOSS CMASS sample. Second, we test our sensitivity to the fiducial cosmology and priors imposed on nuisance parameters by creating a series of noiseless theory vectors from theoretical predictions with different parameters. The second set of tests directly gauges the dependence on the nuisance parameters and their priors by considering whether the true values lie within the credible intervals. The first set of tests allows us to test the dependence on noise, and helps us to understand the sensitivity to changes in the model. It also tests whether the model can account for the nonlinear structure formation encapsulated by the mocks.

We use the 84 Nseries BOSS CMASS mocks presented in \cite{GilMarin20} to test our pipeline.\footnote{See also \url{https://www.ub.edu/bispectrum/bispectrum_public/bispectrum_notes.pdf}.} 
Many different mocks have been created for BOSS, such as the Patchy mocks \cite{Kitaura16}, QPM \cite{White14}, and runA and runPB \cite{Beutler16BAO}. The QPM mocks were created to match LOWZ and CMASS as separate samples, rather than the final BOSS sample, which combines LOWZ and CMASS and then applies overall redshift cuts.
As a result, the Patchy mocks match the combined sample better than QPM (e.g.\ due to QPM not properly matching redshift evolution within LOWZ) \cite{Kitaura16}.
Moreover, both Patchy and QPM
are approximate fast mocks created
to measure covariance matrices from a large number of samples and therefore may not match the real Universe.
Indeed, \cite{Beutler16BAO} finds that the BAO damping is too large in the Patchy mocks, which they attribute
to the approximate nature of the mocks.
As a result, (and following \cite{Beutler16BAO}) we favor using true $N$-body mocks to test our pipeline.
The runA and runPB simulations used in \cite{Beutler16BAO} are only periodic boxes. Hence we use Nseries as the only $N$-body mock available in sky coordinates R.A., declination, and redshift.

The Nseries footprint matches BOSS CMASS NGC. Each mock covers $\sim$4 Gpc$^3$ effective volume (scaling by the sky fraction in NGC vs.\ the total sky fraction, and the effective volume from Table 3 in \cite{Cuesta16}).
This is approximately the same effective volume as bins $z_1$ (3.7 Gpc$^3$), $z_2$ (4.2 Gpc$^3$), and $z_3$ (4.1 Gpc$^3$) (Table 2 in \cite{Alam17}).

These mocks were created from a single halo occupation distribution tuned to be consistent with the BOSS CMASS sample, and applied to 7 independent 2.6\,$h^{-1}$Gpc periodic boxes with the same cosmology at $z = 0.569$. Twelve different orientations were created out of each box, leading to 84 mocks. The boxes have high resolution, with $2048^3$ particles and a mass per particle of $1.5 \times 10^{11}$\,$h^{-1}$ $M_{\odot}$. The input cosmological parameters are $\Omega_m = 0.286$, $\Omega_b = 0.047$, $h = 0.7$, $n_s = 0.96$, $\sigma_8 = 0.82$, and no massive neutrinos. BAO reconstruction is run on the catalogs and then a post-reconstruction catalog is produced.

We generate the noiseless theory vectors with a variety of different cosmological and nuisance parameters, including $\Lambda$CDM cosmologies, cosmologies with varying neutrino mass and numbers of relativistic species, and Early Dark Energy cosmologies.  We describe these below in the subsection for each of the three types of fits.

We test the recovery of the baryon fraction $f_b\equiv\Omega_b/\Omega_m$ (comparing our estimator $\gamma_\text{B}$ to the true value of $f_b$),
the normalised Hubble parameter $h\equiv H_0$\,\Hunit$/100$\,\Hunit\ determined using $\gamma_\text{B}$, the BBN measurement of $\Omega_bh^2$ and the value of $\Omega_m$ from the AP effect (as described in \cite{krolewski}). 


\subsection{Full-shape tests} \label{sec:nbody_tests_full_shape}

We test the credible intervals inferred for $\gamma_\text{B}$ and $h$ derived from the posterior when fitting to either the Nseries mocks or synthetic data vectors from EDE and $\Lambda$CDM cosmologies. We fit the monopole $P_0$, the quadrupole $P_2$, and the hexadecapole $P_4$, measured between $k_{\textrm{min}} = 0.01$ and $k_{\textrm{max}}=0.20$\,$h$\,Mpc$^{-1}$. We use the BOSS $z_3$ NGC covariance, rescaled by the ratio of the effective volume of the total BOSS volume in bins $z_1$ and $z_3$, to the effective volume of $z_3$ NGC (a factor of 2.6 using Table 2 in \cite{Alam17}).
The volume rescaling allows us to compare the shifts between the truth and the credible regions relative to the statistical power of the entire BOSS dataset.
We use the MontePython sampler \cite{Brinckmann:2018cvx,Audren:2012wb}, run the chains
until a Gelman-Rubin convergence criterion of R-1=0.1 is reached \cite{GelmanRubin}, and analyze the chains with GetDist \cite{GetDist}.

In order to ensure that our geometrical constraint on $\Omega_m$ is uncontaminated by the sound horizon, we only include measurements of $\alpha_{\textrm{AP}} \equiv \alpha_{\parallel}/\alpha_{\perp}$ in the fit.\footnote{In our companion paper \cite{krolewski}, we also use uncalibrated BAO information from the isotropic distance scale, which considerably improves the $\Omega_m$ constraint beyond just using $\alpha_{\textrm{AP}}$. However, the mock tests are all run in a single redshift bin because Nseries was only run for CMASS NGC. In a single bin, the only sound horizon free information on $\Omega_m$ comes from $\alpha_{\textrm{AP}}$.} To do this, we use error propagation to transform the covariance between the multipoles $P_\ell(k)$ and the AP parameters into the covariance between $P_\ell(k)$ and $\alpha_{\textrm{AP}}$.
We consider two versions of the data vector used to constrain $\alpha_{\textrm{AP}}$: one using BAO information only, and one using better constraints on $\alpha_{\textrm{AP}}$ from voids \cite{Woodfinden22,Woodfinden23}.
For the voids case, the $\alpha_{\parallel}$ and $\alpha_{\perp}$ covariances are scaled by a factor of approximately 4, corresponding
to the improvement from BAO to void $\alpha_{\textrm{AP}}$ shown in Table 1 of \cite{Woodfinden23}.
This matches the factor of two improvement on $\Omega_m$ precision, from $\sigma_{\Omega_m} \sim 0.052$ from the anisotropic BAO AP information in the $z_1$ and $z_3$ bins, and the $\sigma_{\Omega_m} \sim 0.027$ constraint from \cite{Woodfinden22,Woodfinden23}.

In addition to tests on the Nseries mocks, we also test our parameter recovery using noiseless data vectors as the input data.
We consider two classes of noiseless mocks, EDE cosmologies and $\Lambda$CDM cosmologies.
For the EDE cosmologies,
we use the best-fit EDE model studied in \cite{Hill20} (originally from \cite{Smith20})
$h = 0.7219$, $\omega_b = 0.02253$, $\omega_{cdm} = 0.1306$, $A_s = 2.215 \times 10^{-9}$, $n_s = 0.9889$, $f_{\textrm{EDE}} = 0.122$, $\log_{10}(z_c) = 3.562$, and $\theta_{i,scf} = 2.83$. We consider a slightly different model from \cite{Hill20,Smith20} in that
we set the neutrino mass to zero rather than the minimal mass. We generate the full nonlinear power spectrum using the \textsc{EDE\_class\_pt} code\footnote{\url{https://github.com/Michalychforever/EDE_class_pt}} \cite{Ivanov20EDE}, with $b_1 = 1.78$ and the remaining bias and counterterms set to their fiducial values: $b_2 = -0.12$, $b_{\mathcal{G}_2} = -0.23$, $b_{\Gamma_3} = 0$, $c_0$ = 0, $c_2 = 30$, $c_4 = 0$, $\tilde{c} = 500$, $P_{\textrm{shot}} = 6669.63$ ($h^{-1}$\,Mpc)$^3$.
We also consider 9 variations on this EDE cosmology, with varying $h$ and all other parameters fixed; we fix $\Omega_m$ and $\omega_b$ so that changes in $h$ are entirely propagated into changes in the baryon fraction. We consider variations in $f_b$ between 0.123 and 0.195 (centered on the fiducial value of 0.147), corresponding to variations in $h$ of 0.627 to 0.790 (fiducial value 0.722).

We also consider $\Lambda$CDM and $w_0 w_a$CDM noiseless mocks using cosmologies drawn from the AbacusSummit suite \cite{Maksimova21}.
We choose Abacus cosmologies spanning a wide range in $f_b$ (from 0.138 to 0.176); the cosmological parameters
are summarized in Table~\ref{tab:abacus_lambda_cdm_mocks}.
These mocks were generated with $b_1 = 1.99$, $b_2 = 0.24$, $b_{\mathcal{G}_2} = -0.22$, $b_{\Gamma_3} = 0.54$, 
$c_0 = 0$, $c_2 = 30$, $c_4 = 0$, $\tilde{c} = 0$, $a_0 = 0$, $a_2 = 0$, $P_{\textrm{shot}} = 0$ (i.e.\ shot noise fixed to the true value of $1/\bar{n}$ = 5000 ($h^{-1}$\,Mpc)$^3$).

\begin{table*}[]
    \centering
    \begin{tabular}{c|ccccccccc}
    Name & $h$ & $\omega_b$ & $\omega_{cdm}$ & $A_s$ & $n_s$ & $\alpha_s$ & $N_{\textrm{ur}}$ & $w_0$ & $w_a$ \\
    \hline
    cosm131 & 0.7165 & 0.02237 & 0.1086 & $2.3146 \times 10^{-9}$ & 0.9649 & 0 & 2.0328 & -1 & 0 \\ 
    cosm139 & 0.7164 & 0.02416 & 0.1128 & $2.1681 \times 10^{-9}$ & 0.9732 & 0 & 2.0327 & -1 & 0 \\
    cosm130 & 0.6736 & 0.02237 & 0.1200 & $1.614 \times 10^{-9}$ & 0.9649 & 0 & 2.0328 & -1 & 0 \\
    cosm159 & 0.5967 & 0.02206 & 0.1261 & $1.7977 \times 10^{-9}$ & 0.9673 & 0 & 2.0328 & -1.082 & -0.392 \\
    cosm167 & 0.7050 & 0.02248 & 0.1400 & $1.7321 \times 10^{-9}$ & 0.9715 & -0.017 & 2.889 & -1 & 0 \\
    cosm168 & 0.6070 & 0.02157 & 0.1084 & $2.5378 \times 10^{-9}$ & 0.9275 & 0.016 & 1.1911 & -1 & 0 \\
    cosm172 & 0.6800 & 0.02232 & 0.1161 & $1.8904 \times 10^{-9}$ & 0.9628 & 0.038 & 1.9049 & -1 & 0 \\
    cosm177 & 0.6820 & 0.02239 & 0.1344 & $2.0641 \times 10^{-9}$ & 1.002 & 0.007 & 2.8643 & -0.757 & -0.443 \\
    cosm169 & 0.6474 & 0.02189 & 0.1222 & $1.8067 \times 10^{-9}$ & 0.9898 & 0.038 & 1.9109 & -1 & 0 \\
    cosm181 & 0.5745 & 0.02169 & 0.1112 & $2.0651 \times 10^{-9}$ & 0.9336 & -0.013 & 1.4059 & -0.910 & 0.350
    \end{tabular}
    \caption{AbacusSummit \cite{Maksimova21} cosmologies chosen to test recovery of $\gamma_\text{B}$ in full-shape fits. We choose a wide range of input values of the baryon fraction, beyond testing the recovery of $\gamma_\text{B}$ for the single fixed true cosmology of the Nseries mocks. In addition to the $\Lambda$CDM parameters, the running $\alpha_s$, the number of ultrarelativistic species $N_{\textrm{ur}}$ (with a single
    massive neutrino), and the dark energy equation of state parameters $w_0$ and $w_a$ are also varied.
    \label{tab:abacus_lambda_cdm_mocks}}
\end{table*}

\subsection{Template fits}

As for the full-shape fit tests discussed above, we test the credible intervals inferred for $\gamma_\text{B}$ derived from the posterior when using the template based approach to fit either the Nseries mocks or synthetic data vectors.
Here we only fit for the BAO amplitude and do not combine with the AP parameters, and so therefore only show credible intervals on $\gamma_\text{B}$ and not on $h^{\textrm{dens}}$.
We sample the posterior with Cobaya \cite{Cobaya,CobayaCode} and run until R-1=0.05; the faster likelihood evaluation time allows us to run to a smaller value of R-1 than the full-shape fits, but in both cases we test that the results are unchanged if we use a slightly higher value of R-1.

For the synthetic data vector fits, the default cosmological parameters are the fiducial BOSS cosmology, and the default nuisance parameters are $b_\partial$ = 0 (pre-recon) or $b_\partial = 15$ (post-recon), $\Sigma_{xy}$ = 6\,$h^{-1}$\,Mpc pre-recon or $\Sigma_{sm}$ = 15\,$h^{-1}$\,Mpc post-recon, and $\Sigma_{fog}$ = 4\,$h^{-1}$\,Mpc. We consider varying the derivative bias to $b_\partial$ = -10 (5 post-recon) and $b_\partial = $ 10 (25 post-recon); the smoothing scale to $\Sigma_{xy}$ = 3 and 12\,$h^{-1}$\,Mpc pre-recon ($\Sigma_{sm} = 8$\,$h^{-1}$\,Mpc and $\Sigma_{sm} = 24$\,$h^{-1}$\,Mpc post-recon); and the Finger-of-God damping to $\Sigma_{fog} = 0$ and $\Sigma_{fog} = 8$\,$h^{-1}$\,Mpc. We take the covariance matrix from the BOSS $z_3$ fit (combining both hemispheres), scaled to the combined volume of BOSS $z_1$ and $z_3$. Since our model exactly matches the data for a single set of parameters, and we marginalize over all nuisance parameters, our fits are essentially tests of how sensitive we are to the prior, including projection effects that could potentially drive the marginalized constraints away from the best-fit. 

We also test our sensitivity to the wiggle/no-wiggle split by generating a data vector with the smooth power spectrum generated in a different way from the EH98 fitting function. In this alternative approach, we follow the method of \cite{Bernal20} and fit a smoothing spline to the linear correlation function in two windows at $ 70$\,$h^{-1}$\,Mpc and $ 250$\,$h^{-1}$\,Mpc, smoothly interpolating between the two to remove the BAO bump.

We then use several data vectors generated from different cosmologies to test the sensitivity to the assumption of a fiducial, fixed template. First, we test that our results are robust to the change in cosmological parameters required for an Early Dark Energy fit to the CMB \cite{Hill20}.
We generate a data vector from the best-fit EDE parameters to the CMB: $h = 0.7219$, $\omega_b = 0.02253$, $\omega_{cdm} = 0.1306$, $A_s = 2.215 \times 10^{-9}$, $n_s = 0.9889$,  one massive neutrino with mass 0.06 eV, $f_{\textrm{EDE}} = 0.122$, $\log_{10}(z_c) = 3.562$, and $\theta_{i,scf} = 2.83$. We use the CLASSEDE\footnote{\url{https://github.com/mwt5345/class_ede}} code to create the $z=0$ matter power spectrum. We also consider the impact of two other ``extended'' models that also affect the sound horizon and phase of BAO oscillations; a model changing $\Sigma_{m_\nu}$ to 0.20 eV, and a model changing $N_{\textrm{eff}} = 3.82$ (both drawn from the Planck \citep{Planck:2020} extended-model chains with that value of $N_{\textrm{eff}}$ or $\Sigma_{m_\nu}$, allowing other parameters to shift to compensate; see Table~\ref{tab:template_cosmos} for parameters).

Finally, we consider the impact of uncertainty in the power spectrum template. We aim to keep our results model- and CMB-independent; we want to consider a wider range in cosmological parameters than that probed by the Planck $\Lambda$CDM fits. Therefore, we consider the range in cosmological parameters probed by the BOSS data themselves. We use the parameter constraints from the full-shape fits of \cite{Philcox20}.\footnote{Other full-shape fits have substantially similar errorbars when using similar datasets and cosmological parameters \citep{Chen22,Simon23,DAmico19}; thus our results would be substantially similar if we sampled from chains from these results rather than from \cite{Philcox20}.}
We create 10 test cosmologies drawn from MCMC chains fit to the BOSS power spectrum multipoles only (i.e.\ not including the real-space power spectrum $Q_0$, BAO, or the bispectrum, which are part of their fiducial dataset),
with free $n_s$.

\begin{table}[]
    \centering
    \begin{tabular}{c|ccccc}
    Name & $h$ & $\omega_b$ & $\omega_{cdm}$ & $A_s$ & $n_s$ \\
    \hline
    Cosmology 0 & 0.6912 & 0.02237 & 0.1210 & $2.715 \times 10^{-9}$ & $0.9719$ \\
    Cosmology 1 & 0.7074 & 0.02237 & 0.1472 & $2.683 \times 10^{-9}$ & 0.8391 \\
    Cosmology 2 & 0.6826 & 0.02237 & 0.1184 & $2.678 \times 10^{-9}$ & 0.9762 \\
    Cosmology 3 & 0.688 & 0.02237 & 0.1266 & $2.889 \times 10^{-9}$ & 1.001 \\
    Cosmology 4 & 0.7097 & 0.02237 & 0.1207 & $2.76 \times 10^{-9}$ & 0.9788 \\
    Cosmology 5 & 0.6879 & 0.02237 & 0.1200 & $2.875 \times 10^{-9}$ & 1.009 \\
    Cosmology 6 & 0.6744 & 0.02237 & 0.1308 & $2.943 \times 10^{-9}$ & 0.8072 \\
    Cosmology 7 & 0.6923 & 0.02237 & 0.1282 & $2.556 \times 10^{-9}$ & 0.9033 \\
    Cosmology 8 & 0.7019 & 0.02237 & 0.1236 & $2.993 \times 10^{-9}$ & 0.9459 \\
    Cosmology 9 & 0.6957 & 0.02237 & 0.1336 & $2.875 \times 10^{-9}$ & 0.8835  \\
    $N_{\textrm{eff}} = 3.82$ & 0.72024 & 0.02291 & 0.13110 & $2.168 \times 10^{-9}$ & 0.98974 \\
    $\Sigma_{mu_\nu} = 0.20$ eV & 0.67248 & 0.02259 & 0.11557 & $2.159 \times 10^{-9}$ & 0.96965
    \end{tabular}
    \caption{Cosmologies chosen to test the template fits: parameters for 10 cosmologies drawn from the full-shape fits of \cite{Philcox_Ivanov} to test the sensitivity of $\gamma_\text{B}$ to the choice of template cosmology.
    \label{tab:template_cosmos}}
\end{table}

\section{Results of mock tests} \label{sec:mock-results}

\subsection{Full-shape fits}

\begin{table*}[]
    \centering
    \begin{tabular}{l|cc|cccc|cc}
    \multirow{2}{0cm}{Mock} & \multicolumn{2}{c}{Truth} & 
    \multicolumn{4}{c}{BAO only} & \multicolumn{2}{c}{Add voids} \\
    & 
    $h$ & $f_b$ & $h$ & $n\sigma$ & $\gamma_\text{B}$ & $n\sigma$ & $h$ & $n\sigma$ \\   
    \hline
Nseries & 0.7 & 0.164 & $0.689^{+0.083}_{-0.074}$
 & -0.13 & $0.172^{+0.031}_{-0.027}$
 & 0.30 & $0.689^{+0.057}_{-0.051}$
 & -0.19 \\
$z_3$ covariance & 0.7 & 0.164 & $0.694^{+0.116}_{-0.093}$
 & -0.05 & $0.168^{+0.034}_{-0.032}$
 & 0.13 & $0.698^{+0.080}_{-0.071}$
 & -0.03 \\
$z_3$-NGC covariance & 0.7 & 0.164 & $0.704^{+0.149}_{-0.113}$
 & 0.04 & $0.159^{+0.039}_{-0.035}$
 & -0.13 & $0.694^{+0.108}_{-0.078}$
 & -0.06 \\
$z_3$-SGC covariance & 0.7 & 0.164 & $0.719^{+0.413}_{-0.186}$
 & 0.10 & $0.132^{+0.047}_{-0.087}$
 & -0.68 & $0.778^{+0.291}_{-0.189}$
 & 0.41 \\
$z_3$-NGC + SGC & 0.7 & 0.164 & $0.706^{+0.129}_{-0.099}$ & 0.06 & $0.151^{+0.034}_{-0.029}$ & -0.38 & $0.715^{+0.092}_{-0.078}$ & 0.19 \\
Nseries $+ Q_0$ & 0.7 & 0.164 & $0.698^{+0.081}_{-0.069}$
 & -0.02 & $0.168^{+0.029}_{-0.026}$
 & 0.15 & $0.694^{+0.060}_{-0.054}$
 & -0.11 \\
\cite{Philcox_Ivanov} counterterm prior & 0.7 & 0.164 & $0.694^{+0.096}_{-0.072}$
 & -0.07 & $0.167^{+0.031}_{-0.027}$
 & 0.11 & $0.691^{+0.075}_{-0.054}$
 & -0.16 \\
$2\times$ bias prior & 0.7 & 0.164 & $0.704^{+0.084}_{-0.078}$
 & 0.05 & $0.166^{+0.032}_{-0.026}$
 & 0.07 & $0.698^{+0.068}_{-0.056}$
 & -0.04 \\
$2\times$\cite{Philcox_Ivanov} c.t. prior & 0.7 & 0.164 & $0.727^{+0.093}_{-0.087}$
 & 0.36 & $0.152^{+0.034}_{-0.026}$
 & -0.39 & $0.730^{+0.075}_{-0.072}$
 & 0.59 \\
 \hline
EDE & 0.722 & 0.147 & $0.754^{+0.114}_{-0.098}$
 & 0.43 & $0.132^{+0.038}_{-0.031}$
 & -0.48 & $0.745^{+0.099}_{-0.077}$
 & 0.45 \\
Scale covariance by $10\times$ volume & 0.722 & 0.147 & $0.745^{+0.033}_{-0.035}$
 & 0.31 & $0.138^{+0.014}_{-0.011}$
 & -0.29 & $0.739^{+0.033}_{-0.029}$
 & 0.33 \\
\cite{Philcox_Ivanov} counterterm prior & 0.722 & 0.147 & $0.761^{+0.110}_{-0.090}$
 & 0.53 & $0.125^{+0.032}_{-0.026}$
 & -0.71 & $0.761^{+0.092}_{-0.078}$
 & 0.76 \\
\cite{Philcox_Ivanov} counterterm prior + scale cov.~by $10\times$ vol. & 0.722 & 0.147 & $0.740^{+0.035}_{-0.036}$
 & 0.24 & $0.139^{+0.014}_{-0.013}$
 & -0.26 & $0.736^{+0.035}_{-0.032}$
 & 0.27 \\
\end{tabular}
\caption{Parameter recovery tests for Nseries mocks (first block) and noiseless early dark energy (EDE) theory vector (second block).
Two data combinations are shown for each test: both use $\gamma_\text{B}$ from the BAO amplitude and an $\Omega_b h^2$ prior with a BBN-like width. ``BAO only'' uses $\Omega_m$ constraints from the Alcock-Paczynski ratio ($D_A/D_H$) from BAO only, and ``Add voids'' includes tighter AP ratio constraints from voids. $n \sigma$ deviations are in units of the Nseries errorbar, using the appropriate upper or lower errorbar, except for the cases with increased covariance matrix ($z_3$, $z_3$-NGC, $z_3$-SGC, $z_3$-NGC + SGC), where the errorbar from that fit is used. \label{tab:full-fit-results}}
\end{table*}

\begin{figure*}
\includegraphics[scale=0.8]{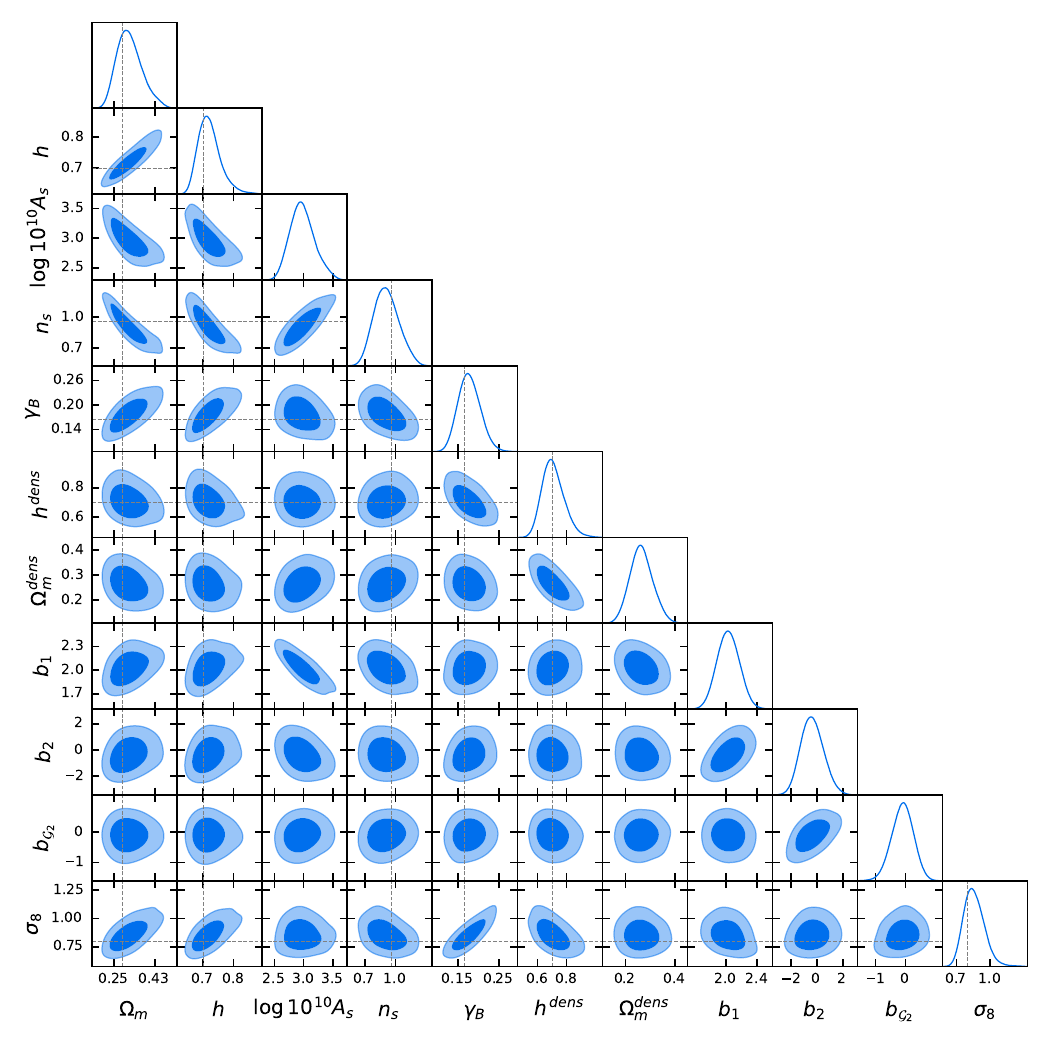}
    \caption{Parameter constraints for the full-shape plus BAO amplitude model fitted to the mean of the 84 Nseries mocks,
 using the covariance for the BOSS $z_3$ NGC region, scaled to the entire BOSS volume. The data vector consists of the power spectrum monopole, quadrupole, and hexadecapole at $0.01 < k < 0.2$\,$h$\,Mpc$^{-1}$, combined with the ratio of the BAO scaling parameters, $\alpha_{\perp}/\alpha_{\parallel}$ and a Gaussian prior on $\omega_b$ with a width equal to current BBN constraints.
 We vary the standard $\Lambda$CDM parameters as well as two extra parameters, $h^{\textrm{dens}}$ and $\Omega_m^{\textrm{dens}}$, which only affect the BAO amplitude and the Alcock-Paczynski parameters. From these varied parameters we derive the BAO amplitude estimate $\gamma_\text{B}$. The galaxy biases $b_1$, $b_2$, and $b_{\mathcal{G}_2}$ are sampled
 in the MCMC chain, whereas $b_{\Gamma_3}$, the counterterms, and the stochastic parameters are marginalized analytically
 and hence not shown.
    \label{fig:triangle_full_shape}}
\end{figure*}

\begin{figure*}
\includegraphics[scale=0.4]{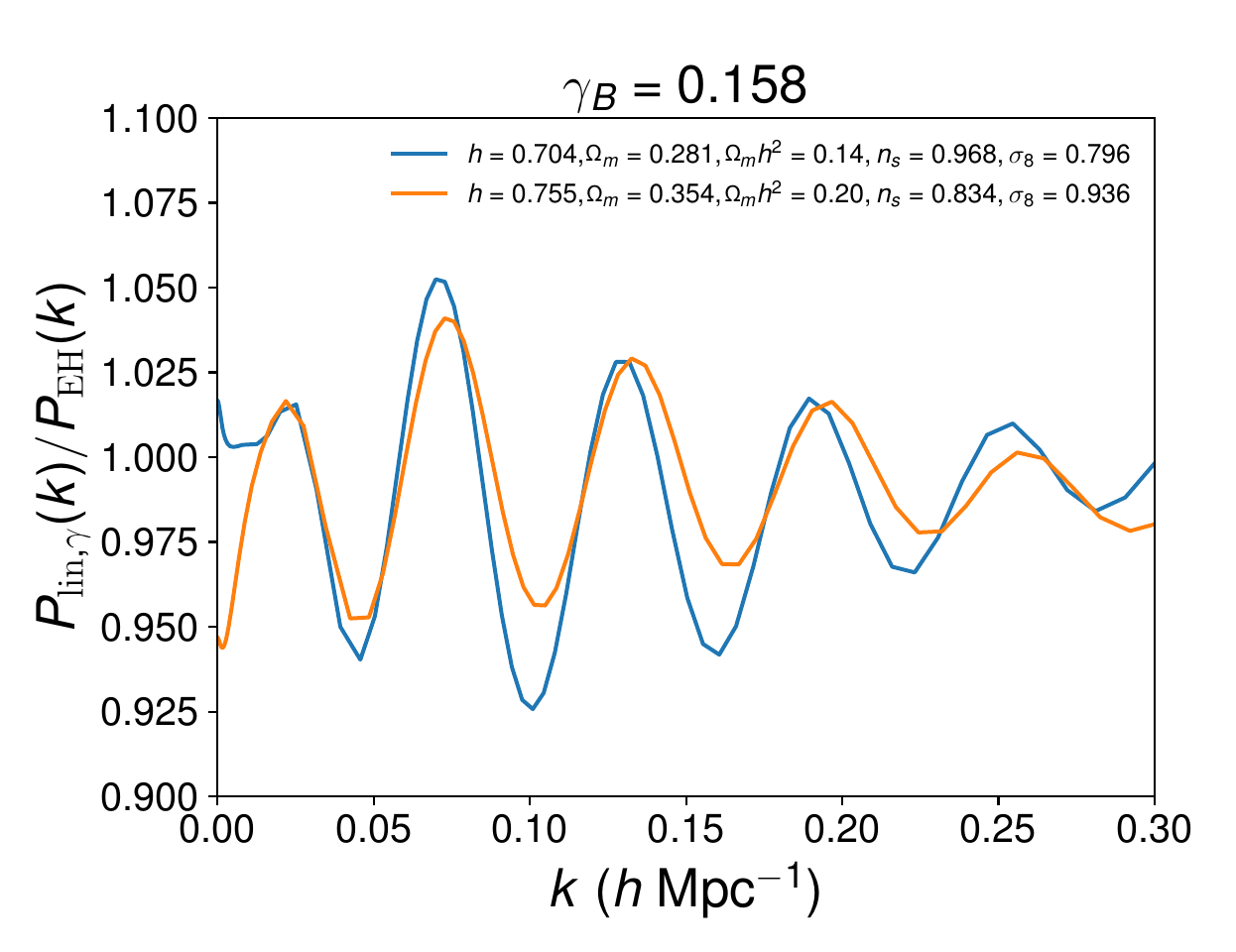}
\includegraphics[scale=1.0]{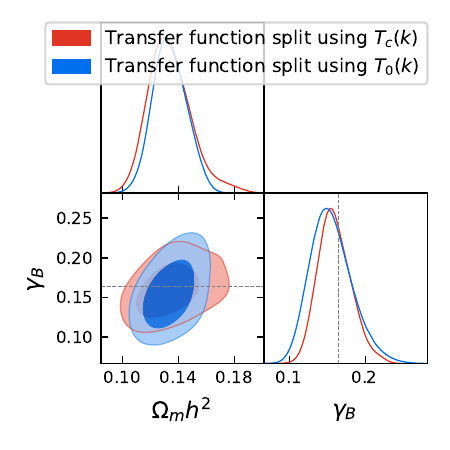}
\caption{\textit{Left:} Comparison between two sets of parameters chosen along the $\Omega_m$-$\gamma_\text{B}$
degeneracy, adjusting both so that $\gamma_\text{B}$ is the same 
($\gamma_\text{B} = 0.158$). We plot the
ratio between the linear power spectrum with adjusted BAO amplitude, $P_{\textrm{lin},\gamma}(k)$, and 
a smooth power spectrum consisting of the Eisenstein \& Hu \cite{Eisenstein-Hu} fitting function plus
a linear correction due to the change in shape of $P_{\textrm{lin},\gamma}$ in the second set where $\gamma_\text{B}$ differs significantly from the true $f_b$.
The BAO wiggles are notably larger for smaller $\Omega_m h^2$ despite the fact that $\gamma_\text{B}$ is the same, leading to a degeneracy between $\Omega_b h^2$ and $\gamma_\text{B}$. \textit{Right:} Using the CDM transfer function $T_c(k)$ (Eq.~17 in \cite{Eisenstein-Hu}) rather than the zero-baryon transfer function $T_0(k)$ (Eq.~29 in \cite{Eisenstein-Hu}) tightens constraints on $\gamma_\text{B}$ by 20\% by bringing in additional shape information, mitigating the $\Omega_m h^2$-$\gamma_\text{B}$ degeneracy.
\label{fig:omh2_gamma_degeneracy}} 
\end{figure*}

\subsubsection{Fits to the Nseries mocks}

The parameters of the true cosmology generally lie well within the 1-$\sigma$ credible interval when fitting to the Nseries mocks, with the marginalised value of $h$ lower than the truth by 0.13$\sigma$, $f_b$ biased high by 0.30$\sigma$, and an 11.4\% $H_0$ constraint, when using $\Omega_m$ information from the AP ratio from BOSS BAO. This is a relatively poor $\Omega_m$ constraint, with $\sigma_{\Omega_m} = 0.052$ (i.e.\ much of the cosmological information from post-reconstruction BAO is lost when marginalizing over the sound horizon and therefore limiting the available information to the AP ratio). We also consider adding constraints from the AP ratio measured using voids, which improves the precision on $\Omega_m$ by a factor of 2; this leads to a 0.19$\sigma$ bias low on $H_0$ and an 8\% $H_0$ measurement.

We show a triangle plot for the posterior from a fit to Nseries in Fig.~\ref{fig:triangle_full_shape}. $\gamma_\text{B}$ and $h^{\textrm{dens}}$ are primarily degenerate with cosmological parameters rather than galaxy biases, though recall that many nuisance parameters are marginalized analytically and hence not shown on Fig.~\ref{fig:triangle_full_shape}. Of the galaxy biases, we see the strongest degeneracies with $b_2$, in line with its contributions to the nonlinear BAO shift \citep{CrocceScoccimarro08,PadmanabhanWhite09,SherwinZaldarriaga12,ChenHowlett24}. The strongest degeneracies are with $\Omega_m$, $\sigma_8$, $h$, and $n_s$. As explained below, these degeneracies arise from the approximation that only the baryon fraction affects the amplitude of the baryon signature. In reality, there are weak dependencies on other parameters, principally $\Omega_m$ and $h$ (Fig.~\ref{fig:lcdm_vs_ede_schematic}). The BAO amplitude is least degenerate with the primordial power spectrum amplitude $A_s$; this is reasonable because the BAO feature is the difference between the wiggle and no-wiggle power spectrum and thus should not depend on power spectrum amplitude. In contrast, $\sigma_8$ mixes the primordial amplitude with $\Omega_m$ and  $n_s$, since it is the $z=0$ amplitude of the small-scale power spectrum. Hence it acquires a degeneracy with the BAO amplitude due to its degeneracies with $\Omega_m$ and $n_s$. Finally, the $n_s$-$\gamma_\text{B}$ degeneracy comes primarily from the stronger degeneracies between $\Omega_m$ and $\gamma_\text{B}$, and $n_s$ and $\Omega_m$ (coming from the impact of $\Omega_m$ on the power spectrum shape).


\subsubsection{Degeneracies between the baryon fraction and $\Omega_m h^2$}
\label{sec:cosmo_degeneracies}

The largest contribution to the degeneracies is the trend between $\Omega_m h^2$ and the power spectrum peak or trough height, shown in Fig.\ 5 of \cite{Eisenstein-Hu}. In other words, the heights of the peaks depend primarily on the baryon fraction, but have an important secondary dependence on $\Omega_m h^2$. This comes primarily from the impact of baryons on the growth rate (which also leads to the baryonic suppression of the power spectrum shape) \citep{HuSugiyama96,Meszaros74}. Cold dark matter, which is not coupled to the photon-baryon oscillations, begins to grown after matter-radiation equality, whereas baryonic perturbations are suppressed until the acoustic oscillations end at the drag epoch. Hence, if $\Omega_m h^2$ is larger, equality occurs earlier, the CDM has more time to grow while the baryons are oscillating, and the BAO amplitude is suppressed. There are also cosmological dependencies through the Silk damping scale and the time-variation of the sound speed (leading to a time-varying oscillation which translates into a time-varying amplitude) \citep{HuSugiyama96}, though the latter depends only on $\omega_b$ and is thus well-constrained given our assumed $\omega_b$ constraint.

We show the $\Omega_m h^2$--$\gamma_\text{B}$ degeneracy in Fig.~\ref{fig:omh2_gamma_degeneracy}. The left panel shows the ratio of the linear power spectrum to the Eisenstein \& Hu \cite{Eisenstein-Hu} fit for two different cosmologies which have been adjusted to have the same baryon fraction, $\gamma_\text{B} = 0.158$. Despite the match in $\gamma_\text{B}$, the BAO wiggles clearly have different amplitude. The cosmology with higher $\Omega_m h^2$ has less prominent BAO wiggles, matching the expected trend. The right panel shows that the degeneracy is present either when using $T_c(k$) or $T_0(k)$ to split the transfer functions, showing that the degeneracy comes from the BAO wiggles themselves and not the baryonic shape change. However, using $T_c(k)$ brings in 20\% stronger constraints on $\gamma_\text{B}$ from the baryonic shape change, which leads to slightly less degeneracy between $\Omega_m h^2$ and $\gamma_\text{B}$. Hence we prefer to use $T_c(k)$ rather than $T_0(k)$ to split the transfer function.

These cosmological parameter degeneracies motivate our further consideration of template-style fits, where the cosmological dependence of the broadband is fixed and marginalized over with other nuisance parameters. The template fits thus assume that \textit{only} the baryon fraction affects the split between transfer functions, since all other cosmological parameters are fixed. As shown in \cite{HuSugiyama96} and described above, the largest additional dependence comes from baryonic suppression of perturbations and the growth rate; at fixed $\omega_b$, the Silk damping scale is only weakly dependent on $\Omega_m h^2$, and the time-dependence of the sound speed, frequency, and thus amplitude only depends on $\omega_b$. Thus, the template fit assumes that matter growth from equality to decoupling matches that in the best-fit $\Lambda$CDM model. One might worry that this assumption limits the applicability of our method to extended models such as Early Dark Energy. However, the early growth history has been constrained by measurements of the early Integrated Sachs-Wolfe (ISW) effect \cite{Hou12,Cabass16,Kable21,Vagnozzi21}, in which the $\Lambda$CDM early-ISW amplitude is allowed to vary freely. In the ISW effect \cite{SachsWolfe67,ReesSciama68}, time-varying gravitational potentials contribute to the CMB anisotropies; since potentials are constant in matter domination, the time variation either happens at early times, in the transition from radiation to matter domination, or at late times, when dark energy becomes non-negligible. The early ISW effect is thus a direct probe of the growth rate at the transition from radiation to matter domination, exactly overlapping the epoch from equality to drag that affects the BAO amplitude. Ref.~\cite{Vagnozzi21} shows that the early ISW amplitude is constrained within 3\% and agrees within 1$\sigma$ of the $\Lambda$CDM prediction. This drives $\Omega_{cdm} h^2$ higher in Early Dark Energy fits to the CMB; if the non-EDE parameters were fixed to their $\Lambda$CDM values, the early ISW amplitude would be over-predicted by 20\%. These strong constraints on the early growth history therefore suggest that a template-based approach is reasonable.

\subsubsection{Nuisance parameter priors}

The dependence on the priors applied to nuisance parameters \citep[e.g.][]{Hamann12,Planck13Profile,Smith21,Herold22,Holm22,Simon23} are a significant concern in this analysis. The three AbacusSummit $\Lambda$CDM noiseless theory vectors (cosm131, cosm139, and cosm130) show significant biases of 0.3--0.5$\sigma$ on $f_b$ and $h$, even though they are generated from the same theory that we use for the fits and thus the true cosmology has $\chi^2 = 0$. One option is to instead perform a frequentist analysis and determine the maximum a posteriori (MAP) point and profile likelihoods. However, this option has several disadvantages: lack of suitable tools compared to tools available for Bayesian analysis (although see Procoli \citep{Karwal24} and  PROSPECT \citep{Holm23}, which were released during the preparation of this manuscript); inability to interpret the frequentist results in a Bayesian framework; and losing the advantages of the analytic marginalization over $b_{\Gamma_3}$, $c_0$, $c_2$, $c_4$, $\tilde{c}$, $P_{\textrm{shot}}$, $a_0$ and $a_2$, which adds 8 extra nuisance parameters and leads to a corresponding increase in the difficulty of the minimization problem.

\begin{figure}
\includegraphics[scale=0.5]{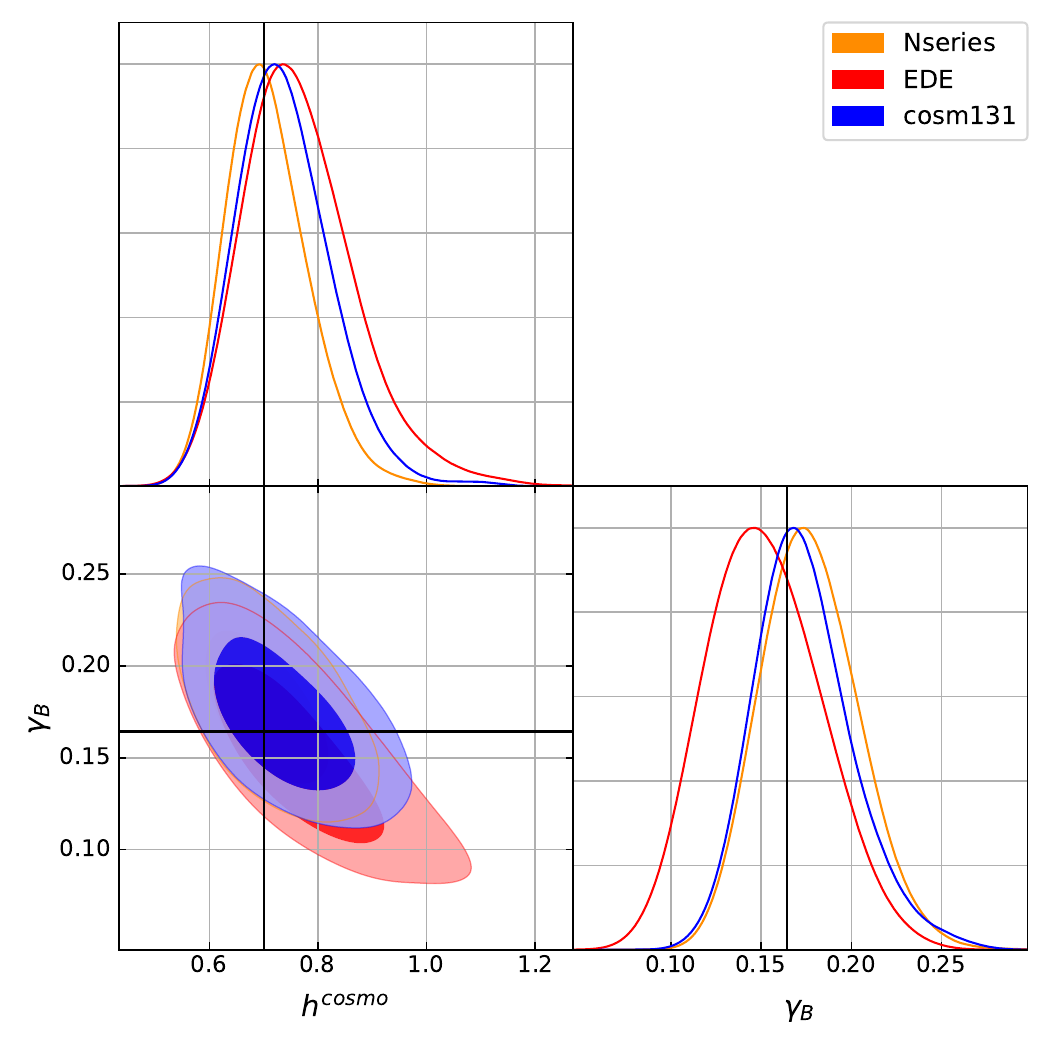}
    \caption{Marginalized posteriors on $h$ and $\gamma_\text{B}$
    from Nseries, EDE, and AbacusSummit cosm131 (a $\Lambda$CDM cosmology) tests. The cosm131 and particularly the EDE 1D posteriors are notably skewed; hence we use the maximum of the 1D posterior rather than the mean or median as a more robust statistic. The EDE and cosm131 posteriors are shifted so that their true cosmology coincides with the Nseries cosmology, $h = 0.7$ and $f_b = 0.164$ (solid black line).
    \label{fig:skewed_posteriors}}
\end{figure}

Part of the difficulty arises from skewed marginalized posteriors (Fig.~\ref{fig:skewed_posteriors}). We mitigate the effect of the skewed posteriors by choosing more robust statistics: we quote the maximum of the marginalized posterior, rather than the mean or the median. We quote the (generally asymmetric) 68\% highest-density interval as the errorbars (identical to the M-HPD statistic in \cite{KIDS20}). The shifts between the mean, median, and maximum are especially dramatic for the noiseless theory vector tests; we find that $h$ is biased high by 0.5--0.6$\sigma$ when using the mean or median for the three $\Lambda$CDM theory vectors, but only 0.3$\sigma$ when using the maximum. The EDE noiseless theory vector has even larger biases, with biases of 0.8--0.9$\sigma$ for the mean and median, but only 0.5$\sigma$ for the maximum. In contrast, the posteriors are less skewed for Nseries, with biases of 0.12$\sigma$ high for the mean, 0.03$\sigma$ high for the median, and 0.13$\sigma$ low for the maximum. This result emphasizes the importance of using a wide variety of test cases, rather than one $N$-body mock run with a single cosmology. If the data prefers different parameters from that mock, the prior effects may be different, and the data may be biased in a different way than the mock.

We also find that shrinking the priors on the counterterms and stochastic parameters improves parameter recovery when fitting the EDE theory vector. Since early dark energy leads to notable changes in the power spectrum compared to $\Lambda$CDM (see Fig.~5 in \cite{Hill20}), parameter biases in the EDE test could indicate model misspecification. Indeed, we find biases on $h$ of 0.53$\sigma$ and 0.76$\sigma$ when using the same counterterm and stochastic term prior as \cite{Philcox_Ivanov} (for the ``BAO only'' and ``add voids'' cases, respectively). However, shrinking the covariance by a factor of 10 (corresponding to 10 times more volume) leads to minimal biases on $h$ (0.24$\sigma$ and 0.27$\sigma$), implying that the biases are due to prior effects, not model misspecification. We cannot shrink the covariance when running on data, but we can reduce the widths of priors, which also mitigates the dependence on the prior. We find that halving the counterterm and stochastic term priors leads to reduced biases of 0.43$\sigma$ and 0.45$\sigma$ on $h$, which are more robust to scaling the covariance down by a factor of 10 (dropping to 0.31$\sigma$ and 0.33$\sigma$). Shrinking these priors does reduce the errors on $h$ and $f_b$ by $\sim10$\%, raising the possibility that our errors are underestimated; however, we consider this a reasonable price to pay for reducing the biases on $h$ and $f_b$.

Given the impact of the nuisance term priors, we test if parameter recovery is different when using covariances corresponding to each separate region, rather than to the combined BOSS volume. This is important because of the lack of realistic $N$-body mocks with full coverage of the BOSS region in all redshifts and hemispheres;\footnote{The PATCHY approximate $N$-body mocks cover the entire BOSS footprint, but  may not correctly describe the Universe on large scales. \cite{Beutler16BAO} find that they underestimate the BAO signal, with a larger damping than calculated from the Zel'dovich approximation and a larger damping than in the BOSS data. \cite{Yu23} find deviations between the Patchy mocks and their perturbative halo model at large scales.} hence we have only one data vector for Nseries (coming from a volume slightly larger than $z_3$ NGC) rather than the necessary four for a fully realistic test. If parameter recovery is less accurate when the covariance is larger, our strategy of testing Nseries with a covariance re-scaled to the entire BOSS volume could underestimate parameter biases. When instead using a covariance scaled to the $z_3$ volume, or the $z_3$ NGC volume, we do find some notable shifts in the parameters, though all biases stay below 0.3$\sigma$. The considerably smaller $z_3$-SGC has considerably larger biases, though these remain a relatively small fraction (0.4--0.55$\sigma$) of the much larger errorbars. However, combining multiple regions in the data should suppress the impact of large fluctuations in the SGC fits. We test this by comparing the results using the $z_3$-scaled covariance to the combined $z_3$ NGC and SGC results. For the BAO-only case, the overall bias on $h$ is similar, though the $z_3$ covariance is biased low and the NGC + SGC case is biased high. The baryon fraction results are somewhat less consistent, with NGC + SGC having a larger bias of almost 0.4$\sigma$. However, when adding the void AP information, the results on $h$ are very similar, though the NGC+SGC combination does have a 7\% larger errorbar.

We also test whether adding the real-space power spectrum estimate $Q_0$ improves our constraints. $Q_0$ is a linear combination of the monopole, quadrupole, and hexadecapole (equal to the $\mu = 0$ moment of the power spectrum) in which redshift-space distortions approximately cancel \cite{Ivanov21}. This allows $Q_0$ to be modelled to $k_{\textrm{max}} = 0.4$\,$h$\,Mpc$^{-1}$ in perturbation theory, in contrast to the $k_{\textrm{max}} = 0.2$\,$h$\,Mpc$^{-1}$ achievable for the multipoles. The higher $k_{\textrm{max}}$ means that $Q_0$ contains more BAO wiggles and thus potentially offers a tighter constraint on the baryon fraction. We find that adding $Q_0$ offers 7--10\% tighter constraints on $h$, though only minimally stronger constraints on $f_b$.
Due to the minimal statistical gains, and balancing against the possibility for extra systematic errors due to the small scales used ($Q_0$ is fit only between $k = 0.2$ $h$ Mpc$^{-1}$ and 0.4 $h$ Mpc$^{-1}$ where it is not covariant with the power spectrum multipoles) we do not include $Q_0$ in our fiducial analysis.

We find similar results when doubling the bias or counterterm priors (matching the counterterm priors of \cite{Philcox_Ivanov}). Fitting to the Nseries mocks and doubling the bias priors leads to similarly tight constraints on $h$ and $f_b$ in the BAO only case, and 20\% wider constraints on $h$ when adding voids. The parameter biases in this case are very small, $<0.1\sigma$. We also test doubling the counterterm prior of \cite{Philcox_Ivanov}. This leads to larger biases of $\sim0.4\sigma$ on $h$ and $f_b$, consistent with the argument that a larger counterterm prior leads to larger biases from prior effects. We therefore use our fiducial counterterm priors (half the width of \cite{Philcox_Ivanov}), since they lead to smaller parameter biases.

\begin{figure*}
\includegraphics[scale=0.6]{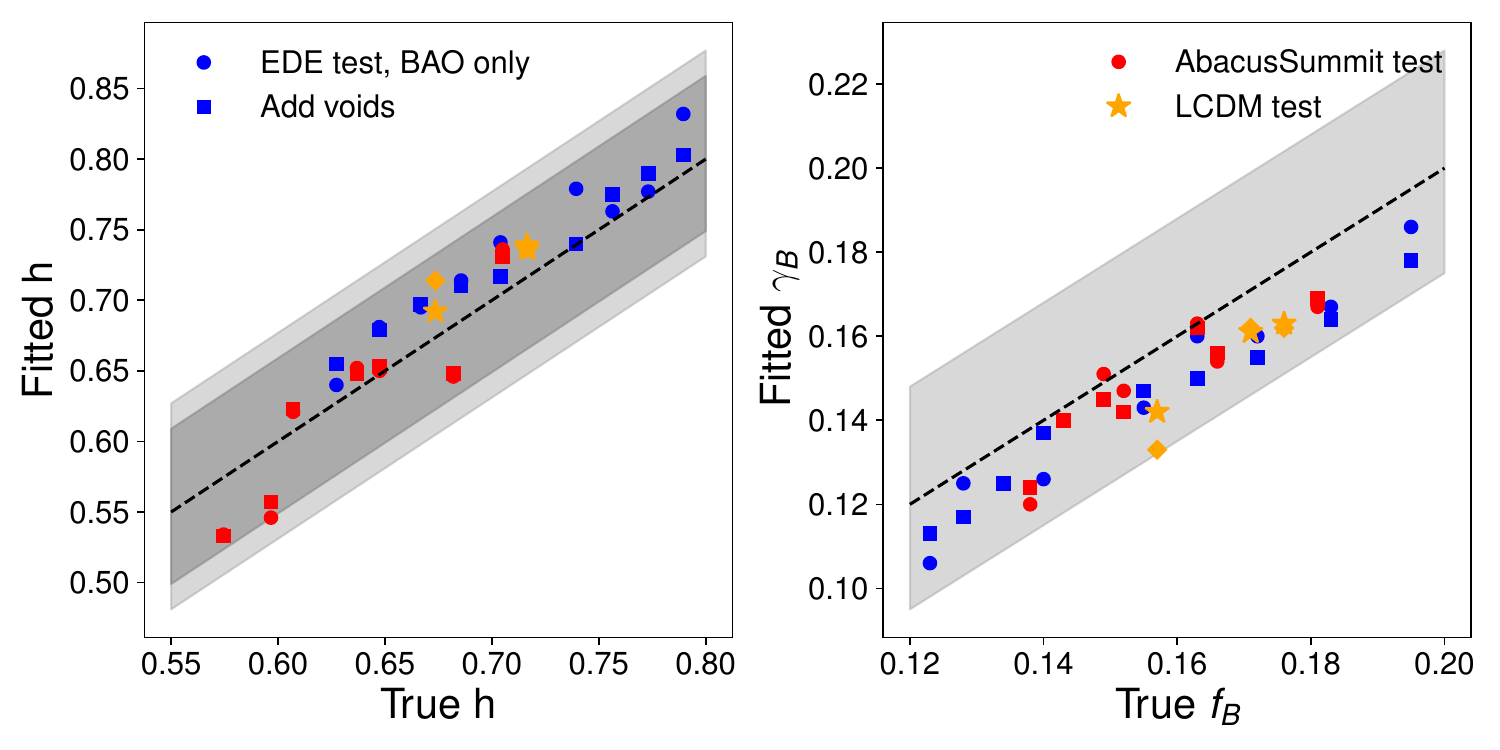}
    \caption{Fitted and true $h$ (\textit{left}) and $\gamma_\text{B}$ (\textit{right}) for noiseless theory vector tests.
    Tests of EDE cosmology are shown in blue, either using Alcock-Paczynski constraints from BAO (circles) or stronger
    AP constraints from voids (squares, with half the $\Omega_m$ error of BAO AP). AbacusSummit tests are shown in red and orange; the 3 AbacusSummit theory vectors generated in a $\Lambda$CDM cosmology are orange stars and diamonds, whereas the other 7 AbacusSummit theory vectors are red. Errorbars are the $h$ and $\gamma_\text{B}$ constraints from the Nseries mock test
    with the covariance rescaled to the BOSS volume; the inner errorbands on the left panel are the errorbars when adding AP information from voids. While both the AbacusSummit and EDE tests are biased, they are both biased by a similar amount, and the $\Lambda$CDM tests are no less biased than the others (even though the best-fit model has $\chi^2=0$ by definition in the $\Lambda$CDM test). Moreover, all biases are a fraction of a $\sigma$, generally $<0.5\sigma$.
    \label{fig:h_fb_scatter_full-shape}}
\end{figure*}

\subsubsection{Fitting noiseless theory vectors}

We compare the true and fitted values of $h$ and $f_b$ for the AbacusSummit and EDE noiseless theory vectors in Fig.~\ref{fig:h_fb_scatter_full-shape}. We find a consistent bias towards
higher $h$ and lower $f_b$. When using only the BAO AP ratio, the mean bias on $h$ is 0.12$\sigma$ (0.37$\sigma$) for AbacusSummit (EDE); the mean bias on $f_b$ is -0.32$\sigma$ (-0.38$\sigma$).
When also adding the voids AP ratio, the mean bias is 0.22$\sigma$ (0.32$\sigma$).
The AbacusSummit noiseless mock tests give significantly smaller biases on $H_0$ than the EDE noiseless mock tests. This is due to the inclusion of 3 $w_0$$w_a$ cosmologies, in which the late-time change to the expansion history biases the AP measurement of $\Omega_m$ and leads to a bias towards smaller values of $h$, rather than the bias towards larger $h$ seen in the other AbacusSummit mock tests and in the EDE tests. Excluding these three mock tests leads to a mean bias on $h$ of 0.22$\sigma$ (0.33$\sigma$) for the BAO AP ratio (voids) case.
While the bias in units of the errorbar is larger in the voids case, the absolute value of the bias is smaller, as would be expected from adding additional data with the correct cosmology.

 The scatter about these biases is relatively tight, 0.2--0.4$\sigma$. 
 The biases in the EDE and AbacusSummit tests are quite similar (excluding the $w_0$$w_a$ cosmologies as mentioned above), all around 0.3--0.5$\sigma$. Thus, they could be entirely consistent with prior volume projection effects. 
Furthermore, the biases on Nseries have the opposite sign, making it difficult to tell whether our method is truly biased towards high $h$.
At the current constraining power of the data, we regard these biases as negligible, and further point out that the combination of the full-shape and template fits offers a powerful consistency check with quite different systematics. For future data, we are encouraged by the results of the volume scaling test (particularly when restoring our counterterm prior to match that of \cite{Philcox_Ivanov}), which suggests that as the data gets more precise, the biases will become smaller. If the bias results from prior effects, we expect it to improve with the statistical error rather than remaining as a systematic floor.

\subsection{Pre-reconstruction template fits}

\begin{table*}[]
    \centering
    \begin{tabular}{l|cccc|cccc|cccc}
    Data & $\gamma_\text{B}$ & $\sigma_{\gamma}$
     & $f_b$ & $n\sigma$ &
      $\alpha_{\perp}$ & $\sigma_{\alpha_{\perp}}$ &
      True &
     $n\sigma$ &
     $\alpha_{\parallel}$ & $\sigma_{\alpha_{\parallel}}$ & True &
     $n\sigma$  \\    
    \hline
Nseries & 0.176 & 0.038 & 0.164 & 0.42 & 0.979 & 0.021 & 0.979 & -0.00 & 1.001 & 0.062 & 0.988 & 0.36 \\
$z_1$+$z_3$ Cov.\  & 0.167 & 0.029 & 0.164 & 0.09 & 0.979 & 0.013 & 0.979 & 0.03 & 1.000 & 0.036 & 0.988 & 0.35 \\
\hline
$\Lambda$CDM & 0.153 & 0.033 & 0.155 & -0.06 & 0.998 & 0.020 & 1.000 & -0.17 & 1.010 & 0.065 & 1.000 & 0.29 \\
EDE & 0.166 & 0.029 & 0.174 & -0.29 & 1.017 & 0.017 & 1.019 & -0.17 & 1.036 & 0.047 & 1.022 & 0.38 \\
$b_\partial = -10$  & 0.158 & 0.033 & 0.155 & 0.09 & 0.996 & 0.022 & 1.000 & -0.30 & 1.011 & 0.064 & 1.000 & 0.31 \\
$b_\partial = 10$  & 0.154 & 0.033 & 0.155 & -0.03 & 0.998 & 0.018 & 1.000 & -0.19 & 1.008 & 0.058 & 1.000 & 0.23 \\
$\Sigma_{xy} = 3$ & 0.149 & 0.030 & 0.155 & -0.21 & 0.997 & 0.012 & 1.000 & -0.23 & 1.007 & 0.030 & 1.000 & 0.20 \\
$\Sigma_{xy} = 12$ & 0.190 & 0.052 & 0.155 & 1.22 & 0.992 & 0.075 & 1.000 & -0.59 & 0.985 & 0.094 & 1.000 & -0.41 \\
$\Sigma_{\textrm{fog}} = 0$  & 0.156 & 0.031 & 0.155 & 0.03 & 0.996 & 0.020 & 1.000 & -0.33 & 1.014 & 0.055 & 1.000 & 0.39 \\
$\Sigma_{\textrm{fog}} = 8$  & 0.151 & 0.033 & 0.155 & -0.14 & 0.998 & 0.022 & 1.000 & -0.14 & 1.010 & 0.078 & 1.000 & 0.28 \\
\small{$\Sigma m_{\nu} = 0.2$eV} & 0.162 & 0.033 & 0.164 & -0.05 & 1.001 & 0.019 & 1.000 & 0.11 & 1.017 & 0.057 & 1.000 & 0.48 \\
$N_{\textrm{eff}} = 3.82$ & 0.152 & 0.031 & 0.149 & 0.11 & 0.990 & 0.019 & 0.994 & -0.30 & 1.005 & 0.058 & 0.999 & 0.16 \\
Spline $P_{\textrm{nw}}$ & 0.156 & 0.032 & 0.155 & 0.04 & 0.997 & 0.020 & 1.000 & -0.20 & 1.016 & 0.062 & 1.000 & 0.43 \\
    \end{tabular}
    \caption{Template fits to the pre-reconstruction correlation function. The top row shows results on the Nseries mocks
    using the BOSS $z_3$ covariance matrix; the second row uses the same data vector but the covariance is scaled to match the combined $z_1$ and $z_3$ volume. Since we fit $\gamma_\text{B}$ separately in the two redshift bins and then combine them,
    we use runs with the $z_3$ covariance to determine biases on $\gamma_\text{B}$, but compare to the errorbar from the scaled covariance runs. The next set shows results from input noiseless theory vectors with varying nuisance parameters, a different choice for how to split $P_w$ and $P_{nw}$, and varying cosmologies (Early Dark Energy, larger neutrino mass, and $N_{\textrm{eff}}$). 
    \label{tab:pre-recon_template_fits}}
\end{table*}

\subsubsection{Fits to the Nseries mocks}

We summarize the results of the pre-reconstruction template fits to the correlation function in Table~\ref{tab:pre-recon_template_fits}. When fitting to the Nseries mocks, the marginalized constraint on $\gamma_\text{B}$ slightly depends on the scaling of the covariance matrix; we see a 0.42$\sigma$ bias on $\gamma_\text{B}$ when using the covariance matrix from BOSS $z_3$, but only a 0.09$\sigma$ bias when scaling this covariance matrix to the entire BOSS volume. These results are quite similar to the full-shape fits to $P(k)$, with $\gamma_\text{B}$ changing from 0.171 using full-shape fits to 0.176 using template fits, and the errorbar on the combined BOSS volume is nearly the same (0.027 vs.\ 0.029).

Our measurements of the BAO shifting parameters are very similar to those presented in \cite{GilMarin20} for $\alpha_\perp$ (they find $0.977 \pm 0.002$, with a much smaller error since they scale the covariance matrix to the Nseries volume), but different for $\alpha_\parallel$ (they find $0.983 \pm 0.0041$). There are significant differences in the pipeline: they use a linear fit with $b_\partial = 0$ and fix the damping parameters; so it is perhaps not surprising that the more poorly constrained $\alpha_{\parallel}$ is different.

\begin{figure*}
\includegraphics[]{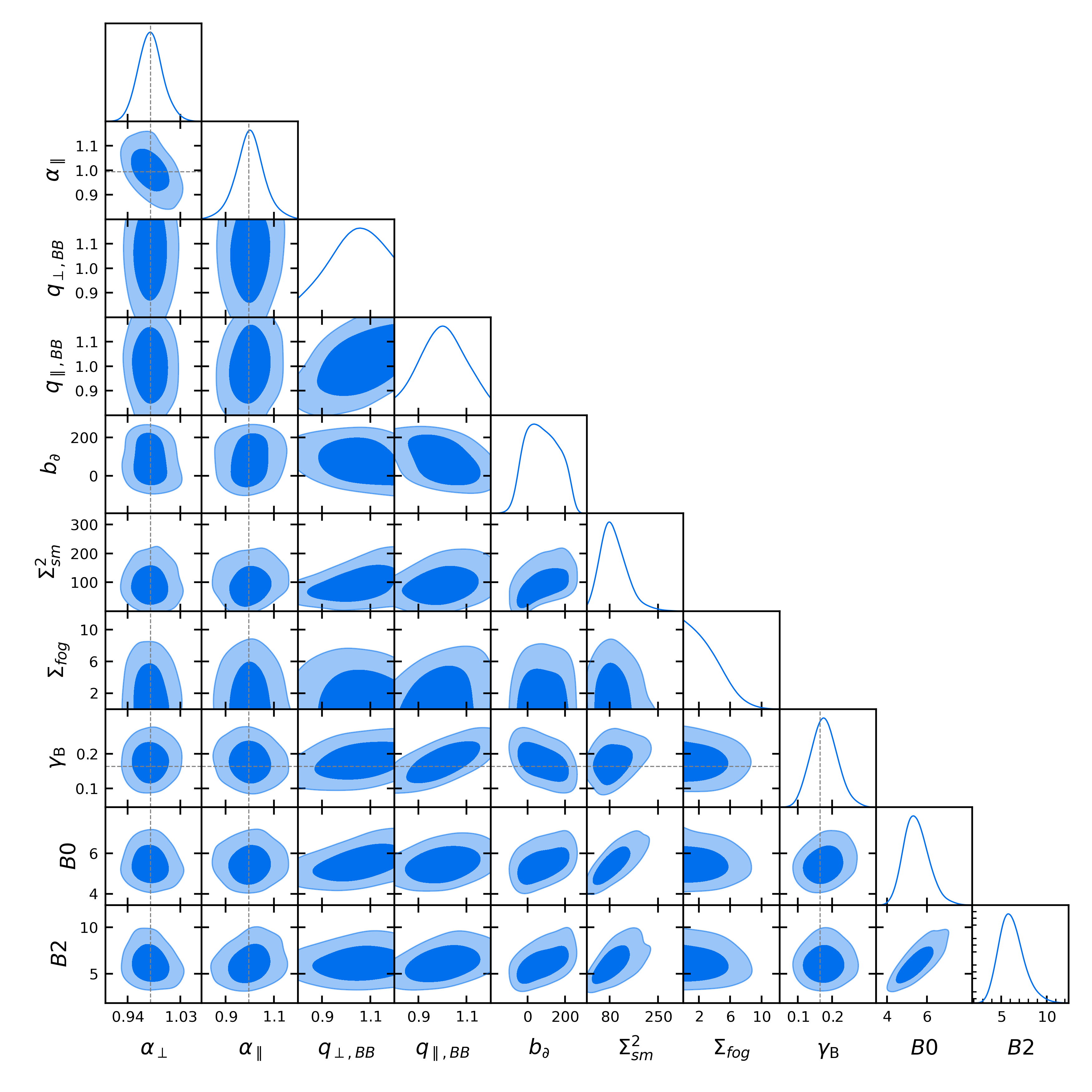}
    \caption{Parameter constraints for the damped BAO template model fitted to the pre-reconstruction correlation function monopole and quadrupole at $50 < r < 150$\,$h^{-1}$\,Mpc,
    averaged over the 84 Nseries mocks. We use the covariance for BOSS $z_3$ (combining NGC and SGC). The polynomial terms are marginalized analytically and hence not shown in this plot.
    \label{fig:triangle_pre_recon}}
\end{figure*}

We show a triangle plot of parameter constraints in Fig.~\ref{fig:triangle_pre_recon}. $\gamma_\text{B}$ is not well correlated with the BAO shift parameters $\alpha_{\parallel}$ and $\alpha_\perp$ or the biases $B_0$ and $B_2$. It is  somewhat degenerate with the broadband shifting $q_{\parallel,BB}$, the derivative bias $b_\partial$ and the smoothing scale $\Sigma_{\textrm{sm}}$. The degeneracy with $b_\partial$ recalls the $b_2$-$\gamma_\text{B}$ degeneracy in the full-shape fits, and the $b_2$ contributions to the nonlinear BAO shift. The simple model that we use cannot generate nonlinear BAO shifts, since these come from one-loop terms. However, the $k^2 P_{\textrm{lin}}$ template, multiplied by $b_\partial$, must have a different scale dependence from $P_{\textrm{lin}}$ and therefore can change the amplitude of the BAO peak. Likewise, increasing the BAO damping $\Sigma_{\textrm{sm}}$ suppresses the BAO peak, and therefore must be compensated by increasing $\gamma_\text{B}$, leading to a $\gamma_\text{B}$-$\Sigma_{\textrm{sm}}^2$ degeneracy.

\subsubsection{Robustness of the modeling}

The amplitude of the baryon signature in the pre-reconstruction template fits is quite robust to extensions to $\Lambda$CDM, even as we always use the same fiducial $\Lambda$CDM model for the template. We find $<0.3\sigma$ biases on $\gamma_\text{B}$ when using an EDE cosmology with a different sound horizon as the input data, or when using a cosmology with a higher neutrino mass or $N_{\textrm{eff}}$, inducing a phase shift in the BAO wiggles. The BAO scaling parameters $\alpha_{\parallel}$ and $\alpha_{\perp}$ are also recovered well ($<0.3\sigma$ bias) in all cases except for $\alpha_{\parallel}$ when $\Sigma m_{\nu} = 0.2$ eV, in which case the recovery is modestly worse, $0.5\sigma$. This is similar to past results showing that the BAO scale is robust to changes in $N_{\textrm{eff}}$ and $\Sigma m_{\nu}$ \citep{ThepsuriyaLewis}.

We also show that the recovered baryon fraction and shift parameters are robust to changing the true values of the nuisance parameters $b_\partial$, $\Sigma_{xy}$, and $\Sigma_{\textrm{fog}}$, except when the damping parameter $\Sigma_{xy}$ becomes very large at 12\,$h^{-1}$\,Mpc. In this case, the BAO amplitude is highly suppressed and poorly detected. The bias on $\gamma_\text{B}$ is still 0.67$\sigma$ even if we measured it relative to the larger $\sigma_{\gamma_\text{B}} = 0.052$ from this fit, rather than $\sigma_{\gamma_\text{B}} = 0.029$ from the Nseries fit with scaled-down covariance. However, it isn't surprising that our model doesn't work well when the BAO shifts are so poorly constrained, and models with such a high damping parameter are robustly excluded by both the data and theoretical considerations anyway. Finally, we test the dependence of our model on inaccuracies in the \cite{Eisenstein-Hu} transfer function fits by replacing the smooth $P_{\textrm{nw}}$ used to construct the mocks with a Fourier-transformed spline fit to two windows in the correlation function on either side of the BAO bump. This leads to negligible changes in the fitted parameters.

\subsubsection{Fitting noiseless theory vectors}

\begin{figure*}
\includegraphics[scale=0.6]{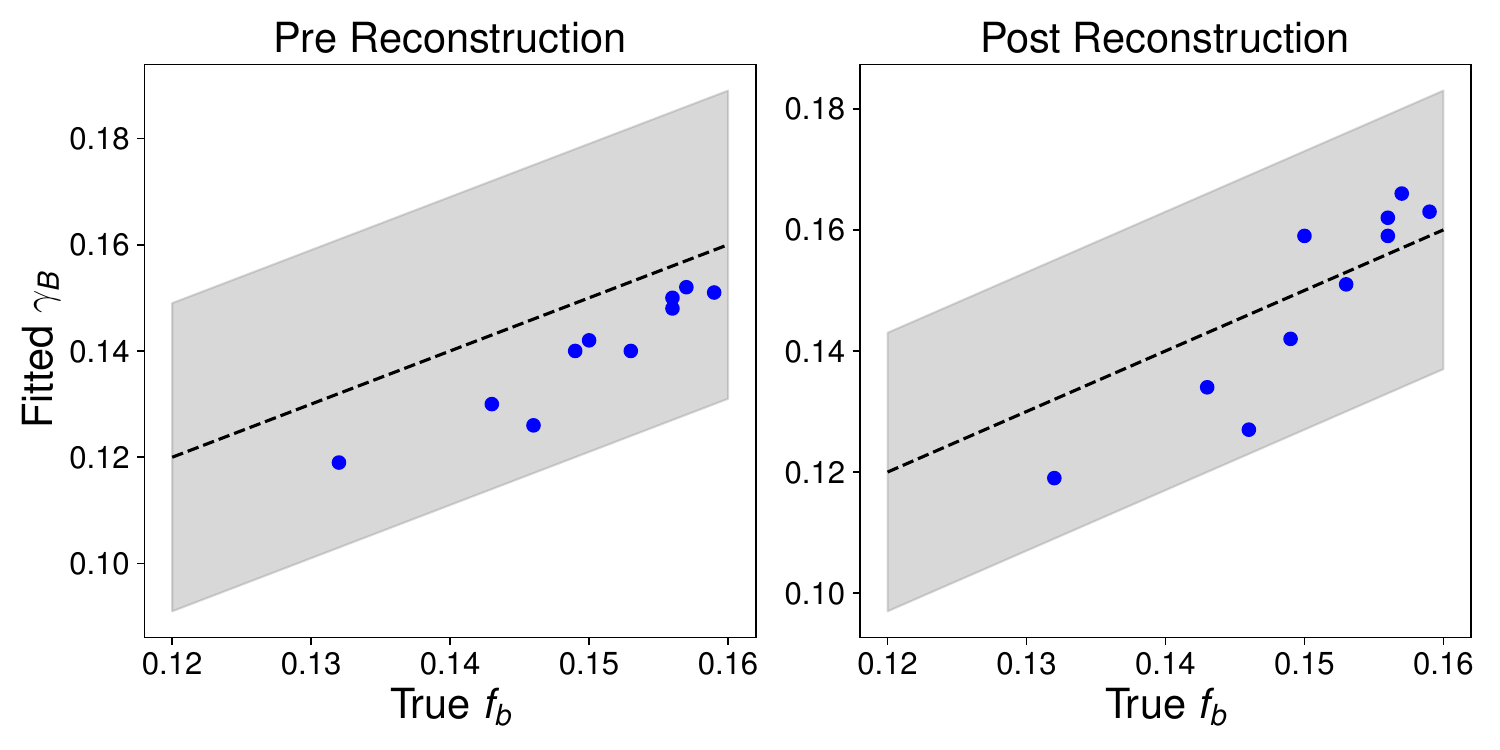}
    \caption{
    Fitted $\gamma_\text{B}$ and true $f_b$ for noiseless theory vector tests with varying $\Lambda$CDM parameters, both for pre-reconstruction template fits (\textit{left}) and post-reconstruction (\textit{right}). Gray errorbands are the 1$\sigma$ errors from Nseries fit with the BOSS $z_3$ covariance scaled to the entire BOSS volume.
    \label{fig:h_scatter_template}}
\end{figure*}

Finally, we test whether the baryon fraction is recovered well when changing the $\Lambda$CDM parameters of the input noiseless theory vector (Fig.~\ref{fig:h_scatter_template}). Here we find small but persistent systemic biases on $\gamma_\text{B}$ (average $-0.36\pm$0.15$\sigma$), $\alpha_\perp$ (average -0.19$\pm$0.13$\sigma$) and $\alpha_{\parallel}$ (average 0.34$\pm$0.10$\sigma$). These biases would be much larger if we only included a constant term in the monopole and not the quadrupole (matching our treatment of the post-reconstruction correlation function), -0.55$\sigma$, -0.46$\sigma$, and 0.61$\sigma$ in $\gamma_\text{B}$, $\alpha_\perp$ and $\alpha_\parallel$. This motivates the inclusion of the extra constant term in the quadrupole. Likewise, the biases would also be larger if we allowed a wider prior on $b_\partial$, $-0.64\sigma$, $-0.33\sigma$, and $0.46\sigma$ in $\gamma_\text{B}$, $\alpha_\perp$ and $\alpha_\parallel$, if we allowed a flat prior on $b_\partial$ between -1000 and 1000. However, shrinking the $b_\partial$ prior too much leads to unacceptably large biases on Nseries; hence we choose a flat prior on $b_\partial$ between -250 and 250 to balance these concerns.

\subsection{Post-reconstruction template fits}

\begin{table*}[]
    \centering
    \begin{tabular}{l|cccc|cccc|cccc}
    Data & $\gamma_\text{B}$ & $\sigma_{\gamma}$
     & $f_b$ & $n\sigma$ &
      $\alpha_{\perp}$ & $\sigma_{\alpha_{\perp}}$ &
      True  &
     $n\sigma$ &
     $\alpha_{\parallel}$ & $\sigma_{\alpha_{\parallel}}$ & True  &
     $n\sigma$  \\    
    \hline
Nseries & 0.171 & 0.030 & 0.164 & 0.31 & 0.978 & 0.012 & 0.979 & -0.09 & 0.995 & 0.025 & 0.988 & 0.43 \\
$z_1$+$z_3$ Cov.\  & 0.170 & 0.023 & 0.164 & 0.27 & 0.978 & 0.008 & 0.979 & -0.12 & 0.994 & 0.016 & 0.988 & 0.42 \\
\hline
$\Lambda$CDM & 0.164 & 0.034 & 0.155 & 0.37 & 0.999 & 0.013 & 1.000 & -0.12 & 1.002 & 0.026 & 1.000 & 0.10 \\
EDE & 0.178 & 0.031 & 0.174 & 0.19 & 1.019 & 0.010 & 1.019 & 0.02 & 1.023 & 0.021 & 1.022 & 0.09 \\
$b_\partial = 5$  & 0.153 & 0.034 & 0.155 & -0.10 & 0.998 & 0.015 & 1.000 & -0.18 & 1.002 & 0.028 & 1.000 & 0.12 \\
$b_\partial = 25$  & 0.176 & 0.036 & 0.155 & 0.89 & 1.000 & 0.012 & 1.000 & -0.04 & 0.999 & 0.023 & 1.000 & -0.04 \\
$\Sigma_{\textrm{sm}} = 8$ & 0.157 & 0.032 & 0.155 & 0.10 & 0.999 & 0.012 & 1.000 & -0.10 & 1.000 & 0.025 & 1.000 & 0.03 \\
$\Sigma_{\textrm{sm}} = 24$ & 0.157 & 0.032 & 0.155 & 0.08 & 1.000 & 0.016 & 1.000 & -0.05 & 1.001 & 0.029 & 1.000 & 0.07 \\
$\Sigma_{\textrm{fog}} = 0$  & 0.163 & 0.034 & 0.155 & 0.34 & 0.999 & 0.013 & 1.000 & -0.11 & 1.001 & 0.025 & 1.000 & 0.08 \\
$\Sigma_{\textrm{fog}} = 8$  & 0.162 & 0.034 & 0.155 & 0.30 & 0.998 & 0.013 & 1.000 & -0.21 & 1.002 & 0.026 & 1.000 & 0.12 \\
\small{$\Sigma m_{\nu} = 0.2$eV} & 0.166 & 0.031 & 0.164 & 0.09 & 1.003 & 0.012 & 1.000 & 0.44 & 1.005 & 0.023 & 1.000 & 0.35 \\
$N_{\textrm{eff}} = 3.82$ & 0.157 & 0.033 & 0.149 & 0.36 & 0.992 & 0.013 & 0.994 & -0.27 & 0.999 & 0.025 & 0.999 & 0.01 \\
Spline $P_{\textrm{nw}}$ & 0.170 & 0.032 & 0.155 & 0.64 & 0.997 & 0.012 & 1.000 & -0.36 & 0.999 & 0.022 & 1.000 & -0.06 \\
    \end{tabular}
    \caption{Template fits to the post-reconstruction correlation function. Format is the same as Table~\ref{tab:pre-recon_template_fits}.
    \label{tab:post-recon_template_fits}}
    \end{table*}

\subsubsection{Fits to the Nseries mocks}

We recover the true values of $\gamma_\text{B}$, $\alpha_{||}$ $\alpha_\perp$ with biases $< 0.4 \sigma$ when fitting to the mean of the 84 Nseries mocks (Table~\ref{tab:post-recon_template_fits}). These results are much less dependent on the priors than the pre-reconstruction results, with shifts of $<0.05\sigma$ in all parameters. We find good agreement between the pre and post-reconstruction results, particularly in the $\alpha$ parameters; as in the pre-reconstruction results, the bias that we measure on $\alpha_{\parallel}$ has opposite sign to \cite{GilMarin20}.

If we used the linear BAO model of Ref.~\cite{SBRS}, rather than the Ref.~\cite{Ding18} model, we would find  a significant bias on $\gamma_\text{B}$ on the Nseries mocks, with  $\gamma_\text{B} = 0.189 \pm 0.035$ (0.71$\sigma$ bias). This is largely due to the absence of the derivative bias, $b_\partial$, which adds power on small scales to partially emulate the effects of nonlinear structure formation. By adding a component $\propto k^2 P_{\textrm{lin}}$, which has a different weighting of the BAO oscillations, $b_\partial$ can cause the amplitude of the BAO peak to change.

We place an informative prior on $b_\partial$, a Gaussian centered at 15 with a standard deviation of 5. This value of $b_{\partial}$ is motivated by the best-fit values of $b_\partial$ in the EFT1 model of ref.~\cite{Ding18}, $b_\partial \sim 15$.\footnote{This result depends mildly on redshift but not on linear bias, even over a large range in mean halo mass (Fig.\ 14 in \cite{Ding18}). This suggests that the $\gamma_\text{B}$ measurement is insensitive to the details of the halo occupation distribution.}
In this model, we find $\gamma_\text{B} = 0.171 \pm 0.030$,
an $0.31\sigma$ offset from the truth (using the reduced errorbar from the fit with the covariance scaled to the full BOSS volume). This result is dependent on the $b_\partial$ prior: if we do not impose any prior on $b_\partial$, we find $\gamma_\text{B} = 0.191 \pm 0.034$ and $b_\partial = -4.0 \pm 12.8$. The $b_\partial$ prior allows us to recover an unbiased value of the baryon fraction, at the price of shifting the nuisance parameters ($\sim 1\sigma$ shifts in $\Sigma_{\textrm{fog}}$ and $\Sigma_{sm}$) and a slightly worse fit ($\Delta \chi^2 = 3$). Essentially, the $b_\partial$ prior is a simulation-based constraint on the impact of nonlinear structure formation on the baryon fraction.

\begin{figure*}
\includegraphics[]{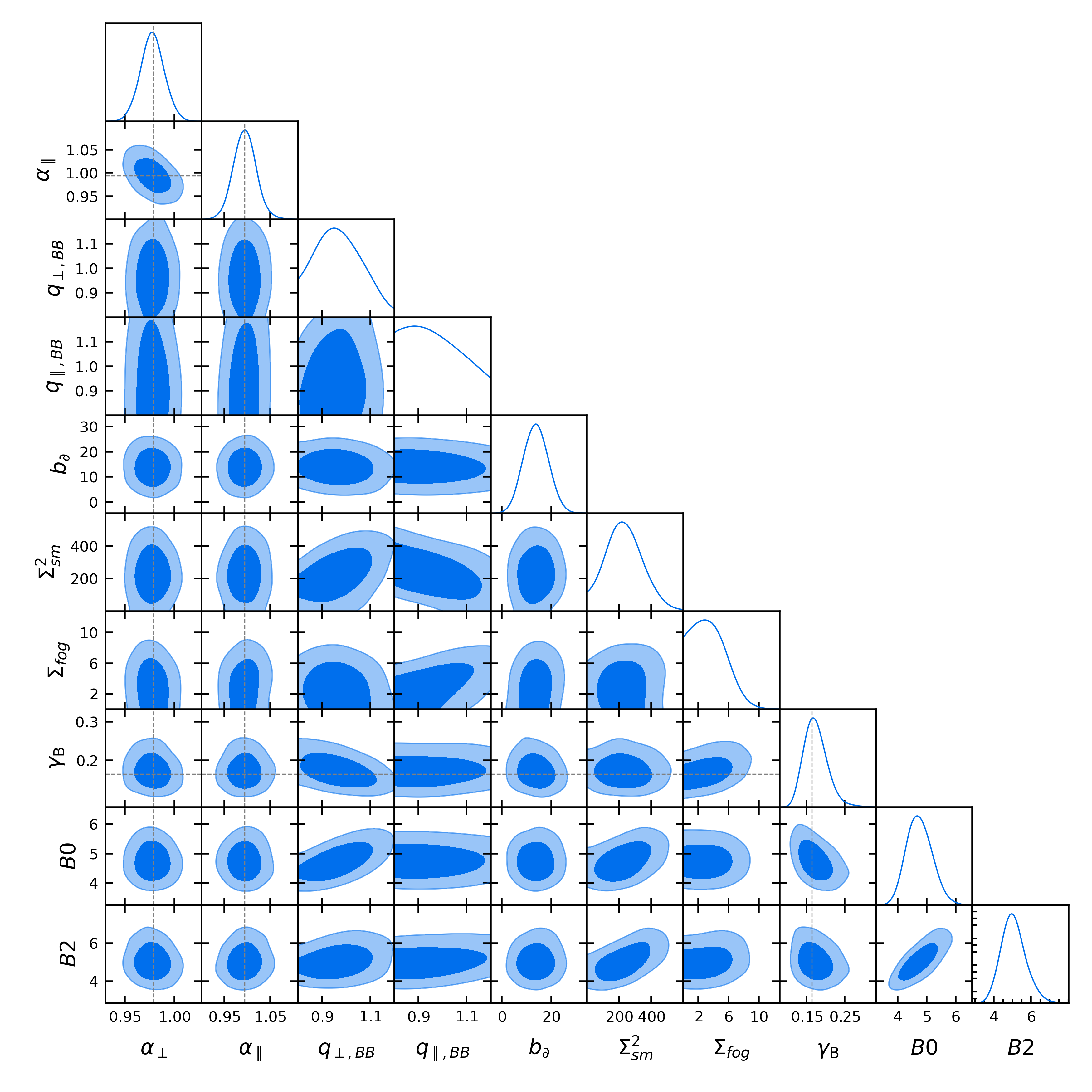}
    \caption{Same as Fig.~\ref{fig:triangle_pre_recon}, but for the post-reconstruction correlation function averaged over the 84 Nseries mocks.
    \label{fig:triangle_post_recon}}
\end{figure*}

We show a triangle plot of parameter constraints in Fig.~\ref{fig:triangle_post_recon}. The degeneracies are similar to the pre-reconstruction case in Fig.~\ref{fig:triangle_pre_recon}, but generally less prominent. While $\gamma_\text{B}$ appears to be much less degenerate with $b_\partial$ in the post-reconstruction case, we note that this is largely due to the restrictive prior placed on $b_\partial$ post-reconstruction.

\subsubsection{Robustness of the modeling}

We use a considerably more restricted set of broadband polynomials than the standard BAO method used in \cite{Ross2017,SBRS}. Rather than the 6 polynomials used in that model (3 for each multipole), we use a single constant term for the monopole. We also vary four other physically motivated broadband terms\footnote{Three of these parameters are truly free, while $b_\partial$ is prior-dominated.}: 2 shift parameters $q_{||,BB}$ and $q_{\perp,BB}$, $\Sigma_{fog}$ and $b_{\partial}$.  The reason for this is the strong degeneracy between higher-order polynomials and the BAO amplitude, which can partially mimic the BAO bump and distort the size of the broadband that the BAO bump is compared with to constrain the BAO amplitude.
Simultaneously fitting $\gamma_\text{B}$ and the 6 polynomial terms leads to a nearly unconstrained $\gamma_\text{B} = 0.31 \pm 0.18$. Despite the weak constraint on $\gamma_\text{B}$, the BAO feature is still detected at high significance (8.3$\sigma$),\footnote{Defined as the square-root in the difference of $\chi^2$ between the best-fit model with $\gamma = 0$ (no BAO) and fixing $\gamma = 0.1643$.} due to the highly non-Gaussian shape of the $\gamma_\text{B}$ posterior. Most of the loss of constraining power comes from the $1/r^2$ term. We furthermore find that the other two polynomial terms are not favored in fits to data (i.e.\ they do not improve the goodness of fit by more than expected for adding extra degrees of freedom). However, we keep the constant term in the monopole to marginalize over the potential impact of observational systematics, which produce nearly constant changes in the monopole on the large scales that we fit ($r > 50$ $h^{-1}$\,Mpc). This is shown in Fig.\ 6 in \cite{Ross2017}, which demonstrates that turning the stellar density weights off changes the monpole by an additive constant on large scales, and does not affect the quadrupole.\footnote{Fig.\ 6 in \cite{Ross2017} shows a large scale-dependent impact of seeing weights in the LOWZE3 region; however, this is $<10\%$ of the total BOSS area and the impact of seeing weights are negligible across the entire footprint.}

As in the pre-reconstruction case, we find good recovery (generally $<0.3\sigma$) of $\gamma_\text{B}$, $\alpha_{\parallel}$ and $\alpha_{\perp}$ for the beyond-$\Lambda$CDM models considered: early dark energy, massive neutrinos, and extra relativistic degrees of freedom. We also find that $\gamma_\text{B}$ is insensitive to changes in the nuisance parameters, with the exception of $b_\partial$. Large values of $b_\partial$ create a bias towards high $\gamma_\text{B}$. Given the $b_\partial$-$\gamma_\text{B}$ degeneracy discussed before and noted for the pre-reconstruction case, this is not surprising.



\subsubsection{Fitting noiseless theory vectors}

The post-reconstruction fits are less sensitive to changing the $\Lambda$CDM parameters than the pre-reconstruction fits (Fig.~\ref{fig:h_scatter_template}). We find a mean bias of $-0.09\sigma$ and a standard deviation of $0.40\sigma$ across the ten cosmologies tested. Additionally, $\alpha_\perp$ and $\alpha_\parallel$ are considerably less biased than in the pre-reconstruction case, with biases of $-0.12\sigma$ and $0.10\sigma$, respectively. The considerably weaker dependence on the fiducial cosmology in the post-reconstruction case may be due to the sharper BAO peak, which is easier to distinguish from the broadband and hence less degenerate with the cosmological parameters that affect its shape.
The scatter when changing cosmologies is much larger if we instead scale the entire correlation function by $\alpha_{||}$ and $\alpha_\perp$ (dropping the two extra broadband shift parameters), with a 0.74$\sigma$ scatter in $\gamma_\text{B}$.\footnote{This can be mitigated by adding the ShapeFit parameter $m$ \cite{Brieden21} to marginalize over the impact of the fiducial cosmology. A relatively narrow prior on $m$ between -0.1 and 0.1 encompasses the range in cosmological parameters constrained by the BOSS full shape fits, reducing the scatter in $\gamma_\text{B}$ to 0.31$\sigma$ without adding significant prior volume effects. However, in our fiducial method with 4 scaling parameters, adding $m$ as an additional parameter does not reduce the scatter in $\gamma_\text{B}$ when varying cosmology, and generally has minimal impact on the results. For simplicity (and due to the arbitrariness of the chosen priors in $m$ to avoid prior effects), we therefore do not vary $m$ in our baseline model.} While the baseline constraints on Nseries are very similar whether we use the standard two scaling parameters or the new method with four scaling parameters, we favor using four scaling parameters because it decouples $\gamma_\text{B}$  from the broadband changing with the fiducial cosmology.


\subsection{Summary of mock tests}
\label{sec:summary-mock-tests}

\begin{figure}
    \includegraphics[width=0.5\textwidth]{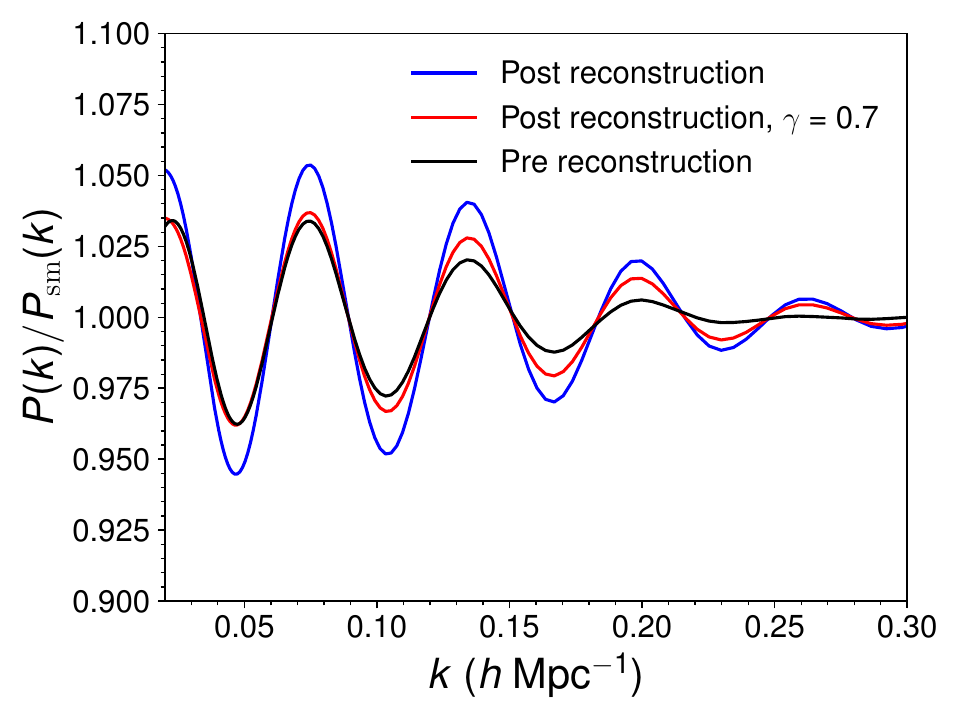}
    \caption{The oscillatory portion of the power spectrum, $P(k)/P_{\textrm{nw}}(k)$, for a pre-reconstruction model, and a post-reconstruction model with the default $\gamma_\text{B}$ and with $\gamma_\text{B} = 0.7$. The effects of reconstruction (or equivalently, reducing the BAO damping) and changing $\gamma_\text{B}$ are distinct; reconstruction sharpens the higher-$k$ peaks whereas $\gamma_\text{B}$ affects all peaks equally.
    Therefore, reconstruction does not lead to tighter constraints on $\gamma_\text{B}$.
    \label{fig:gamma_vs_damping}}
\end{figure}

We have demonstrated three separate methods to fit the amplitude of the baryon signature that recover similar results from Nseries mocks: $0.172^{+0.031}_{-0.027}$ for full-shape perturbative fitting to $P(k)$; $0.167 \pm 0.029$ for pre-reconstruction template fitting to $\xi(r)$; and $0.170 \pm 0.023$ for post-reconstruction template fitting, compared to the true value of $f_b = 0.164$. The constraining power of the post-reconstruction fits is slightly better, with a 20\% reduction in the errorbar. However, this is largely attributable to the informative prior on $b_\partial$; if we instead place the same broad flat prior on $b_\partial$ that is used for the pre-reconstruction fits, $\sigma_{\gamma_\text{B}}$ = 0.026. Hence, despite the sharper BAO peak post-reconstruction, the constraints on $\gamma_\text{B}$ are not tighter. This is more readily appreciated in Fourier space, where we see that changing $\gamma_\text{B}$ changes the amplitudes of all peaks, whereas reconstruction changes the damping scale and thus changes the relative peak heights (Fig.~\ref{fig:gamma_vs_damping}). Since $\gamma_\text{B}$ is not very degenerate with the damping parameters, sharpening the BAO peaks therefore does not improve $\gamma_\text{B}$ constraints much.

We also find that all methods are able to recover the correct baryon fraction if the true cosmology is an EDE cosmology rather than $\Lambda$CDM. When considering the best-fit EDE model to Planck from \cite{Hill20} as the truth, we find biases in $\gamma_{\text{B}}$ of $-0.43\sigma$, $-0.29\sigma$, and $0.19\sigma$ for full-shape, pre-reconstruction, and post-reconstruction fits, respectively. Moreover, in the full-shape case, we test parameter recovery on EDE cosmologies with varying $f_b$, and find
 biases $\sim 0.3\sigma$ that are similar to the biases seen if the true model is $\Lambda$CDM, and hence may be attributable to prior effects.

We place a systematic error of 0.013 on $\gamma_\text{B}$ (equivalent to $\sim0.5\sigma$) measured in all three methods, when combining all redshift bins. This roughly encapsulates the biases seen on the fits to Nseries; the effects of prior  effects on the full-shape fits; and biases seen when varying the true cosmology and baryon fraction. We do not shift the central value of $\gamma_\text{B}$, since the Nseries tests are generally biased high whereas the varying cosmology (noiseless theory vector) tests are biased low, i.e. we do not trust the Nseries power spectrum at the level where we can use the fits to set shifts on recovered parameters.

\section{Baryon fraction from BOSS data}
\label{sec:baofit_data}

\begin{table*}[]
    \centering
    \small
    \begin{tabular}{l|ccccc}
    Model & $\chi^2$ &  $\gamma_{\rm B}$ & BAO significance ($n\sigma$) & $\alpha_{\perp}$ & $\alpha_{\parallel}$ \\
    \hline
    \textbf{$0.2 < z < 0.5$} & & & &   \\
    \hline
    Pre-Recon & 38 &  $0.156 \pm 0.038$ & 6.3 & $0.979 \pm 0.021$ & $1.051 \pm 0.041$ \\
    Post-Recon & 40 & $0.175 \pm 0.037$ & 8.3 & $0.985 \pm 0.012$ & $1.021 \pm 0.026$ \\
    Full-Shape & 210 & $0.157^{+0.031}_{-0.033}$ \\
    \hline
    \textbf{$0.4 < z < 0.6$} & & & &   \\
    \hline
    Pre-Recon & 43 & $0.156 \pm 0.040$ & 6.3 & $1.00 \pm 0.026$ & $1.022 \pm 0.061$ \\
    Post-Recon & 25 & $0.162 \pm 0.032$ & 8.4 & $0.993 \pm 0.011$ & $0.986 \pm 0.020$ \\  
    \hline
    \textbf{$0.5 < z < 0.75$} & & & &   \\
    \hline
    Pre-Recon & 29 & $0.177 \pm 0.044$ & 6.6 & $0.990 \pm 0.024$ & $0.964 \pm 0.053$ \\
    Post-Recon & 31 & $0.204 \pm 0.042$ & 8.7 & $0.994 \pm 0.012$ & $0.953 \pm 0.019$ \\ 
    Full-Shape & 188 & $0.145^{+0.031}_{-0.032}$ \\
    \hline
    Combined \\
    \hline
    Pre-Recon & & $0.153 \pm 0.029$ \\
    Post-Recon & & $0.173 \pm 0.029$ \\
    Full-Shape & & $0.154 \pm 0.023$ \\
    \end{tabular}
    \caption{Results from fits to the BOSS pre-reconstruction and post-reconstruction correlation functions (using the template method) and unreconstructed power spectrum (using the full shape method). The template fits are split by redshift bin, and we show both $\gamma_\text{B}$ and the Alcock-Paczynski parameters, which are jointly fit. We also show the combined results for $\gamma_\text{B}$, including all bins with the covariance computed from 1000 Patchy mocks. We have 28 (29) degrees of freedom for the pre- (post-)  reconstruction fits, and 195 degrees of freedom (230 data points and 35 parameters) for the full-shape fits.
    \label{tab:results}}
\end{table*}





\subsection{BOSS data}
\label{subsec:data_description}

We use the BOSS \cite{Dawson13} galaxy catalogs released in the 12th Data Release (DR12) of SDSS \cite{Alam15,Alam17}, part of the SDSS-III program \cite{Eisenstein11}.
BOSS galaxies are selected \cite{Reid16} from 5-band \cite{Fukugita96} SDSS imaging \cite{Gunn98,Gunn06} across 14,555 deg$^2$ observed using the BOSS spectrograph \cite{Smee13} on the SDSS telescope \cite{Gunn06} with high redshift success rate \cite{Bolton12}.
BOSS contains 1.4 million galaxies, targeted in two regions of the sky (North Galactic Cap and South Galactic Cap) and with two distinct
galaxy selections, LOWZ and CMASS. LOWZ and CMASS were then merged together into the ``CMASSLOWZTOT'' sample \cite{Reid16},\footnote{Publicly available at \url{data.sdss.org/sas/dr12/boss/lss}.} and simple redshift cuts
were used to define 3 bins: $z_1$ at $0.2 < z < 0.5$, $z_2$ at $0.4 < z < 0.6$, and $z_3$ at $0.5 < z < 0.75$.

To facilitate comparison with past results, we use publicly available correlation functions\footnote{\url{https://github.com/ashleyjross/BAOfit/tree/master/exampledata/Ross_2016_COMBINEDDR12}} and power spectra.\footnote{\url{https://github.com/oliverphilcox/Spectra-Without-Windows}} The correlation functions
used for the template fits are measured with and without BAO reconstruction \cite{eisenstein07}. The BOSS team used the ``Rec-Iso'' convention (i.e.\ the quadrupole is suppressed after running reconstruction), with reconstruction implemented following the algorithm of \cite{Padmanabhan12}. The post-reconstruction correlation function is used for the fiducial BAO fitting result of BOSS, due to the increased signal-to-noise of the BAO peak and reduction in nonlinear systematics that shift the peak. The covariance matrix is measured from 1000 approximate MultiDark-Patchy mocks \citep{Kitaura16,Rodriguez-Torres16}, created using fast, approximate $N$-body simulations and cut to the BOSS sky area and redshift cuts. For the pre-reconstruction correlation functions, we measure the correlation functions and power spectra ourselves from the available pair-counts of the DR12 catalogs and Patchy mocks.\footnote{We found poor agreement with the pre-reconstruction BAO fits in Table 6 of \cite{Ross2017} (even when using their method and code),
if we used the publicly available correlation functions and covariance matrices of \cite{Satpathy17}. Re-measuring the correlation function did not change the results much, whereas re-measuring the covariance matrix led to much better agreement, though it was still not perfect.}
We use the correlation function monopole ($\xi_0$) and quadrupole ($\xi_2$) measured in 20 bins of width 5\,$h^{-1}$\,Mpc between 50\,$h^{-1}$\,Mpc and 150\,$h^{-1}$\,Mpc. Following \cite{Ross2017}, for the post-reconstruction fits, we average $\chi^2$ over fits to data with five different choices of bin center separated by 1\,$h^{-1}$\,Mpc (i.e.\ data starts at 50, 51, 52, 53, and 54\,$h^{-1}$\,Mpc). For the pre-reconstruction fits, we use a single bin center with no offset.

For the full-shape fits, we use the publicly available power spectra of \cite{Philcox_Ivanov}, measured using a window-free quadratic estimator \cite{Philcox21}. While these power spectra are slightly different from those released by the BOSS collaboration (which are convolved with the survey window), we use the new window-free measurements for compatibility with the full-shape pipeline of \cite{Philcox_Ivanov},\footnote{\url{https://github.com/oliverphilcox/full_shape_likelihoods}} which did not exist until after the BOSS data were released. We use the power spectrum monopole ($P_0$), quadrupole ($P_2$) and hexadecapole ($P_4$) measured between $k = 0.01$\,$h$\,Mpc$^{-1}$
and $k = 0.20$\,$h$\,Mpc$^{-1}$ in bins of $\Delta k = 0.005$\,$h$\,Mpc$^{-1}$. Unlike the correlation functions, the power spectrum measurements are split by Galactic cap, since the number density and galaxy selection is slightly different between the two Galactic caps, potentially leading to different best-fit values of the galaxy bias parameters. This change in the nuisance parameters is more relevant for full-shape fitting, which requires accurate modelling of the nuisances to extract cosmological information from the shape of the galaxy power spectrum. Furthermore, the power spectra were only measured for the non-overlapping redshift bins at $0.2 < z < 0.5$ and $0.5 < z < 0.75$, reducing the need to estimate the covariance between all three redshift bins. This would be quite noisy due to the large size of the combined data vector, which is comparable to the number of mocks available.

Covariance matrices are again computed from 1000 Patchy mocks with matching survey geometry. By default, we combine the power spectrum measurements with the post-reconstruction Alcock-Paczynski parameters and directly fit cosmological parameters to the combined dataset. There is a nontrivial covariance between the post-reconstruction AP parameters and the pre-reconstruction power spectrum; we also use the 1000 Patchy mocks to measure this covariance. Unlike in \cite{Philcox_Ivanov}, we use the consensus (Fourier space \cite{Beutler16BAO} plus configuration space \cite{Ross2017}) AP parameters released by the BOSS team, in which NGC and SGC are combined. This allows us to measure
both the covariance between $P_\ell$, $\alpha_{\parallel}$ and $\alpha_{\perp}$, and between $P_\ell$ and $F_{\textrm{AP}} \equiv D_M/D_H$ (which is required for our explicitly sound horizon independent measurements).

In both cases we account for the bias when inverting a noisy covariance matrix by using the Gaussian approximation prescription of Ref.~\cite{Percival22}, which updates the Ref.~\cite{Hartlap07} treatment used in Ref.~\cite{Ross2017} and Ref.~\cite{Philcox_Ivanov}.

\subsection{Results}

\begin{figure*}
\includegraphics[scale=0.8]{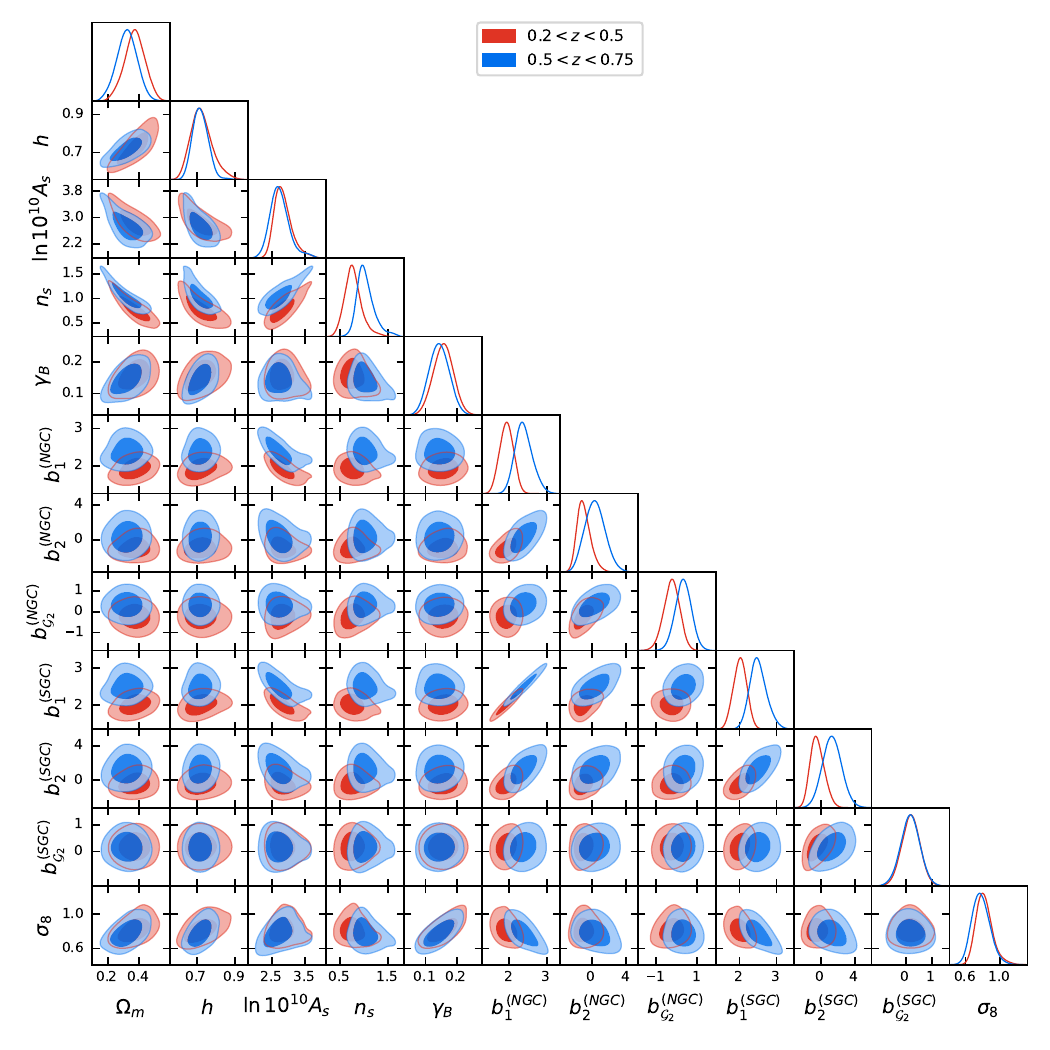}
\includegraphics[scale=0.4]{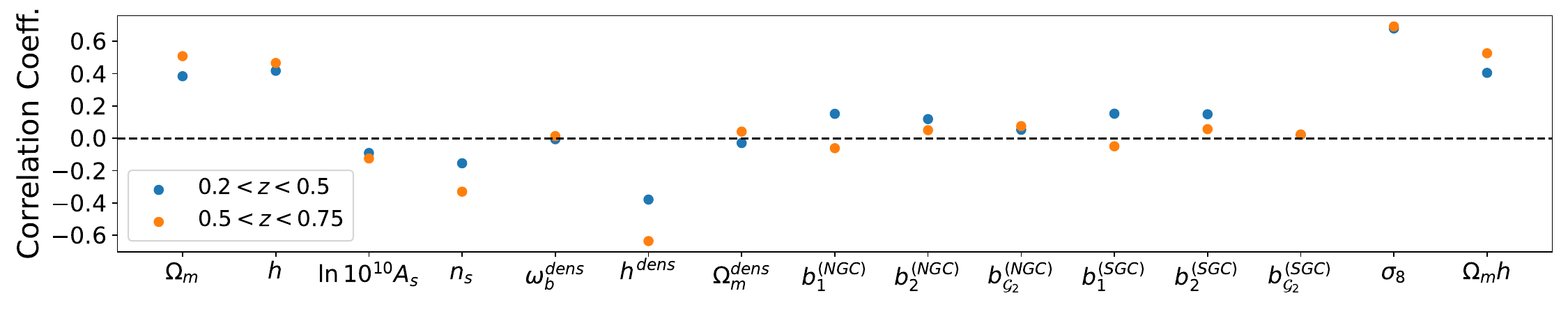}
    \caption{Parameter constraints on BOSS data, using the full-shape method and comparing the two bins, $0.2 < z < 0.5$ and $0.5 < z < 0.75$. The bottom panel shows correlation coefficients between the parameter $\gamma_\text{B}$ and the other parameters of the model; the strongest degeneracies are with the cosmological parameters.
    \label{fig:triangle_full_shape_data}}
\end{figure*}

\begin{figure*}
\includegraphics[scale=0.8]{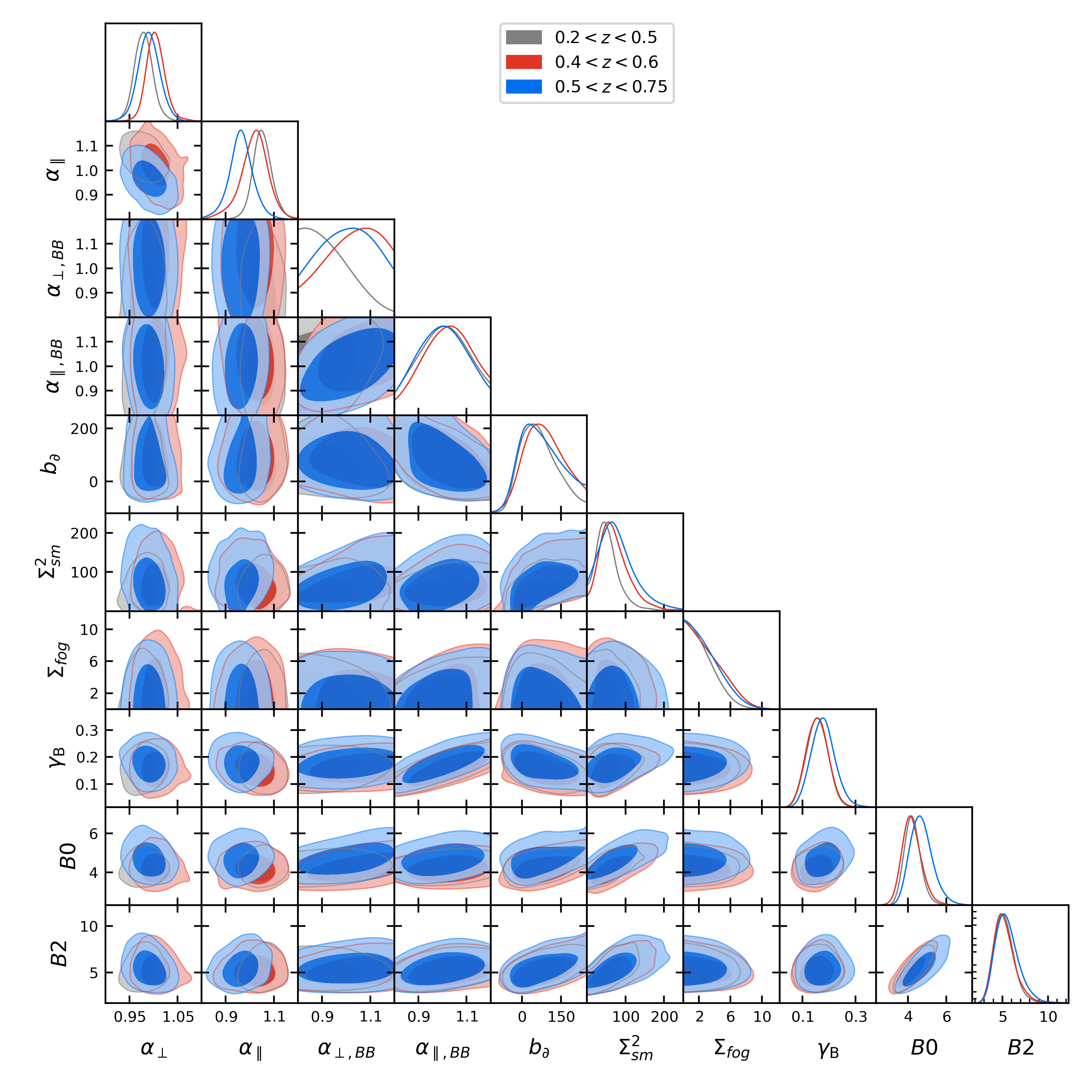}
    \caption{Parameter constraints on BOSS data, from the pre-reconstruction template fits in three bins.
    \label{fig:triangle_prerecon_data}}
\end{figure*}

\begin{figure*}
\includegraphics[scale=0.8]{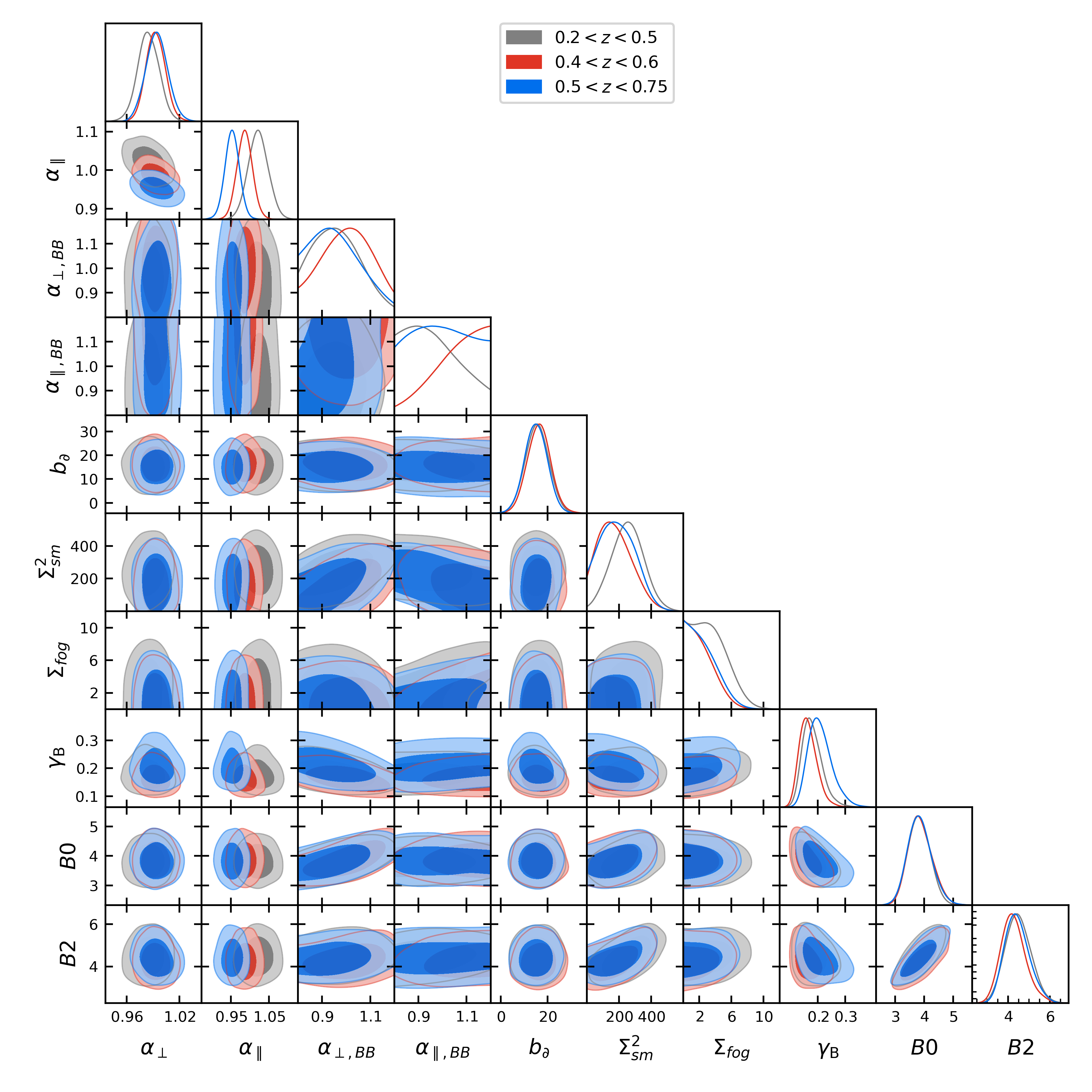}
    \caption{Parameter constraints on BOSS data, from the post-reconstruction template fits in three bins.
    \label{fig:triangle_postrecon_data}}
\end{figure*}

\begin{figure*}
\includegraphics[scale=0.5]{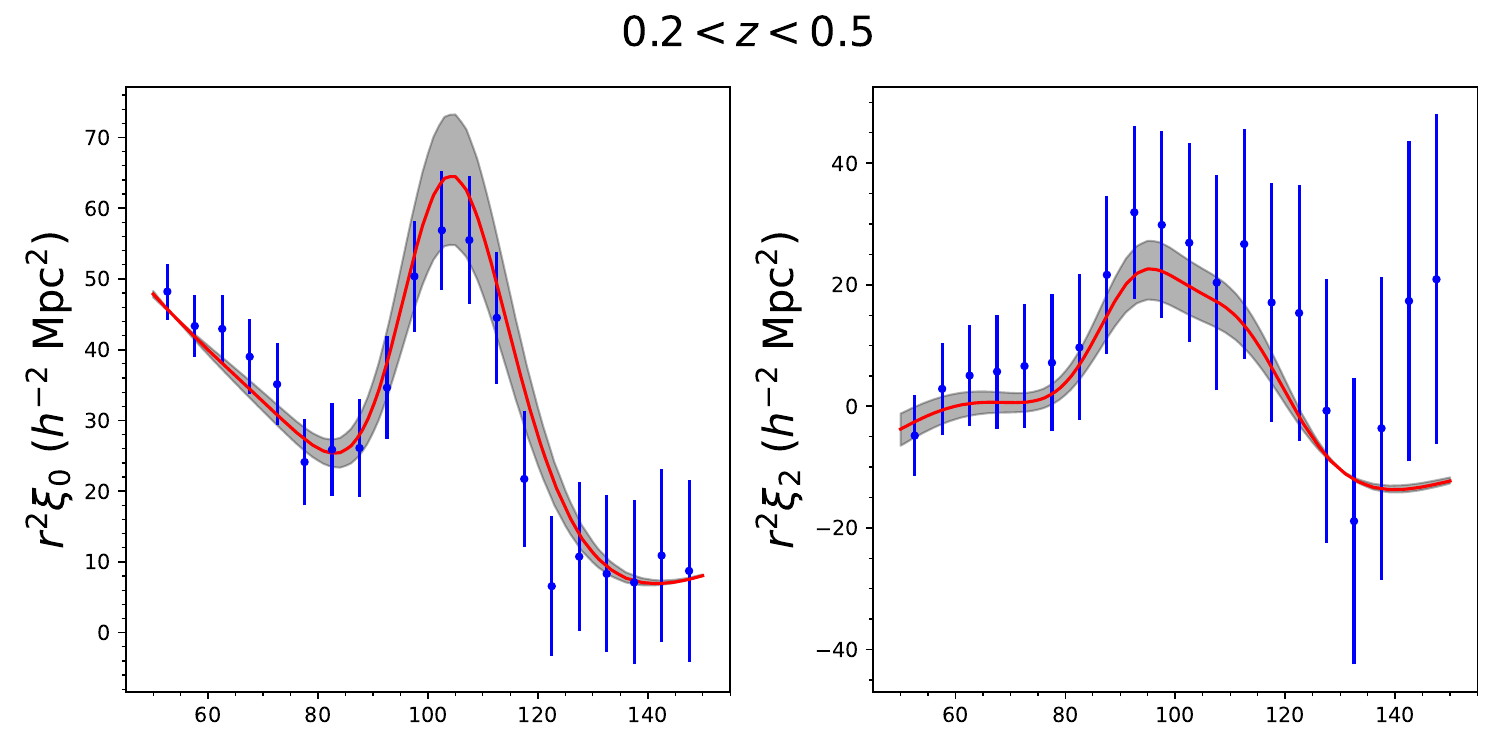}
\includegraphics[scale=0.5]{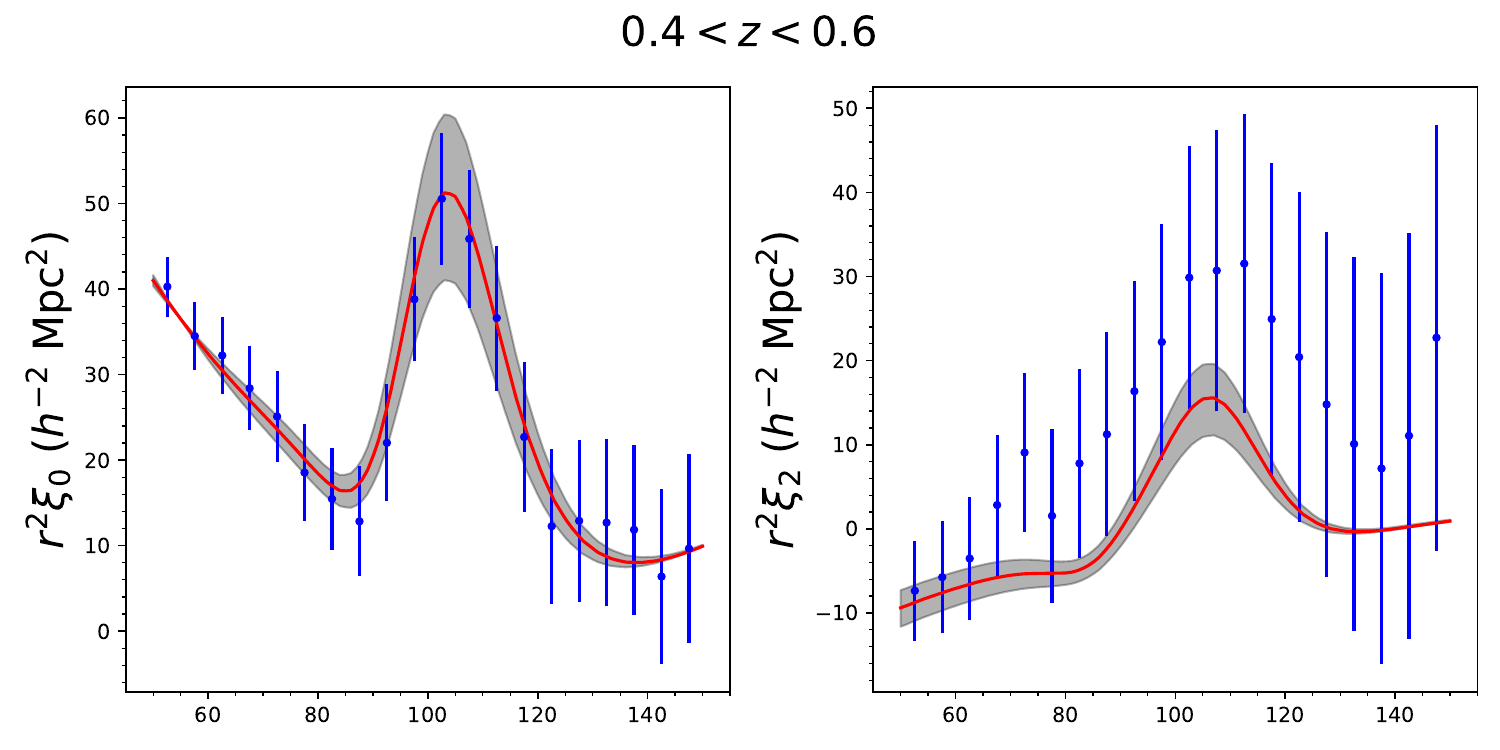}
\includegraphics[scale=0.5]{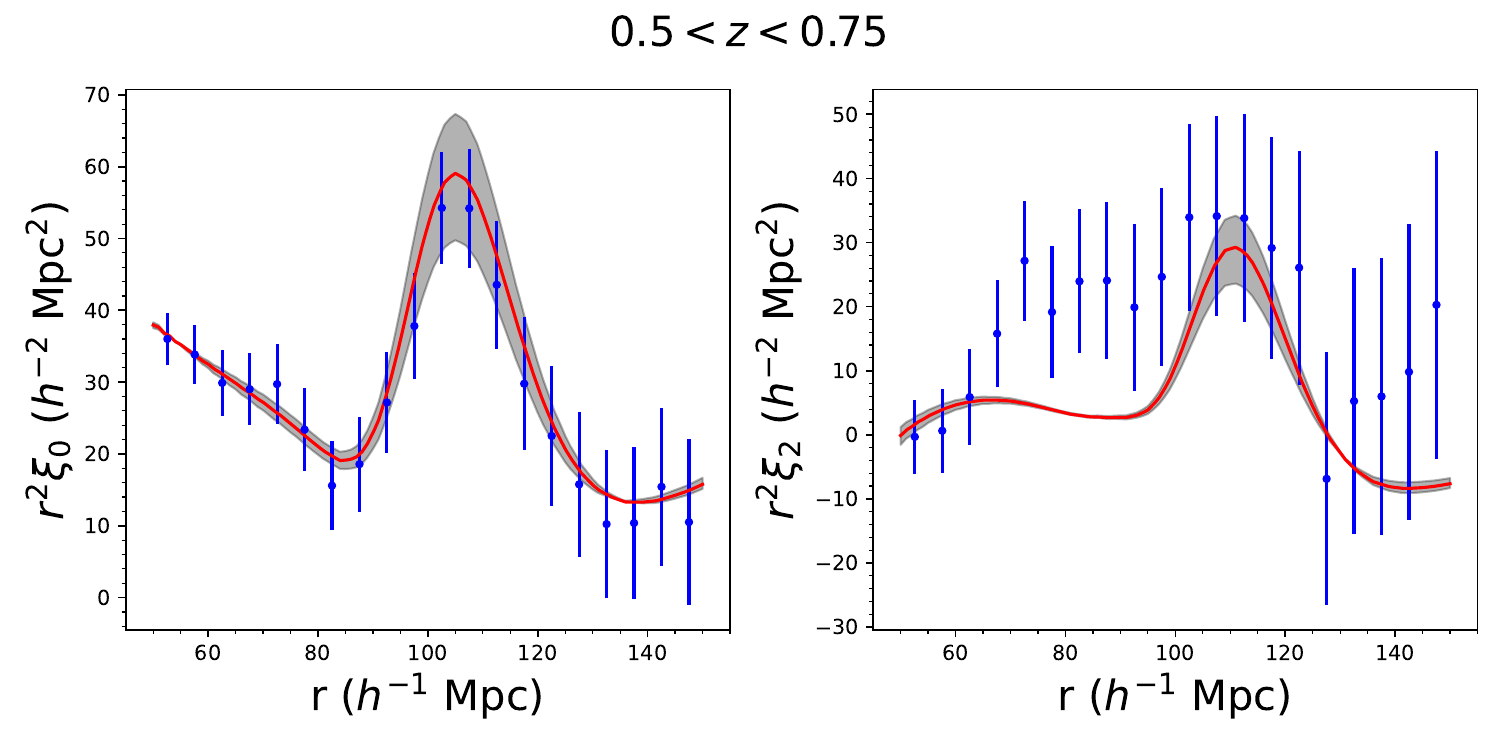}
    \caption{BOSS post-reconstruction data and best fit model. Gray shaded line shows the impact of changing $\gamma_\text{B}$ by 1$\sigma$ from its best fit value.
    \label{fig:data_vs_model_post}}
\end{figure*}

\begin{figure*}
\includegraphics[scale=0.5]{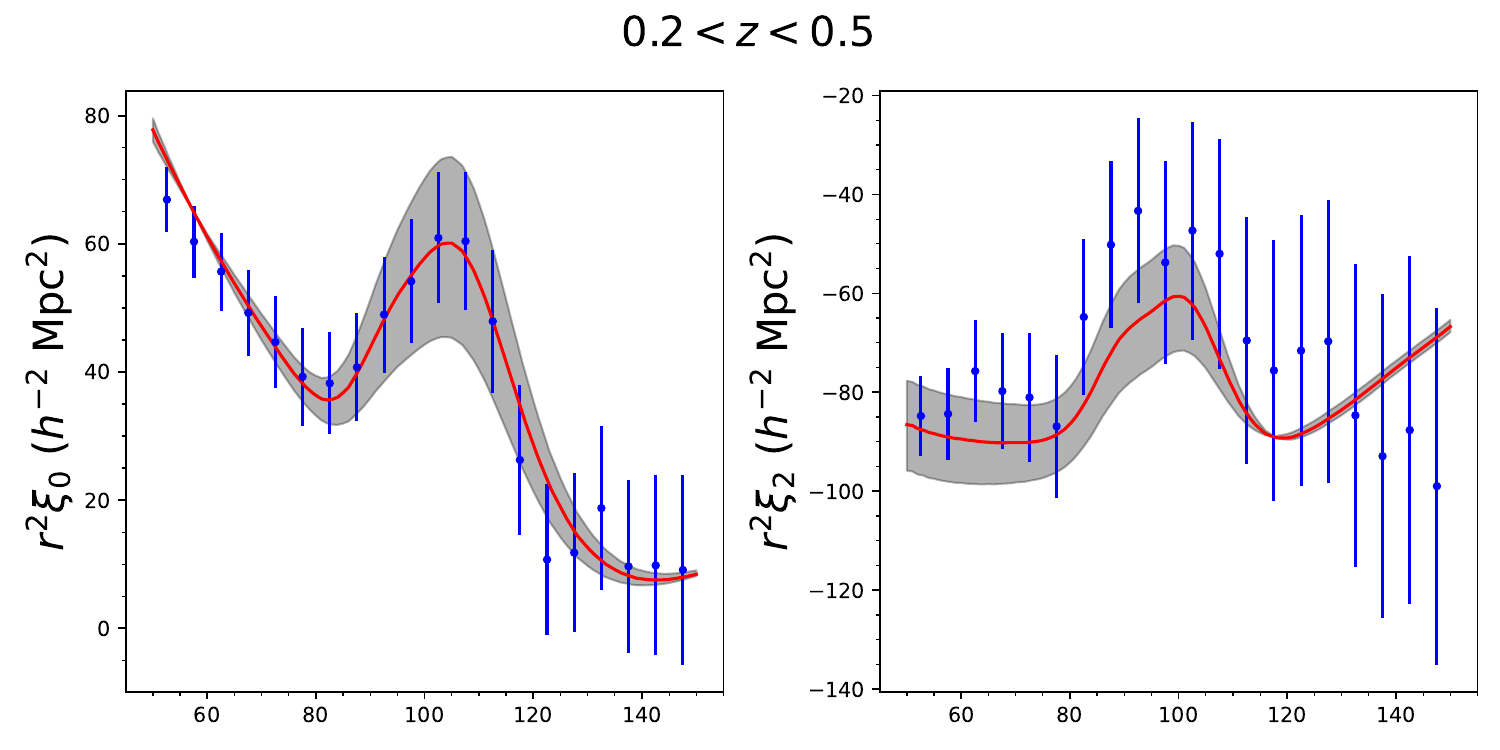}
\includegraphics[scale=0.5]{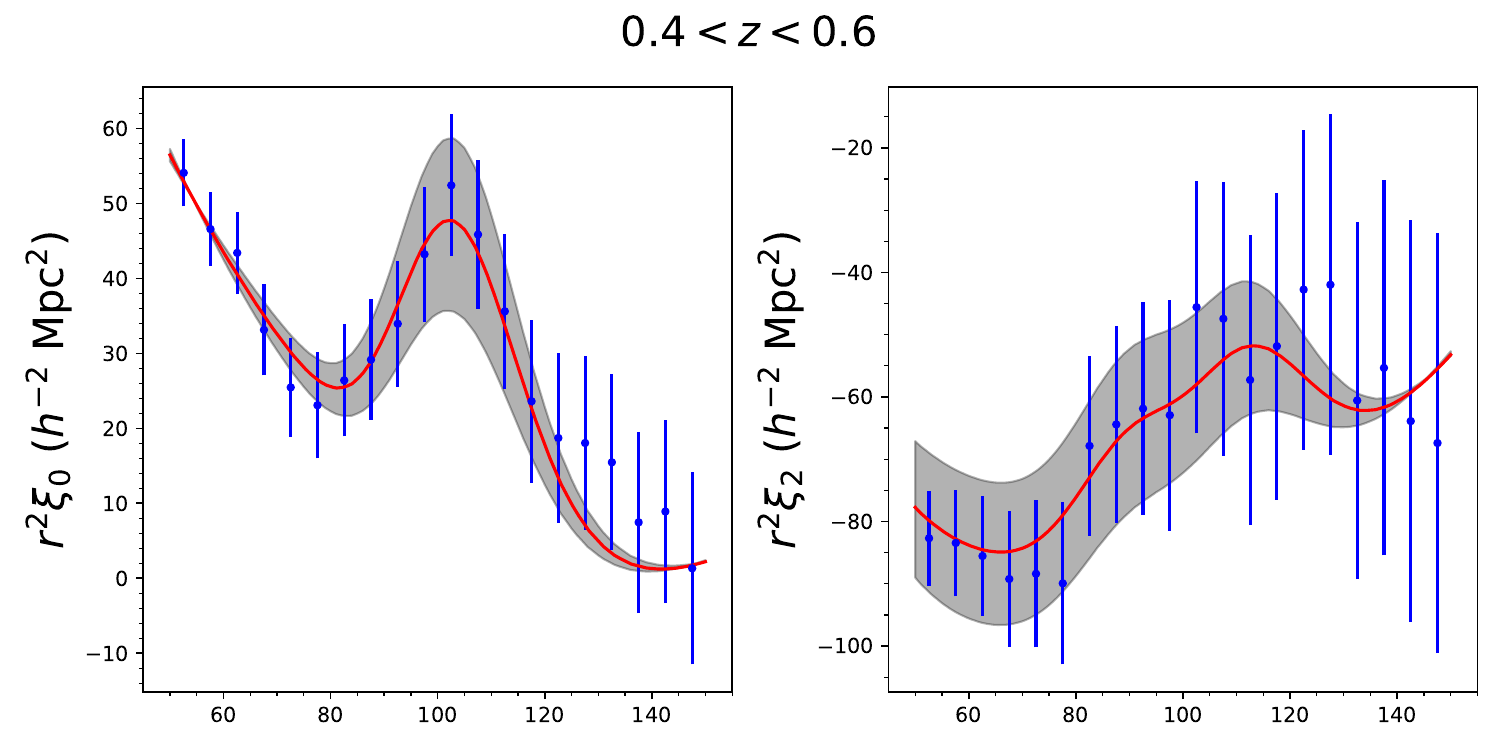}
\includegraphics[scale=0.5]{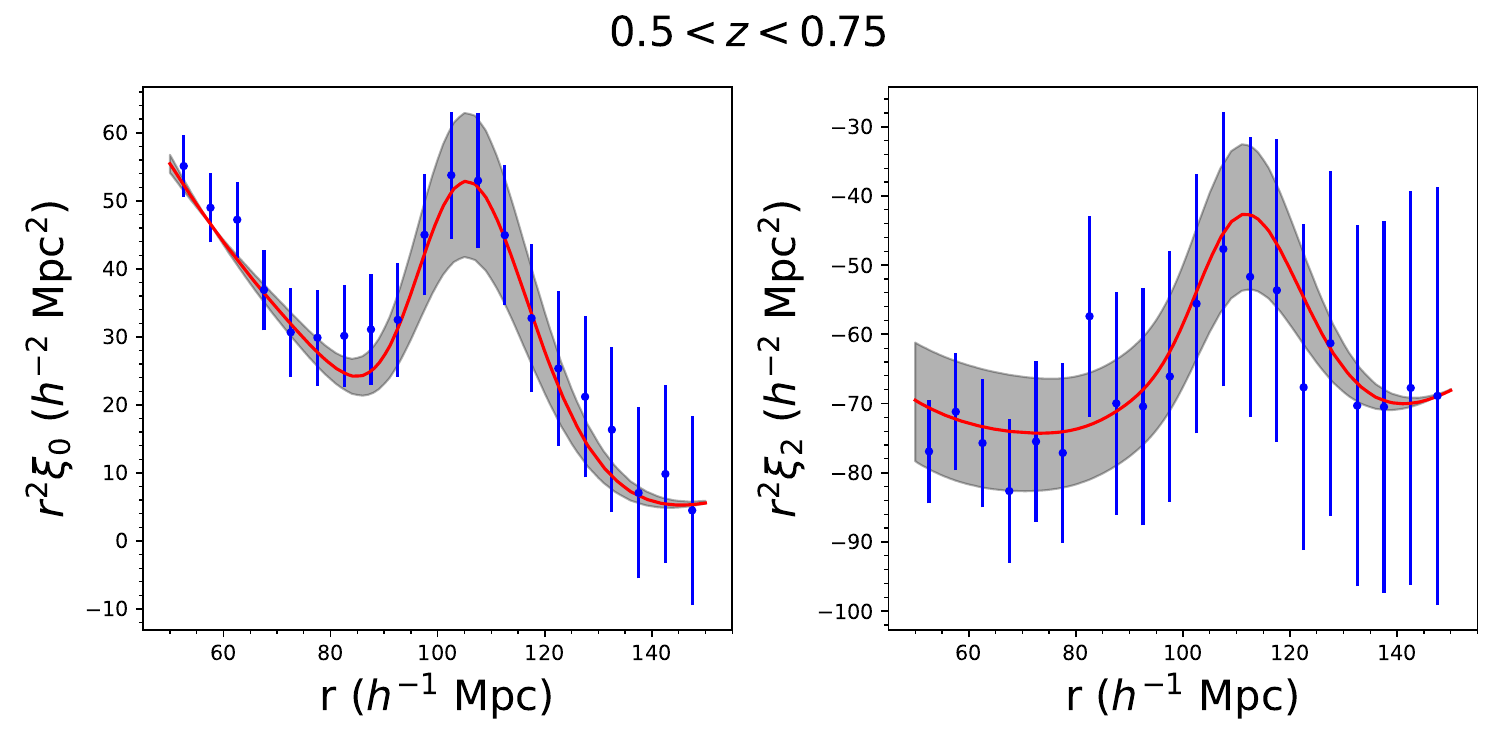}
    \caption{Same as Fig.~\ref{fig:data_vs_model_post}, but for the pre-reconstruction data.
    \label{fig:data_vs_model_pre}}
\end{figure*}

\begin{figure*}
\includegraphics[scale=0.5]{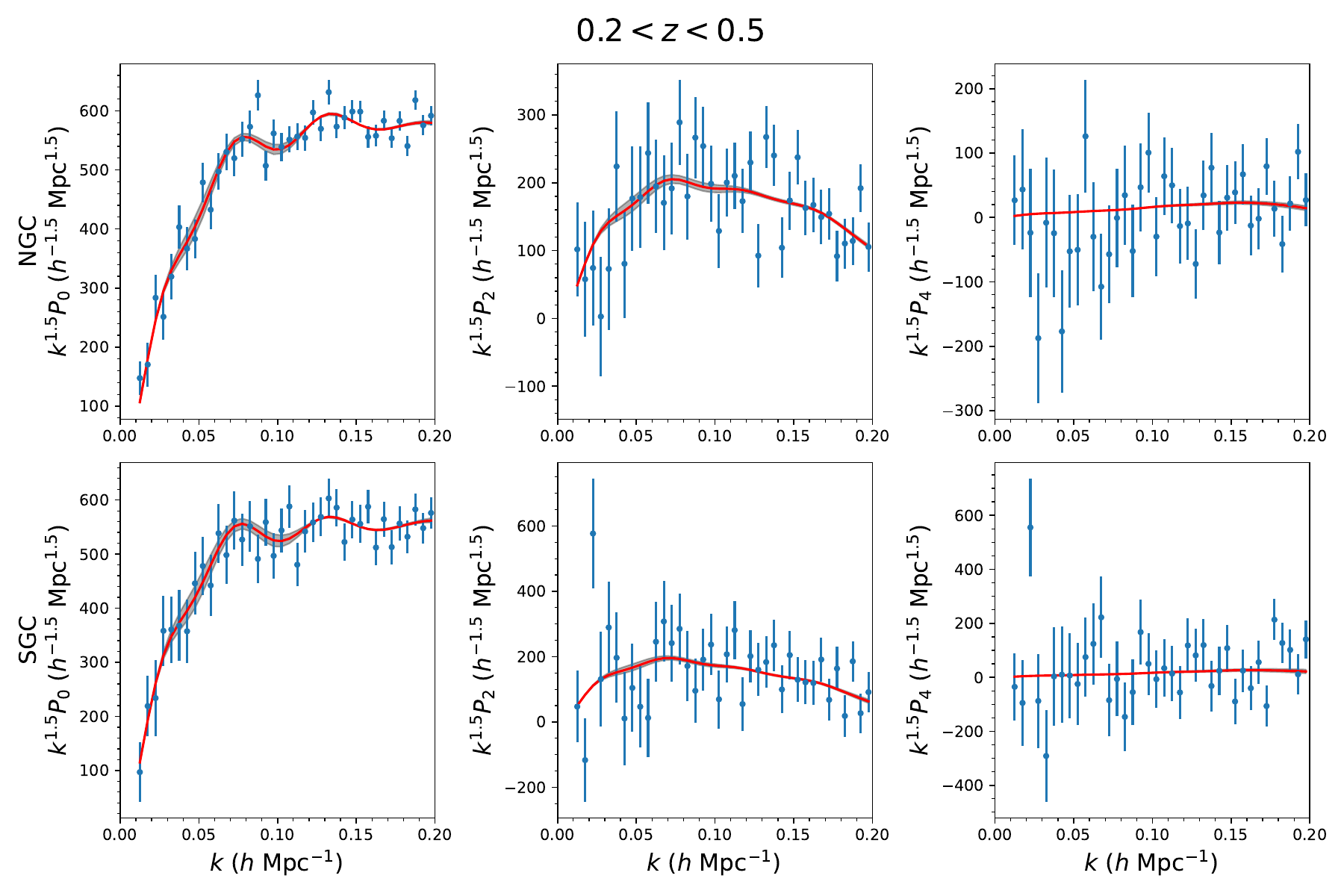}
    \caption{Comparison of power spectrum data, best-fit model (red) and 1$\sigma$ range in $\gamma_\text{B}$ (shaded), for the first redshift bin.
    \label{fig:data_vs_model_full_shape_z1}}
\end{figure*}

\begin{figure*}
\includegraphics[scale=0.5]{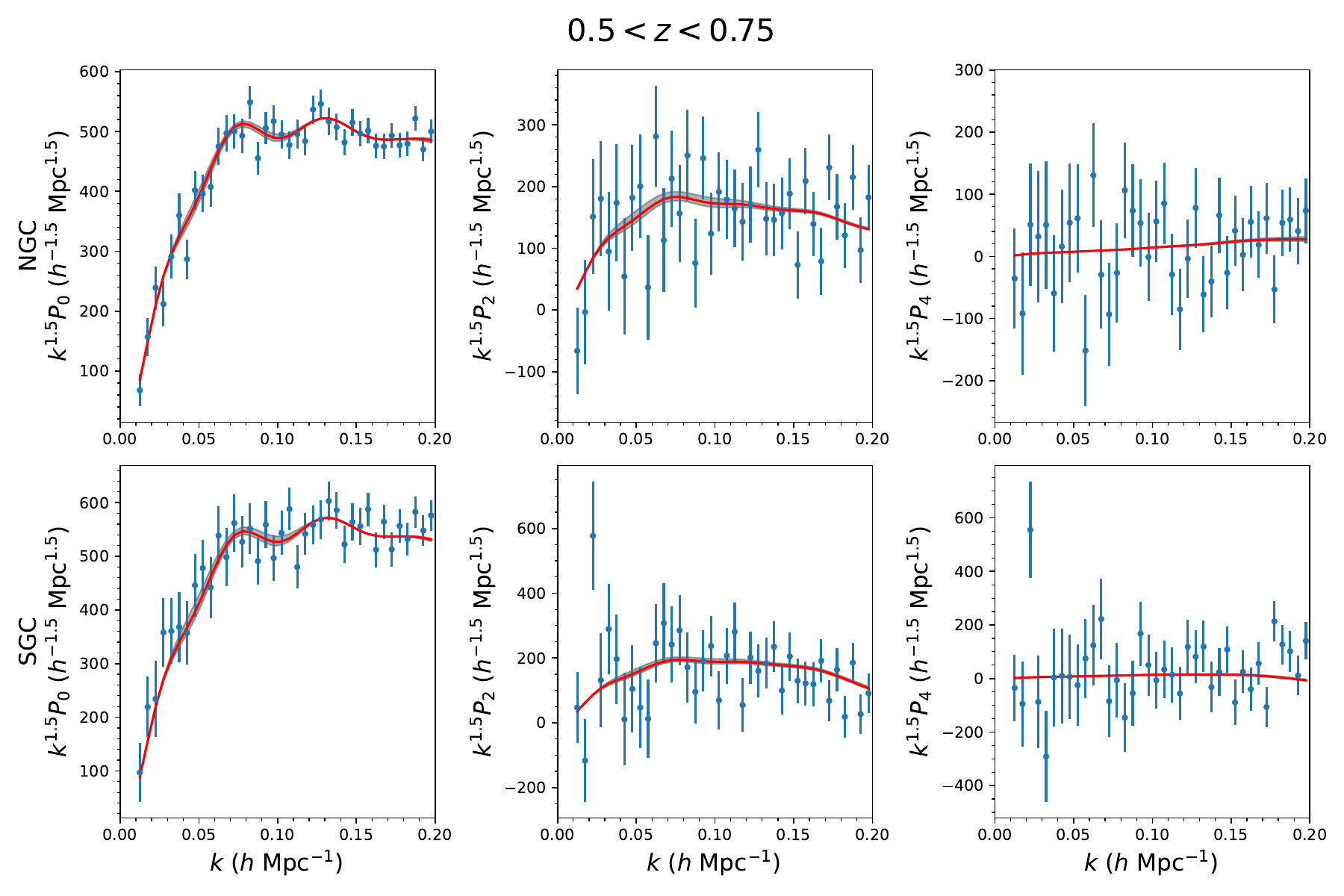}
    \caption{Comparison of power spectrum data, best-fit model (red) and 1$\sigma$ range in $\gamma_\text{B}$ (shaded), for the second redshift bin.
    \label{fig:data_vs_model_full_shape_z2}}
\end{figure*}

Our results are summarized in Table~\ref{tab:results}. We show both the individual bin results and the combination of all three bins, with the proper covariance computed from mocks.
The pre-reconstruction combined $\gamma_\text{B}$ is smaller than $\gamma_\text{B}$ in any of the individual bins. 
This is because we impose a flat prior on $H_0$ when combining the bins, which is not equivalent to a flat prior on $f_b = \omega_b / (\Omega_m h^2)$; since $f_b$ is proportional to $1/h^2$, a flat prior on $H_0$ is necessarily a non-flat prior on $f_b$. We apply this prior because it is appropriate for an $H_0$ measurement, and because the full-shape fits already use a flat $H_0$ prior rather than a flat $f_b$ prior.

We show contours of constant posterior levels in Figs.~\ref{fig:triangle_full_shape_data},~\ref{fig:triangle_prerecon_data},~\ref{fig:triangle_postrecon_data}, for the data fit with the full shape model and pre- and post-reconstruction template fits.
We compare the clustering signal to the best-fit models in Figs.~\ref{fig:data_vs_model_post},~\ref{fig:data_vs_model_pre},~\ref{fig:data_vs_model_full_shape_z1},~\ref{fig:data_vs_model_full_shape_z2}.
While the post-reconstruction quadrupoles in the second and third redshift bins look like poor fits, the significance of the deviation
is overstated in these plots due to the strong correlations
between bins. In particular, for the second bin, the quadrupole has $\chi^2 = 8.9$ with 15 data points (i.e.\ the model fits the data very well). The seemingly highly anomalous run of high points in the quadrupole between 65 and 100\,$h^{-1}$\,Mpc in the third redshift bin only has $\chi^2 = 8$ over 8 data points (though it would have $\chi^2 = 19$ if the covariance were diagonal). Overall, the worst-fit bin
(pre-reconstruction $0.4 < z < 0.6$) havs a $p$-value of 0.035.
We conclude that the model fits the data well.

We find $\gamma_\text{B}$ agrees well between the pre-reconstruction and post-reconstruction fits, with offsets of 0.5$\sigma$, 0.15$\sigma$, and 0.61$\sigma$. This is similar to the agreement between pre- and post-reconstruction
$\alpha_{\parallel}$ and $\alpha_{\perp}$, either as determined by our fits or by those of \cite{Ross2017}.
Our constraints on the AP parameters agree well with \cite{Ross2017}: the central values agree with \cite{Ross2017} to within 0.3$\sigma$, although some of the errorbars are different
($\sim$30\% tighter constraints on post-reconstruction $\alpha_{\perp}$ and 50\% looser constraints on pre-reconstruction $\alpha_{\parallel}$).
The reasons for these differences are explored in detail in Appendix~\ref{sec:compare_to_ross}. The smaller errorbar on post-reconstruction $\alpha_{\perp}$ is primarily due to a change in the power spectrum template used to fit the BAO.\footnote{We thank Ashley Ross for finding this discrepancy between the current version of the \texttt{LSSAnalysis} code and the pipeline used for BOSS, which corresponds to this commit: \url{https://github.com/ashleyjross/LSSanalysis/commit/6ab0faa9f4ce9ec740fe66e871c7e250d0fd720f}.}
The larger error on pre-reconstruction $\alpha_{\parallel}$ in $z_2$ and $z_3$ comes from a variety of sources, with similar contributions
from freeing the damping parameters, defining $\alpha_{\parallel}$ so that it only shifts the oscillatory component of the transfer function, rather than the entire power spectrum, and the updates in the BAO templates.

We use a considerably more restricted set of polynomials than \cite{Ross2017}, but we find no evidence that our model needs any more polynomial terms.
For the post-reconstruction correlation function, adding the $1/r$ and $1/r^2$ polynomials (and the constant term to the quadrupole) improves
$\chi^2$ by 2 (3, 4) for the first (second, third) redshift bins. Accounting for the additional 5 free parameters leads to $\Delta$AIC = -8 (-7, -6) using the Akaike Information Criterion, i.e.\ the default model with the single constant term in the monopole is favored.
Likewise, we find improvements in $\chi^2$ of 4 (0, 1) for the three bins in the pre-reconstruction fits, yielding $\Delta$AIC = -4 (-8, -7).

We find good agreement in $\gamma_\text{B}$ between pre-reconstruction, post-reconstruction, and full shape results in the first redshift bin, but the agreement is less good in the third bin, in particular with an offset of 0.059 between the post-reconstruction and full-shape results.

We compare the observed $\gamma_{\text{B}}$ discrepancies between pre-reconstruction, post-reconstruction and full shape to the discrepancies measured in the 84 Nseries mocks.
The volume of each mock is similar to the volume of each redshift bin, and thus the differences from the mocks can be compared to the differences observed in all bins.
We find 43, 67 and 32 (52\%, 81\%, and 39\%) of the mocks have a larger absolute value difference between
the pre and post-reconstruction fits than the differences in data of 0.019, 0.006, and 0.027 for the three redshift bins. Hence these differences are totally in line with the expected scatter from the mocks.
Likewise, we find 80 (28) (96\%, 34\%) of the 84 mocks have larger absolute differences between the pre-reconstruction and full-shape fits than the 0.002 and 0.032 observed.
Finally, we find 48 (5) (58\%, 6\%) of the 84 mocks have larger absolute differences
between post-reconstruction and full-shape than the 0.018 and 0.059 observed.
While the difference between post-reconstruction and full-shape fits in the $0.5 < z < 0.75$ is large, it is not out of line with the differences observed in mocks.

As mentioned above, we find some differences in the derived $\gamma_\text{B}$ credible regions between data and mocks.
Most notably, the template fit credible regions are somewhat larger than the expected 0.038 (pre-recon) or 0.030 (post-recon).
The full-shape fits in individual redshift bins have similar errors between data and mocks (0.032), but as shown in our companion paper \cite{krolewski}, the 
error on the combined fit is considerably smaller on data (0.021) than in mocks using the combined covariance (0.029).
We individually fit the 84 Nseries mocks with a covariance matrix scaled to the entire BOSS volume.
We find a mean error of 0.0285; 6 (5) mocks (7\%, 6\%) have a lower (upper) error smaller than the observed combined error of 0.021. 
For the pre-reconstruction template fits, we find 38 (23, 7) (46\%, 28\%, 8\%) of the mocks have larger errors than observed.
For the post-reconstruction, template fits, 10 (29, 3) (12\%, 35\%, 3.6\%) have larger errorbars than observed. 

\section{Discussion and conclusions}  \label{sec:conclusions}

We have developed a model to constrain the baryon fraction using the amplitude of the baryon signal by splitting the transfer function into components arising from cold dark matter and baryons. We describe the transfer function split and apply it to both full-shape and template-based BAO models in  Section~\ref{sec:BAO-amp}.
In Sections~\ref{sec:mock-tests} and~\ref{sec:mock-results}, we describe the results of testing our model on the mean of the 84 Nseries mocks, as well as noiseless theory vectors generated in early dark energy (EDE) or $\Lambda$CDM cosmology. On this extensive series of mock tests, 
we find biases on the recovered baryon fraction of $<0.3\sigma$ for the full-shape and post-reconstruction fits, and $<0.4\sigma$ for the pre-reconstruction fits compared to the errors derived fitting the BOSS data.
When fitting for an explicitly sound horizon independent Hubble constant, in combination with the physical baryon abundance $\Omega_b h^2$ and the matter density $\Omega_m$ from geometrical probes (excluding isotropic BAO distance information that requires a calibrated value of the sound horizon),
we find biases of less than $<0.2\sigma$.
To test that the results are independent of the sound horizon, we fit for the baryon fraction using an early dark energy cosmology as the truth, which has true $H_0 = 72.19$\,\Hunit\ and sound horizon $r_s = 138$\,Mpc \cite{Smith20}.
We find credible regions with biases of $0.4\sigma$ (0.3$\sigma$, 0.2$\sigma$) on the baryon fraction for full shape, pre-reconstruction, and post-reconstruction fits.
We also tested that we recovered the correct cosmology when varying the input $f_b$ within both EDE and $\Lambda$CDM cosmologies.
We find an average bias of 1.67\,\Hunit\ on $H_0$ in the $\Lambda$CDM cosmologies (or those with a running spectral index or extra ultrarelativistic species), and 1.8\,\Hunit\ on the EDE cosmologies (i.e.\ our recovered values of $H_0$ are biased high from the truth).

\begin{figure}
    \centering

    \includegraphics[scale=0.6]{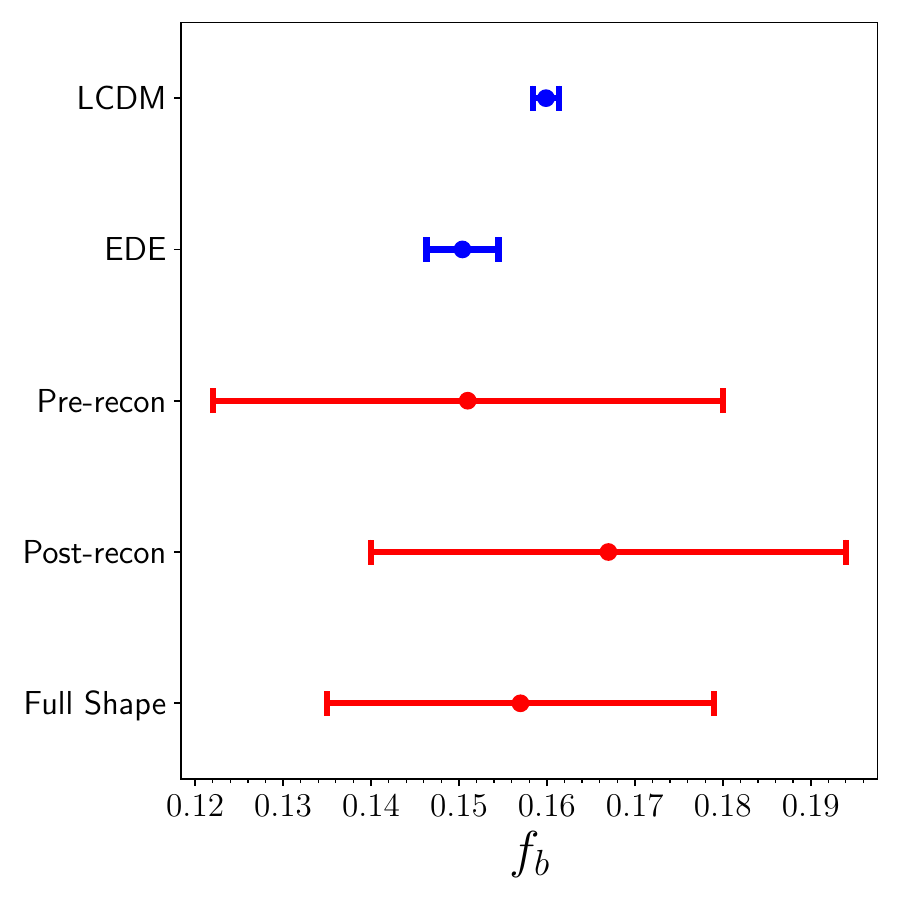}
    \caption{Comparison of our $f_b$ measurement (red), using the three fitting methods, to $\Lambda$CDM models fit to Planck and Early Dark Energy $f_b$ models fit to the combination of Planck, BAO, supernovae, SH0ES, and RSD.}
    \label{fig:fb_comparison}
\end{figure}

When fitting to data, we find consistent
values of the baryon fraction between our three methods, with discrepancies consistent
with the scatter seen in mocks.
When combining all redshift bins, we find 
$f_b = 0.153 \pm 0.029$ for pre-reconstruction template fits, $f_b = 0.173 \pm 0.027$ for the post-reconstruction template fits, 
and $f_b = 0.154 \pm 0.022$ for the full-shape fits.
We estimate the systematic error as 0.013 for all three methods following the findings summarized in Sec.~\ref{sec:summary-mock-tests}.
The baryon fraction differs between $\Lambda$CDM and Early Dark Energy fits to the cosmic microwave background, due to the higher cold dark matter density required in the EDE cosmology. The higher matter density is required to match the observed
imprint of the early ISW effect, which is modified due to the altered growth history around recombination from early dark energy \cite{Vagnozzi21}. Using Planck TTTEEE and lensing, BAO, Type Ia supernovae, RSD, and SH0ES, $f_b = 0.1599 \pm 0.0015$ in $\Lambda$CDM and $f_b = 0.1504 \pm 0.0040$ in EDE \cite{Hill20}. Our measurement of the baryon fraction is consistent
with both constraints; indeed, it will require an order-of-magnitude improvement on the baryon fraction measured from the BAO to differentiate between the models (Fig.~\ref{fig:fb_comparison}).

Each of our three fitting methods has its own benefits and drawbacks. The full-shape fitting method is by far the most computationally expensive, with many more free parameters and a considerably more complicated model than the template-fit BAO models considered.
However, it offers considerably more flexibility than the template-fit models, with a full set of nonlinear and bias parameters
and variations in cosmological parameters. It allows us to consider cases where the early growth history deviates from that given in the template cosmology, changing the relationship between $\gamma_\text{B}$ and the baryon fraction in a way that depends on $\Omega_m h^2$. On the other hand, the model is quite sensitive to the choice of prior on nuisance parameters, leading to biases after marginalization even when the data under consideration is a noiseless theory model, and hence the best-fit has $\chi^2 = 0$.
In contrast, the template-fit methods are much cheaper to evaluate and quicker to interpret. However, the effects of nonlinearity and higher-order bias are represented only in an ad hoc way, with a $k^2$-dependent bias parameterized by $b_\partial$. Since the prior on $b_\partial$ is informative in the post-reconstruction case, it would be preferable to have a more complete model of bias and nonlinearity. Biases resulting from prior marginalization effects are less prevalent than in the full-shape case due to the smaller number of nuisance parameters, but prior choice is no less important.
Last, the BAO peak is sharper in the post-reconstruction case and it is thus potentially possible to extract more information about the BAO amplitude than in the pre-reconstruction or full-shape cases; however, our fits suggest that this extra information is modest, reducing the uncertainty on the baryon fraction
by 7\% on data and 20\% on mocks.
Nevertheless, we prefer to use the post-reconstruction template fits rather than pre-reconstruction,
since reconstruction removes nonlinear peak-shifting that could bias the BAO amplitude.

\subsection{Future potential of this method}

In light of the Hubble tension, constraints on $H_0$ from spectroscopic galaxy surveys have been an intense focus of interest \cite{Alam-eBOSS:2021}. The tightest constraints achieve $\sigma_{H_0} \sim 0.4$\,\Hunit\ 
\citep{Alam17,Cuceu19,Blomqvist19,Alam-eBOSS:2021}
by combining BAO
with precise measurements of $r_d$ (depending on $\omega_b$) and $\Omega_m$ from Planck.
However, this measurement cannot test for systematics in the CMB dataset or changes due to modified early-time physics. Excluding Planck information and calibrating $r_d$ with $\omega_b$ from BBN leads to $\sigma_{H_0} \sim 1.0$\,\Hunit\  
\cite{Wang17,Schoeneberg19,Schoeneberg22}.
Full-shape fits bring extra information from the shape of the matter power spectrum, and lead to similar values of $H_0$ with a slightly more accurate measurement ($\sigma_{H_0} \sim 0.8$\,\Hunit) if a Planck prior on $n_s$ is used
\cite{Ivanov19,DAmico19,Colas19,Troster20,Wadekar20,Chen21,Philcox20,DAmico21Hub,DAmico22,Zhang22,Ivanov23}. In the absence of such a prior, the constraining power is similar to BAO + BBN; the additional shape information does not contribute much to shrinking the $H_0$ error.

All of these probes assume $\Lambda$CDM at recombination, either in the fits to Planck or in the extrapolation from the $\omega_b$ value at BBN to the sound horizon at the drag epoch.
Changes to early-time physics (most notably EDE, but also additional relativistic degrees of freedom, or varying electron mass \cite{H0Olympics}) can modify the relationship between $\omega_b$ and $r_d$.
Hence, Ref.~\cite{BaxterSherwin21} proposed a sound horizon independent measurement of $H_0$, based on the equality scale, which depends on $\Omega_m h^2$, and additional probes of $\Omega_m$ such as supernovae or weak lensing.
Ref.~\cite{BaxterSherwin21} used large-scale structure information from CMB lensing, and additional work
applied this method to the galaxy power spectrum \cite{Philcox21c,Philcox22,Farren22}.
The implementation of this method used a perturbative $\Lambda$CDM fit to the BOSS power spectrum, adding an extra isotropic BAO scaling parameter to explicitly marginalize over the 
sound horizon and decouple it from the rest of the cosmological parameters.
When combining the BOSS power spectrum with Pantheon supernovae and CMB lensing, \cite{Philcox22} obtain $H_0 = 64.8^{+2.2}_{-2.5}$\,\Hunit.
Without the BOSS power spectrum, ACT and Planck CMB lensing and Pantheon+ alone give a similar result, $H_0 = 64.9\pm2.8$\,\Hunit \cite{Madhavacheril24}.
However, Ref.~\cite{Smith22} point out that early dark energy also alters the shape of the power spectrum by changing the early expansion history (see also \cite{Kable24}). Therefore, the sound horizon marginalized constraints on $H_0$, when analyzed in an early dark energy cosmology,
shift towards higher values of $H_0$ by $\Delta H_0 = 3.8 (4.9)$\,\Hunit\ (depending on whether $\Omega_m$ is constrained from supernovae or uncalibrated BAO and CMB), and the error nearly doubles.
Similar shifts are seen in a cosmology with extra relativistic degrees of freedom, $\Delta N_{\textrm{eff}}$. On the other hand, if the 
sound horizon is modified via a time-varying electron mass, only the time of recombination is affected, not the early growth history,
and the constraints on $H_0$ are unchanged.

The approach of 
Ref.~\cite{Brieden23} is similar. Restricting their dataset to galaxy surveys only, plus a BBN prior on $\omega_b$, they combine uncalibrated BAO with constraints on the power spectrum shape from the ShapeFit compression \citep{Brieden21,Brieden22}, interpreted with $\Lambda$CDM, to obtain $H_0 = 70.1^{+1.9}_{-2.1}$\,\Hunit. This result is similarly model dependent. Allowing for a wide prior on $n_s$, freeing the neutrino mass, or adding $\Delta N_{\textrm{eff}}$ broadens the constraint to $\sigma_{H_0} = 2.5$ (3.2, 2.5)\,\Hunit, while leaving the central value largely unchanged except for $\nu\Lambda$CDM, which increases $H_0$ to 73.3\,\Hunit.
Last, Ref.~\cite{Kalus23} focus on measuring only the turnover scale, rather than the entire shape of the power spectrum, and linking it to $\Omega_m h^2$ and therefore $H_0$ with supernovae or uncalibrated BAO; they obtain $H_0 = 72.9^{+10.0}_{-8.6}$\,\Hunit\ from eBOSS quasars when combining with BAO. Since they only use the equality scale (which is close to the maximum scale of the survey) rather than the shape of the power spectrum, they obtain larger errorbars.

Our method probes different, more simple physics than the methods above, and currently yields a larger error on $H_0$ than Refs.~\cite{Philcox22} or \cite{Brieden23},
with $\sigma_{H_0} = 5.4$\,\Hunit\ from BOSS data alone. These errors are dominated by uncertainty on $f_b$ and do not improve much with better measurements of $\Omega_m$. However, our method is complementary in other ways and, in particular, is less model-dependent than the results of \cite{Philcox22,Smith22}. 

There will be a bias in our recovered $H_0$ if the model does not have enough freedom to fit the true shape of the underlying baryon or CDM transfer functions, if there are strong prior dependencies on the parameters of those shapes biasing the model, or if the posterior is biased away from the true matter-radiation equality scale (i.e finds the wrong $\Omega_mh^2$). The latter changes the time when CDM fluctuations can start to grow while baryon perturbations are frozen until decoupling, changing the amplitude of the BAO as discussed in Ref.~\cite{Eisenstein-Hu}. We find that if the universe truly followed an EDE cosmology, but we fit only $\Lambda$CDM models, our $H_0$ constraint would be biased high by 2\,\Hunit.
This may be due to limitations in the modelling, but is also consistent with the shifts seen in $\Lambda$CDM noiseless theory vectors, which are due entirely to prior effects as the true cosmology has a best-fit $\chi^2 = 0$ by definition, in this case.

In a related test, \cite{Smith22} fixed their data and changed the theory model from $\Lambda$CDM to EDE, when fitting a full EFT model, finding an increase in $H_0$ by a larger amount, 4--5\,\Hunit. For the tests we have performed for all of the three methods used, presented in Tables~\ref{tab:full-fit-results}, \ref{tab:pre-recon_template_fits} \&~\ref{tab:post-recon_template_fits}, we find smaller changes in $H_0$ ($\sim$2 \Hunit) at a level below the statistical error, which is approximately $\Delta H_0\pm5$\,\Hunit.

Moreover, the BAO amplitude information, which dominates our measurement of the baryon signature, is independent of the ShapeFit compression of \cite{Brieden23}. ShapeFit allows for the broad-band power spectrum shape to deviate from a template linear power spectrum, parameterized by $m$, which gives the width of a $\tanh$ function modulated power law that multiplies the template power spectrum. Our measurement primarily picks up the effect of changes in the BAO amplitude, which is explicitly not included in the broad-band shape information of ShapeFit.\footnote{Fig.\ 5 shows that we do receive some degeneracy-breaking power from the shape of the power spectrum, but this only modestly improves our $\gamma_\text{B}$ and thus $H_0$ constraints.}

Ongoing and future spectroscopic surveys will radically improve our measurement of the BAO feature. DESI covers an effective volume of 60 Gpc$^3$ \cite{DESISV},\footnote{Note that the forecasts presented in Section 6 of \cite{DESISV} supersede the older forecasts of \cite{DESI2016}.} nearly an order-of-magnitude larger than BOSS at 7.8 Gpc$^3$ \cite{Alam17}. We show in Table~\ref{tab:full-fit-results} that increasing the volume by a factor of 10 improves $\gamma_\text{B}$ constraints by $\sqrt{10}$. Our method will continue to be dominated by the $f_b$ uncertainty rather than $\Omega_m$, implying that DESI will allow us to reach a precision of 2\,\Hunit: sufficient to differentiate SH0ES and CMB/LSS measurements at 3$\sigma$ if our results agree with one or the other. The effective volume of Euclid is similar, 55 Gpc$^{3}$ \cite{Euclid20}.\footnote{Calculated from their Table 3 and Eq.~90 using $P(k=0.14,\mu=0.6)\sim 4000$\,$h^{-3}$\,Mpc$^3$.} This yields an $H_0$ precision of 1.6\,\Hunit\ from the combination of Euclid and DESI.
The high-precision redshift sample of SphereX \cite{Spherex14} will also cover a similar effective volume to DESI and Euclid, though at lower redshift. While the survey strategy has not yet been defined, the Roman space telescope high latitude survey could also achieve a high-precision baryon fraction measurement, with \cite{Eifler21} presenting a survey strategy that achieves an effective volume of 30 Gpc$^3$. Further into the future, Stage 5 spectroscopic surveys \citep{Stage5} such as MegaMapper \citep{MegaMapper} or MSE \cite{MSE19} will achieve similar or tighter precision at even higher redshifts. At the same time, $H_0$ measurements using $r_d$ or full-shape information will also continue to improve, with $H_0$ errors expected to be $<0.5$\,\Hunit\ \cite{Farren22,Ivanov23b}. The very high precision of future surveys allow for exquisite tests of consistency and model-independence between different $H_0$ probes in large-scale structure, which should be highly informative about new physics solutions to the $H_0$ tension.

\acknowledgments

We thank Ashley Ross and Mariana Vargas Magana for useful conversations and for helping to dig out old data. 

AK was supported as a CITA National Fellow by the Natural Sciences and Engineering Research Council of Canada (NSERC), funding reference \#DIS-2022-568580.

WP acknowledges support from the Natural Sciences and Engineering Research Council of Canada (NSERC), [funding reference number RGPIN-2019-03908] and from the Canadian Space Agency.

Research at Perimeter Institute is supported in part by the Government of Canada through the Department of Innovation, Science and Economic Development Canada and by the Province of Ontario through the Ministry of Colleges and Universities.

This research was enabled in part by support provided by Compute Ontario (computeontario.ca) and the Digital Research Alliance of Canada (alliancecan.ca).


\bibliographystyle{JHEP}
\bibliography{References}





\appendix

\section{Results on fits to $f_b$-varying noiseless theory vectors}

Here we show the credible regions of $h$ and $\gamma_\text{B}$ for the tests with varying true $f_b$, either using AbacusSummit cosmologies or using EDE cosmologies, in Tables~\ref{tab:full-shape-noiseless},~\ref{tab:pre-recon_noiseless} and~\ref{tab:post-recon_noiseless}. These are plotted in Figs.~\ref{fig:h_fb_scatter_full-shape} and~\ref{fig:h_scatter_template}.

\begin{table*}[]
    \centering
    \begin{tabular}{l|cc|cccc|cc}
    \multirow{2}{0cm}{Mock} & \multicolumn{2}{c}{Truth} & 
    \multicolumn{4}{c}{BAO only} & \multicolumn{2}{c}{Add voids} \\
    & 
    $h$ & $f_b$ & $h$ & $n\sigma$ & $\gamma_\text{B}$ & $n\sigma$ & $h$ & $n\sigma$ \\   
    \hline
cosm131 & 0.717 & 0.171 & $0.740^{+0.093}_{-0.080}$
 & 0.31 & $0.158^{+0.026}_{-0.022}$
 & -0.42 & $0.734^{+0.068}_{-0.057}$
 & 0.33 \\
cosm139 & 0.716 & 0.176 & $0.743^{+0.090}_{-0.081}$
 & 0.36 & $0.164^{+0.025}_{-0.025}$
 & -0.39 & $0.739^{+0.068}_{-0.054}$
 & 0.45 \\
cosm130 & 0.674 & 0.157 & $0.691^{+0.093}_{-0.077}$
 & 0.23 & $0.142^{+0.032}_{-0.028}$
 & -0.48 & $0.689^{+0.078}_{-0.063}$
 & 0.29 \\
cosm159 & 0.597 & 0.149 & $0.551^{+0.065}_{-0.057}$
 & -0.55 & $0.148^{+0.032}_{-0.031}$
 & -0.03 & $0.554^{+0.068}_{-0.063}$
 & -0.75 \\
cosm167 & 0.705 & 0.138 & $0.731^{+0.098}_{-0.083}$
 & 0.35 & $0.124^{+0.026}_{-0.026}$
 & -0.45 & $0.736^{+0.086}_{-0.075}$
 & 0.61 \\
cosm168 & 0.607 & 0.166 & $0.619^{+0.062}_{-0.057}$
 & 0.16 & $0.159^{+0.022}_{-0.025}$
 & -0.23 & $0.617^{+0.051}_{-0.041}$
 & 0.20 \\
cosm172 & 0.637 & 0.181 & $0.647^{+0.074}_{-0.060}$
 & 0.14 & $0.169^{+0.027}_{-0.025}$
 & -0.39 & $0.650^{+0.056}_{-0.044}$
 & 0.25 \\
cosm177 & 0.682 & 0.143 & $0.649^{+0.081}_{-0.065}$
 & -0.40 & $0.139^{+0.029}_{-0.025}$
 & -0.13 & $0.649^{+0.068}_{-0.060}$
 & -0.58 \\
cosm169 & 0.647 & 0.152 & $0.653^{+0.083}_{-0.068}$
 & 0.08 & $0.143^{+0.031}_{-0.026}$
 & -0.29 & $0.655^{+0.063}_{-0.057}$
 & 0.16 \\
cosm181 & 0.575 & 0.163 & $0.536^{+0.057}_{-0.048}$
 & -0.47 & $0.159^{+0.030}_{-0.028}$
 & -0.13 & $0.535^{+0.053}_{-0.042}$
 & -0.70 \\
 \hline
EDE001 & 0.790 & 0.123 & $0.815^{+0.135}_{-0.104}$
 & 0.34 & $0.111^{+0.026}_{-0.026}$
 & -0.39 & $0.803^{+0.116}_{-0.086}$
 & 0.25 \\
EDE002 & 0.773 & 0.128 & $0.797^{+0.125}_{-0.102}$
 & 0.32 & $0.119^{+0.026}_{-0.027}$
 & -0.29 & $0.791^{+0.101}_{-0.084}$
 & 0.35 \\
EDE003 & 0.756 & 0.134 & $0.778^{+0.122}_{-0.098}$
 & 0.30 & $0.121^{+0.030}_{-0.027}$
 & -0.42 & $0.760^{+0.104}_{-0.072}$
 & 0.08 \\
EDE004 & 0.739 & 0.140 & $0.773^{+0.108}_{-0.098}$
 & 0.46 & $0.128^{+0.030}_{-0.028}$
 & -0.39 & $0.766^{+0.101}_{-0.078}$
 & 0.53 \\
EDE006 & 0.704 & 0.155 & $0.743^{+0.102}_{-0.087}$
 & 0.53 & $0.138^{+0.033}_{-0.029}$
 & -0.55 & $0.733^{+0.090}_{-0.077}$
 & 0.57 \\
EDE007 & 0.686 & 0.163 & $0.721^{+0.095}_{-0.077}$
 & 0.47 & $0.146^{+0.031}_{-0.027}$
 & -0.55 & $0.706^{+0.086}_{-0.066}$
 & 0.39 \\
EDE008 & 0.667 & 0.172 & $0.706^{+0.095}_{-0.075}$
 & 0.53 & $0.150^{+0.035}_{-0.029}$
 & -0.71 & $0.689^{+0.080}_{-0.060}$
 & 0.43 \\
EDE009 & 0.647 & 0.183 & $0.677^{+0.092}_{-0.077}$
 & 0.41 & $0.166^{+0.040}_{-0.036}$
 & -0.55 & $0.668^{+0.077}_{-0.057}$
 & 0.41 \\
EDE010 & 0.627 & 0.195 & $0.653^{+0.090}_{-0.065}$
 & 0.35 & $0.177^{+0.036}_{-0.033}$
 & -0.58 & $0.643^{+0.078}_{-0.057}$
 & 0.31 \\

\end{tabular}
\caption{Full-shape fit parameter recovery tests for 
noiseless theory vectors with varying $f_b$ in $w_0 w_a$CDM cosmologies drawn from AbacusSummit (first block); and noiseless theory vectors with varying $f_b$ and all other parameters fixed to the EDE cosmology (second block). 
The format is the same as Table~\ref{tab:full-fit-results}.
\label{tab:full-shape-noiseless}}
\end{table*}

\begin{table*}[]
    \centering
    \begin{tabular}{l|cccc|cccc|cccc}
    Data & $\gamma_\text{B}$ & $\sigma_{\gamma}$
     & $f_b$ & $n\sigma$ &
      $\alpha_{\perp}$ & $\sigma_{\alpha_{\perp}}$ &
      True &
     $n\sigma$ &
     $\alpha_{\parallel}$ & $\sigma_{\alpha_{\parallel}}$ & True &
     $n\sigma$  \\    
    \hline
Cos.\ 0 & 0.148 & 0.026 & 0.156 & -0.29 & 0.986 & 0.014 & 0.988 & -0.16 & 1.003 & 0.041 & 0.991 & 0.32 \\
Cos.\ 1 & 0.119 & 0.025 & 0.132 & -0.45 & 0.988 & 0.018 & 0.993 & -0.40 & 1.000 & 0.057 & 0.982 & 0.50 \\
Cos.\ 2 & 0.151 & 0.026 & 0.159 & -0.26 & 0.992 & 0.015 & 0.995 & -0.22 & 1.010 & 0.044 & 0.997 & 0.34 \\
Cos.\ 3 & 0.142 & 0.023 & 0.150 & -0.27 & 0.993 & 0.013 & 0.996 & -0.22 & 1.003 & 0.038 & 0.994 & 0.27 \\
Cos.\ 4 & 0.150 & 0.026 & 0.156 & -0.22 & 0.966 & 0.013 & 0.968 & -0.17 & 0.984 & 0.036 & 0.978 & 0.17 \\
Cos.\ 5 & 0.152 & 0.024 & 0.157 & -0.18 & 0.988 & 0.013 & 0.990 & -0.19 & 1.004 & 0.038 & 0.993 & 0.29 \\
Cos.\ 6 & 0.126 & 0.023 & 0.146 & -0.70 & 1.010 & 0.016 & 1.015 & -0.35 & 1.022 & 0.048 & 1.004 & 0.52 \\
Cos.\ 7 & 0.140 & 0.026 & 0.149 & -0.30 & 0.990 & 0.017 & 0.993 & -0.23 & 1.003 & 0.050 & 0.991 & 0.34 \\
Cos.\ 8 & 0.140 & 0.024 & 0.153 & -0.45 & 0.976 & 0.013 & 0.979 & -0.24 & 0.993 & 0.035 & 0.984 & 0.26 \\
Cos.\ 9 & 0.130 & 0.022 & 0.143 & -0.48 & 0.990 & 0.015 & 0.994 & -0.30 & 1.001 & 0.043 & 0.989 & 0.35 \\
    \end{tabular}
    \caption{Template fits to noiseless theory vectors for the pre-reconstruction correlation function, using ten $\Lambda$CDM noiseless theory vectors with varying parameters, all fit with the same fixed-cosmology template.
    \label{tab:pre-recon_noiseless}}
\end{table*}

\begin{table*}[]
    \centering
    \begin{tabular}{l|cccc|cccc|cccc}
    Data & $\gamma_\text{B}$ & $\sigma_{\gamma}$
     & $f_b$ & $n\sigma$ &
      $\alpha_{\perp}$ & $\sigma_{\alpha_{\perp}}$ &
      True  &
     $n\sigma$ &
     $\alpha_{\parallel}$ & $\sigma_{\alpha_{\parallel}}$ & True  &
     $n\sigma$  \\    
    \hline
Cos.\ 0 & 0.159 & 0.025 & 0.156 & 0.12 & 0.987 & 0.010 & 0.988 & 
-0.06 & 0.992 & 0.022 & 0.991 & 0.07 \\
Cos.\ 1 & 0.119 & 0.019 & 0.132 & -0.58 & 0.991 & 0.013 & 0.993 & -0.26 & 0.983 & 0.028 & 0.982 & 0.10 \\
Cos.\ 2 & 0.163 & 0.026 & 0.159 & 0.18 & 0.994 & 0.010 & 0.995 & -0.13 & 0.999 & 0.020 & 0.997 & 0.09 \\
Cos.\ 3 & 0.159 & 0.025 & 0.150 & 0.37 & 0.996 & 0.010 & 0.996 & -0.02 & 0.994 & 0.022 & 0.994 & 0.03 \\
Cos.\ 4 & 0.162 & 0.024 & 0.156 & 0.24 & 0.967 & 0.010 & 0.968 & -0.10 & 0.981 & 0.019 & 0.978 & 0.19 \\
Cos.\ 5 & 0.166 & 0.026 & 0.157 & 0.41 & 0.990 & 0.010 & 0.990 & -0.05 & 0.994 & 0.020 & 0.993 & 0.06 \\
Cos.\ 6 & 0.127 & 0.018 & 0.146 & -0.84 & 1.013 & 0.011 & 1.015 & -0.23 & 1.007 & 0.023 & 1.004 & 0.19 \\
Cos.\ 7 & 0.142 & 0.023 & 0.149 & -0.29 & 0.992 & 0.012 & 0.993 & -0.06 & 0.992 & 0.023 & 0.991 & 0.07 \\
Cos.\ 8 & 0.151 & 0.021 & 0.153 & -0.10 & 0.978 & 0.009 & 0.979 & -0.09 & 0.985 & 0.019 & 0.984 & 0.11 \\
Cos.\ 9 & 0.134 & 0.020 & 0.143 & -0.39 & 0.992 & 0.011 & 0.994 & -0.22 & 0.991 & 0.022 & 0.989 & 0.12 \\
    \end{tabular}
    \caption{Template fits to noiseless theory vectors for the post-reconstruction correlation function, with the same format as Table~\ref{tab:pre-recon_noiseless}.
    \label{tab:post-recon_noiseless}}
    \end{table*}

\section{Differences between our BAO fits and the BOSS DR12 correlation function fits}
\label{sec:compare_to_ross}

We present a detailed comparison of our results and those of the BOSS team \cite{Ross2017} in Table~\ref{tab:compare_to_ross}. Overall, we find that our model fits the correlation functions just as well as the model used in BOSS, with shifts in the AP parameters of $<0.2\sigma$ post-reconstruction and $<0.45\sigma$ pre-reconstruction. Our errorbars are 20--30\% tighter post-reconstruction; similar for pre-reconstruction $\alpha_{\perp}$, and up to 50\% larger for pre-reconstruction $\alpha_{\parallel}$.
For the post-reconstruction fits, this change in errorbars is largely driven by the updated templates; running the BOSS analysis code with the updated templates yields similar errorbars to our baseline constraints and all other choices.
For pre-reconstruction $z_2$ and $z_3$, the relatively large downward shifts of 0.36$\sigma$ and 0.45$\sigma$ in $\alpha_{\perp}$ are driven by our choice to vary the damping parameters, $\gamma_\text{B}$, and the broadband shift parameters, which lead to 0.23$\sigma$ and 0.38$\sigma$ shifts, respectively.
The large increase in pre-reconstruction $\alpha_{\parallel}$ errorbars comes from a variety of sources.

\begin{table*}[]
    \centering
    \small
    \begin{tabular}{l|cccc}
    Redshift bin and model & $\chi^2$ & $\gamma_{\rm B}$   & $\alpha_{\perp}$ & $\alpha_{\parallel}$ \\
    \hline
    \textbf{z1} & & & &   \\
    \hline
    Pre-Recon & 38 &  $0.156 \pm 0.038$ & $0.979 \pm 0.021$ & $1.051 \pm 0.041$ \\
    Pre-Recon, add $1/r$ and $1/r^2$ & 34 &  & $0.974 \pm 0.018$ & $1.057 \pm 0.035$ \\
    Pre-Recon, add $1/r$ and $1/r^2$, fix $\Sigma$, $\gamma_\text{B}$, $q_{\parallel}$, $q_{\perp}$ & 49 &  & $0.972 \pm 0.024$ & $1.047 \pm 0.060$ \\
    Ross et al.\ (2017) + updated templates & 36 & & $0.983 \pm 0.020$ & $1.053 \pm 0.038$ \\
    Ross et al.\ (2017) & 37 & & $0.983 \pm 0.022$ & $1.051 \pm 0.037$ \\
    \hline
    Post-Recon & 40 & $0.175 \pm 0.037$ & $0.985 \pm 0.012$ & $1.021 \pm 0.026$ \\
    Post-recon, add $1/r$ and $1/r^2$ & 38 & & $0.982 \pm 0.012$ & $1.026 \pm 0.027$ \\
    Post-recon, add $1/r$ and $1/r^2$, fix $\Sigma$, $\gamma_\text{B}$, $q_{\parallel}$, $q_{\perp}$ & 39 & & $0.983 \pm 0.012$ & $1.027 \pm 0.026$ \\
    Ross. et al.\ (2017) + updated templates & 39 & & $0.984 \pm 0.013$ & $1.026 \pm 0.027$   \\
    Ross et al.\ (2017) & 42 & & $0.988 \pm 0.015$ & $1.022 \pm 0.027$ \\
    \hline
    \textbf{z2} & & & &   \\
    \hline
    Pre-Recon & 43 & $0.156 \pm 0.040$ & $1.00 \pm 0.026$ & $1.022 \pm 0.061$ \\
    Pre-Recon, add $1/r$ and $1/r^2$ & 43 &  & $1.001 \pm 0.018$ & $1.019 \pm 0.050$ \\
    Pre-Recon,  add $1/r$ and $1/r^2$, fix $\Sigma$, $\gamma_\text{B}$, $q_{\parallel}$, $q_{\perp}$ & 45 &  & $1.006 \pm 0.022$ & $1.027 \pm 0.054$ \\
    Ross et al.\ (2017) + updated templates & 42 & & $1.007 \pm 0.020$ & $1.021 \pm 0.048$ \\
    Ross et al.\ (2017) & 42 & & $1.008 \pm 0.022$ & $1.024 \pm 0.042$ \\
    \hline
    Post-Recon & 25 & $0.162 \pm 0.032$ & $0.993 \pm 0.011$ & $0.986 \pm 0.020$ \\  
    Post-Recon, add $1/r$ and $1/r^2$  & 22 &  & $0.992 \pm 0.011$ & $0.987 \pm 0.020$ \\
    Post-recon, add $1/r$ and $1/r^2$, fix $\Sigma$, $\gamma_\text{B}$, $q_{\parallel}$, $q_{\perp}$ & 26 & & $0.992 \pm 0.011$ & $0.988 \pm 0.022$ \\
    Ross et al.\ (2017) + updated templates & 24 & & $0.994 \pm 0.012$  & $0.988 \pm 0.021$ \\
    Ross et al.\ (2017) & 30 & & $0.994 \pm 0.014$ & $0.984 \pm 0.023$ \\
    \hline
    \textbf{z3} & & & &   \\
    \hline
    Pre-Recon & 29 & $0.177 \pm 0.044$ & $0.990 \pm 0.024$ & $0.964 \pm 0.053$ \\
    Pre-Recon, add $1/r$ and $1/r^2$ & 28 &  & $0.991 \pm 0.023$ & $0.960 \pm 0.056$ \\
    Pre-Recon, add $1/r$ and $1/r^2$, fix $\Sigma$, $\gamma_\text{B}$, $q_{\parallel}$, $q_{\perp}$ & 33 &  & $0.999 \pm 0.025$ & $0.961 \pm 0.046$ \\
    Ross et al.\ (2017) + updated templates  & 28 & & $0.997 \pm 0.023$ & $0.958 \pm 0.042$ \\
    Ross et al.\ (2017) & 28 & & $1.001 \pm 0.024$ & $0.953 \pm 0.035$ \\
    \hline
    Post-Recon & 31 & $0.204 \pm 0.042$ & $0.994 \pm 0.012$ & $0.953 \pm 0.019$ \\  
    Post-Recon, add $1/r$ and $1/r^2$ & 27 &  &  $0.993 \pm 0.013$ & $0.955 \pm 0.019$ \\
    Post-Recon, add $1/r$ and $1/r^2$, fix $\Sigma$, $\gamma_\text{B}$, $q_{\parallel}$, $q_{\perp}$ & 30 & &  $0.992 \pm 0.013$ & $0.957 \pm 0.023$ \\
    Ross et al.\ (2017) + updated templates & 28 & & $0.995 \pm 0.014$  & $0.957 \pm 0.021$ \\    
    Ross et al.\ (2017) & 32 & &  $0.995 \pm 0.016$ & $0.958 \pm 0.023$ \\
    \end{tabular}
    \caption{Differences between our template-fitting results on the AP parameters and the BOSS results from Ross et al.\ (2017) \cite{Ross2017}.
    Within each redshift bin and pre or post-recon block, the top line is our baseline result and the bottom line is the one from Ross et al.\ (2017).
    We then vary our method by adding polynomials $1/r$ and $1/r^2$ to the monopole and quadrupole; fixing the extra parameters that we allow to vary (damping, $\gamma_\text{B}$, and broadband shifts); and then comparing to re-running the publicly available BOSS code with updated templates. $\gamma_\text{B}$ is only varied in our baseline fits and the line below it, and hence not shown for other cases; for the case with the $1/r$ and $1/r^2$ terms, we do not show the $\gamma_\text{B}$ constraint as it is very poor due to degeneracies with the polynomial parameters.
    \label{tab:compare_to_ross}}
\end{table*}



\end{document}